\documentclass [review]{elsarticle} 

\usepackage{lineno,hyperref}
\modulolinenumbers[5]
\usepackage{float}
\usepackage{graphicx}
\usepackage{subfig}
\usepackage{mathrsfs}
\usepackage{graphics}
\usepackage{float}
\usepackage{amsmath}
\usepackage{amsfonts}
\usepackage{optidef}
\usepackage[subnum]{cases}

\usepackage{xcolor}

\usepackage{xcolor}
\hypersetup{
    colorlinks,
    linkcolor={red!50!red},
    citecolor={blue!50!blue},
    urlcolor={blue!80!blue}
}

\usepackage{optidef}
\usepackage{graphicx}
\usepackage{float}
\usepackage{graphicx}
\usepackage{subfig}
\usepackage{mathrsfs}
\usepackage{float}
\usepackage{amsmath}
\usepackage{amsfonts}
\usepackage{booktabs}
\usepackage{algorithm}
\usepackage[noend]{algpseudocode}
\usepackage{multirow}
\algdef{SE}[DOWHILE]{Do}{doWhile}{\algorithmicdo}[1]{\algorithmicwhile\ #1}

\usepackage[letterpaper, portrait, margin=1in]{geometry}

\usepackage{optidef}

\usepackage{amsthm}

\theoremstyle{definition}

\makeatletter
\newcommand*{\centerfloat}{%
  \parindent \z@
  \leftskip \z@ \@plus 1fil \@minus \marginparwidth
  \rightskip \leftskip
  \parfillskip \z@skip}
\makeatother

\usepackage{setspace}
\usepackage{natbib}
\usepackage{algorithm}
\usepackage[noend]{algpseudocode}
\algdef{SE}[DOWHILE]{Do}{doWhile}{\algorithmicdo}[1]{\algorithmicwhile\ #1}

\usepackage{appendix}

\usepackage{array}
\usepackage{threeparttable}
\usepackage{multirow}
\usepackage{rotating}
\usepackage{makecell}

\usepackage{subfig}

\usepackage{url}

\newsavebox{\measurebox}

\journal{}

\biboptions{authoryear}

\begin{document}
\begin{frontmatter}

\title{Modeling Epidemic Spreading through Public Transit using Time-Varying Encounter Network}


\author[label1]{Baichuan Mo\corref{coauthornote}}\ead{baichuan@mit.edu}
\cortext[coauthornote]{Both authors contributed to this paper equally.}
\author[label2]{Kairui Feng\corref{coauthornote}}\ead{kairuif@princeton.edu}
\author[label3]{Yu Shen\corref{mycorrespondingauthor}}\ead{yshen@tongji.edu.cn}
\author[label4]{Clarence Tam}\ead{clarence.tam@nus.edu.sg}
\author[label5]{Daqing Li}\ead{daqingl@buaa.edu.cn}
\author[label6]{Yafeng Yin}\ead{yafeng@umich.edu}
\author[label7]{Jinhua Zhao}\ead{jinhua@mit.edu}
\address[label1]{Department of Civil and Environmental Engineering, Massachusetts Institute of Technology, Cambridge, MA 02139}
\address[label2]{{Department of Civil and Environmental Engineering, Princeton University, Princeton, NJ 08540}}
\address[label3]{{Key Laboratory of Road and Traffic Engineering of the Ministry of Education, Tongji University, Shanghai 201804, China}}
\address[label4]{Saw Swee Hock School of Public Health, National University of Singapore, Singapore 117549}
\address[label5]{School of Reliability and Systems Engineering, Beihang University, Beijing 100191, China}
\address[label6]{Department of Civil and Environmental Engineering, University of Michigan, Ann Arbor, MI 48108}
\address[label7]{Department of Urban Studies and Planning, Massachusetts Institute of Technology, Cambridge, MA 02139}
\cortext[mycorrespondingauthor]{Corresponding author}

\begin{abstract}

Passenger contact in public transit (PT) networks can be a key mediate in the spreading of infectious diseases. This paper proposes a time-varying weighted PT encounter network to model the spreading of infectious diseases through the PT systems. Social activity contacts at both local and global levels are also considered. We select the epidemiological characteristics of coronavirus disease 2019 (COVID-19) as a case study along with smart card data from Singapore to illustrate the model at the metropolitan level. A scalable and lightweight theoretical framework is derived to capture the time-varying and heterogeneous network structures, which enables to solve the problem at the whole population level with low computational costs. Different control policies from both the public health side and the transportation side are evaluated. We find that people’s preventative behavior is one of the most effective measures to control the spreading of epidemics. From the transportation side, partial closure of bus routes helps to slow down but cannot fully contain the spreading of epidemics. Identifying ”influential passengers” using the smart card data and isolating them at an early stage can also effectively reduce the epidemic spreading.

\end{abstract}


\end{frontmatter}


\section{Introduction}\label{intro}

Infectious diseases spread through repetitive social contacts \citep{smieszek2009mechanistic, smieszek2009models}, such as at schools \citep{salathe2010high, litvinova2019reactive}, conferences \citep{stehle2011simulation} or workplaces \citep{potter2015modeling}. Past studies proved that human mobility networks---like air transportation \citep{hufnagel2004forecast, colizza2006role, balcan2009multiscale} or waterways \citep{mari2012modelling, gatto2012generalized}---reconstruct the geometry distance by transporting pathogens or contagious individuals more effectively to widespread locations \citep{brockmann2013hidden}, leading to the outbreak of epidemics. Recently, the outbreak of coronavirus disease 2019 (COVID-19) confirmed the strong connection between human mobility network and disease dynamics. The first case of COVID-19 was reported in Wuhan, China, at the beginning of Dec. 2019, and has then quickly spread to the rest of China through airlines and high-speed rail networks during the Spring Festival travel season \citep{wu2020nowcasting}. Besides the transmissions of pathogens to destination local communities via human mobility network, the densely populated urban public transit (PT) network, however, may also become a key mediate in the spreading of influenza-like epidemics with public transport carriers being the location of transmission \citep{sun2013understanding}. 

The PT system in large metropolitan areas plays a key role in serving the majority of urban commuting demand between highly frequented locations, as human trajectories present a high degree of spatiotemporal regularity following simple, reproducible patterns \citep{gonzalez2008understanding}. By the end of 2017, the annual patronage of urban metro systems worldwide increased from 44 billion in 2013 to 53 billion, and in Asia, the systems carry more than 26 billion passengers a year \citep{UITP2018World}. The urban PT system is often framed as a key solution for building sustainable cities with concerns of environment, economy, and society's effectiveness \citep{miller2016public}. But the indoor environment created by crowded metro carriages or buses can also make an infected individual easily transmit the pathogen to others via droplets or airborne routes \citep{xie2007far, yang2009transmissibility}. In recent years, scholars began to turn their attention to the spreading of the epidemic through the urban PT network. 

Rooted in people’s daily behavior regularity, individuals with repeated encounters in the PT network are found to be strongly connected over time, resulting in a dense contact network across the city \citep{sun2013understanding}. Such mobility features lead to great risks for the outbreak of infectious diseases through bus and metro networks to the whole metropolitan area \citep{sun2014efficient, liu2019investigating}. Based on the contact network developed by \cite{sun2013understanding}, a variation of human contact network structures has been proposed to characterize the movement and encounters of passengers in the PT system, which are then used to model the epidemic spreading among passengers \citep{bota2017identifying, bota2017modeling, hajdu2019discovering, el2019mobility}. 

However, previous studies focusing on the human contacts in PT systems often use a static passenger contact network, discarding the time-varying nature of encounters. The aggregation of the time-varying edges into a static version of the network offers useful insights but can also introduce bias \citep{perra2012random, coviello2016predicting}. Recently, \citet{muller2020using} simulated complete individual mobility trajectories based on mobile phone data and calculated the spreading risks of COVID-19 over dynamic encounters in transportation and buildings. They found that successful contact tracing can reduce the reinfection rate by about 30 to 40\%, which suggests the importance in considering time-varying human interactions for epidemic modeling. However, their research mainly focused on testing the effects of general control strategies at the government level (e.g., closing schools, limiting outdoor activities). From the transportation perspective, the operation strategies by transport authorities (especially PT agencies) may also play an important role in controlling epidemics. Recently, a variety of epidemic control strategies in PT systems have been implemented to respond to the outbreak of COVID-19 since late January 2020. For example, in Wuhan, almost all PT services have been shut down since Jan. 24th. In Wuxi, another Chinese big city, except the 22 arterial bus routes kept running with shortened operation hours, all other PT services (roughly 92\% of bus routes) were suspended since Feb. 1st. In Milan, Italy, the PT services were still in operation, but the suspension of PT has been officially proposed with the rapid surge of COVID-19 cases in the Lombardy area. The impacts of these strategies and other possible PT operation strategies (e.g., distributing passengers' departure time, limiting maximum bus load), however, have seldom been carefully explored. 

To fill these gaps, this study proposes a time-varying weighted PT encounter network (PEN) to model the spreading of the epidemic through urban PT systems. The social activity contacts at both local and global levels are also considered. We select the epidemiological characteristics of COVID-19 as the case study along with high-resolution smart card data from Singapore to illustrate the model at the metropolitan level. Different control policies from both the public health side and the transportation side are evaluated. 

In this work, we do not attempt to reproduce or predict the patterns of COVID-19 spreading in Singapore, where a variety of outbreak prevention and control measures have been implemented \citep{MOHCOVID19} and make most of epidemic prediction models invalid. Instead, since the PT systems in many cities share the similar contact network structure despite the differences in urban structures, PT network layouts and individual mobility patterns \citep{qian2020scaling}, this study aims to employ the smart card data and the PT network of Singapore as proximity to the universal PEN to better understand the general spatiotemporal dynamics of epidemic spreading over the PT system, and to evaluate the potential effects of various measures for epidemic prevention in the PT systems, especially from the PT operation angle. 

The main contribution of this paper is threefold:
\begin{itemize}
\item Propose a PT system-based epidemic spreading model using the smart card data, where the time-varying contacts among passengers at an individual level are captured. 
\item Propose a novel theoretical solving framework for the epidemic dynamics with time-varying and heterogeneous network structures, which enables to solve the problem for the whole population with low computational costs. 
\item Evaluate various potential epidemic control policies from both public health side (e.g., reducing infectious rate) and transportation side (e.g., distributing departure time, closing bus routes)

\end{itemize}

The rest of the paper is organized as follows. In Section \ref{method}, we elaborate on the methodology of establishing contact networks and solving the epidemic transmission model. Section \ref{case_study} presents a case study using the smart card data in Singapore to illustrate the general spatiotemporal dynamics of epidemic spreading through the PT system. In Section \ref{conclusion}, conclusions are made and policy implications are offered.






\section{Methodology} \label{method}
\subsection{Network representation} \label{network_repre}
Epidemic spreading is usually modeled based on individual's contact network. The contact network is a undirected graph where the node is individual and the edge shows the potential infection intensity between corresponding individuals. The majority of previous studies investigated the epidemic process in a static network, where the spreading of the disease is virtually frozen on the time scale of the contagion process. However, static networks are only approximations of the real interplay between time scales. Considering daily mobility patterns, no individual is in contact with all the friends simultaneously all the time. On the contrary, contacts are changing in time, often on a time scale that is shorter than the whole spreading process. Real contact networks are thus inherently dynamic, with connections appearing, disappearing, and being rewired with different characteristic time scales, and are better represented in terms of a temporal or time-varying network. Therefore, modeling the epidemic process on PT should be based on a time-varying contact network. 

Although we focus on the contagion process through PT, passengers' social-activity (SA) contacts besides riding the same vehicles are not neglectable. In this study, two components of the contact network are considered: 1) a PEN that is designated to capture the interaction of passengers on PT, 2) and an SA contact network that captures all other interactions among people.

\subsubsection{PT encounter network}
PT encounters are defined as those who have ever stayed in the same vehicle in a PT system. PT passengers' encounter patterns have been studied by \cite{sun2013understanding} through an encounter network, which is an undirected graph with each node representing a passenger and each edge indicating the paired passengers that have stayed in the same vehicle. The network is constructed by analyzing the smart card data, which includes passengers' tap-in/tap-out time, location, and corresponding bus ID. We assume passengers staying in a same vehicle are close enough to trigger infections. Since PEN provides direct encounter information of passengers, it is an ideal tool to investigate the epidemic spreading through PT.   

Extending the work by \cite{sun2013understanding}, we propose a time-varying weighted PEN to model the epidemic process. We first evenly divide the whole study period into different time intervals $t=1,...,T$. The length of each interval is $\tau$. For a specific time interval $t$, consider a weighted graph $G_t(\mathcal{N},\mathcal{E}_t,\mathcal{W}_t)$, where $\mathcal{N}=\{i: i = 1,..,N\}$ is the node set with each node representing an individual, $N$ is the total number of passengers in the system; $\mathcal{E}_t$ is the edge set and $\mathcal{W}_t$ is the weight set.  The edge between $i$ and $j$ ( $i,j \in \mathcal{N} $), denoted as $e_{ij}^{t}$, exists if $i$ and $j$ have stayed in the same vehicle during the time interval $t$. The weight of $e_{ij}^{t}$, denoted as $w_{ij}^{t}$, is defined as $w_{ij}^{t} = \frac{d_{ij}^t}{\tau}$, where $d_{ij}^t$ is the duration of $i,j$ staying in the same vehicle during time interval $t$. By definition, we have $0 \leq w_{ij}^{t} \leq 1$. The weight is used to capture the fact that epidemic transmission is related to the duration of contact.

\subsubsection{Social-activity contact network}\label{normal_cont_net}
In addition to contacts during rides on the PT, passengers may also contact each other during their daily social activities. Given the heterogeneity of passengers' spatial distributions, people may have various possibilities to contact with different people. However, capturing the real connectivity of passengers in social activities requires a richer dataset (e.g., mobile phone, GPS data), which is beyond the scope of this research. In this study, we made the following assumptions to build the SA contact network.
\begin{itemize}

\item \textbf{Global interaction}:  Passengers may interact with any other individuals in the system during a time interval of $t$ with a uniform probability of $\theta^g$.

\item \textbf{Local interaction}: Passengers with same origins or destinations of PT trips may interact with each other during time interval $t$ with a uniform probability $\theta^{\ell}$. Since local interaction is more intense than global interaction, we have $\theta^{\ell} > \theta^g$.
\end{itemize}

For the global interaction, we assume that the contact time for all connected individuals is $\tau$ for a specific time interval if there are no PT and local contacts between them. Otherwise, the contact time should be subtracted by the PT and local contact duration (CD) at that time interval. For the local interaction, the contact time calculation is illustrated by the following example. Consider passenger $i$ with PT trip sequence $\{(O^i_{t_1},D^i_{t_2}), (O^i_{t_3},D^i_{t_4})\}$, where $t_k$ is the time when the passenger board or alight the vehicles. $O^i_{t_k}$ and $D^i_{t_{k'}}$ are the trip origin and destination, respectively. The trip sequence is defined as a sequence of consecutive PT trips where every adjacent trip pair has an interval of fewer than 24 h (e.g., $t_3 - t_2 <$ 24 h). We call the interval between two consecutive PT trips (e.g., $[t_2, t_3]$) as \emph{activity time} hereafter. Since passengers may not stay in the same place between two consecutive trips, we may have $D^i_{t_2} \neq O^i_{t_3}$. We further assume that from time $t_2$ to $t_3$, the passenger spends half of the activity time at $D^i_{t_2}$ and half of the activity time at $O^i_{t_3}$. 

Suppose passenger $j$ has a trip sequence $\{(O^j_{t'_1},D^j_{t'_2}), (O^j_{t'_3},D^j_{t'_4})\}$, and $D^j_{t'_2} = O^i_{t_3}$; and the overlapping time between intervals [$t_2$, $t_3$] and [$t'_2$, $t'_3$] are \emph{not} zero. This means passengers $i$ and $j$ may have local contact because they have stayed in the same place $D^j_{t'_2} = O^i_{t_3}$ (by definition, the probability of having local contact is $\theta^{\ell}$). Recall that we assume that passengers spend half of the activity time at a specific origin or destination. If they have a local contact, then the CD between passengers $i$ and $j$ is calculated as \emph{half} of the overlapping time between interval [$t_2$, $t_3$] and interval [$t'_2$, $t'_3$]. This calculation gives us the \emph{total} CD of $i$ and $j$ at the local interaction level. For example, if $t_2<t'_2<t_3<t'_3$, the total local CD between $i$ and $j$ is $\frac{1}{2}(t_3 - t'_2)$. Analogizing to the PEN, the total local CD can be mapped to each time interval. For example, if $t^*$ is the time boundary for time interval $t$ and time interval $t+1$, and $ t^* -\tau < t'_2< t^* < t_3 < t^* + \tau$. Denote the local CD between $i$ and $j$ for time interval $t$ as $\tilde{d}_{ij}^{\ell,t}$ ($0 \leq \tilde{d}_{ij}^{\ell,t} \leq \tau$). Then we have $\tilde{d}_{ij}^{\ell,t} = t^* - t'_2$ and $\tilde{d}_{ij}^{\ell,t+1} = t_3 - t^*$.

We denote the SA contact network as $\tilde{G}_t(\mathcal{N},\tilde{\mathcal{E}}_{t}^{g}, \tilde{\mathcal{E}}{t}^{\ell},\tilde{\mathcal{W}}^g_t,\tilde{\mathcal{W}}^\ell_t$), where $\tilde{\mathcal{E}}_{t}^{g}$ is the edge set of global interaction; $\tilde{\mathcal{E}}{t}^{\ell}$ is the edge set of local interaction. The edge of global interaction between any $i$ and $j$, denoted as $\tilde{e}_{ij}^{g,t}$, exists with probability $\theta^{g}$ for all $i,j \in \mathcal{N}$. When $i$ and $j$ share the same PT trip origins or destinations during time interval $t$, the edge of local interaction between $i$ and $j$ ( $\tilde{e}_{ij}^{\ell,t}$) exists with probability $\theta^{\ell}$. $\tilde{\mathcal{W}}^g_t$ and $\tilde{\mathcal{W}}^{\ell}_t$ are the weight set for global and local interaction edges, respectively. By the discussion above, we have $\tilde{{w}}^{\ell,t}_{ij} = \frac{\tilde{d}_{ij}^{\ell,t}}{\tau}$ for all $\tilde{{w}}^{\ell,t}_{ij} \in \tilde{\mathcal{W}}^{\ell}_t$  and $\tilde{{w}}^{g,t}_{ij} = 1 - \tilde{{w}}^{\ell,t}_{ij} - {w}^{t}_{ij}$ for all $\tilde{{w}}^{g,t}_{ij} \in \tilde{\mathcal{W}}^g_t$. By definition, the contacts from three sub-networks (local, global, and PT) are mutually exclusive. 

It is worth noting that though some macroscopic simplification is applied, the SA contact network proposed in this paper is capable to deal with passengers' full activity trajectories. Once the exact travel data of each individual is given, we can construct the contact networks based on their locations and contact duration. Therefore, though the local and global contacts in this study are simplified, the proposed framework is general and still works for more detailed data sources.

\subsubsection{Examples}
\begin{figure}[H]
\centering
\includegraphics[width=5 in]{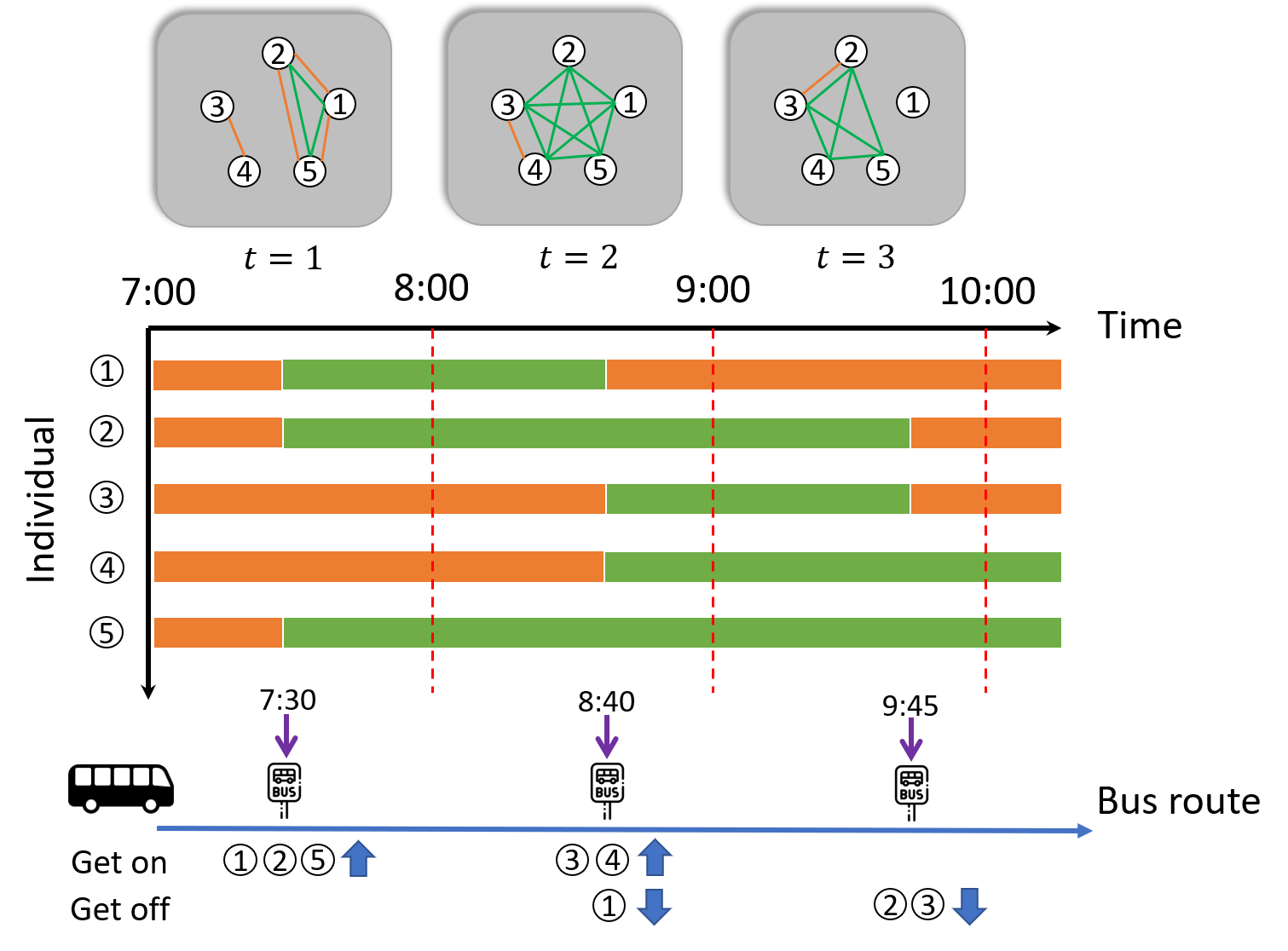}
\caption{Network representation of a five-passenger system}
\label{fig_network_example}
\end{figure}

To illustrate the proposed epidemic contact network, we present a five-passenger system with a single bus route ($N = 5$) in Figure \ref{fig_network_example}. We consider the time period from 7:00 to 10:00 with $\tau = 1$ h. For illustrative purpose, we neglect the global interaction and set the local interaction insensitivity $\theta^{\ell} = 1$. The top of the graph shows the overall time-varying contact networks (i.e. PEN+local) for three time intervals ($t=1,2,3$). The orange (green) edges represent there are SA (PT) contacts between corresponding passengers. The middle of the graph shows the duration of passenger's activity. The orange (green) bar indicates the duration at local activities (at PT trips). The bottom of the graph shows the passengers trajectories along the bus route. 

In the time interval $t=1$, at 7:30, passengers 1,2, and 5 board the bus; since they share the same origin and also are in the same bus during $t=1$, they are connected by the edges of PEN (colored green) and edges of local SA interaction network (colored orange). Accordingly, we have $d_{12}^1 = d_{25}^1 = d_{15}^1 = 0.5$ h. The weights are calculated as $w_{12}^1 = w_{25}^1 = w_{15}^1 = 0.5/1 = 0.5$. In the meanwhile, from the trajectories at $t=2$, we noticed passengers 3 and 4 also share the same origin. Thus, we also have an SA contact edge between $3$ and $4$ at $t=1$. It is worth noting that before 7:00, passengers 1, 2, 5 and passengers 3, 4 may also have the local SA contacts because they share the same origin at 7:30 and we assume they have stayed around the origin from last bus trips (see Section \ref{normal_cont_net} for more details). 

The local CD for passengers 1,2, and 5 at time interval $t=1$ is $\tilde{d}_{12}^{\ell,1} = \tilde{d}_{25}^{\ell,1} = \tilde{d}_{15}^{\ell,1} = \frac{1}{2}\times0.5 = 0.25$ h. The $\frac{1}{2}$ comes from the assumption that these passengers only spend half of their time around this bus station (see Section \ref{normal_cont_net}). Hence, the corresponding weights for the SA contact network are $\tilde{w}_{12}^{\ell,1} = \tilde{w}_{25}^{\ell,1} = \tilde{w}_{15}^{\ell,1} = 0.25/1 = 0.25$. Similarly, we have $\tilde{d}_{34}^{\ell,1} = \frac{1}{2}\times1 = 0.5$ and $\tilde{w}_{34}^{\ell,1} = 0.5$. The weights for $t=2$ and $t=3$ are calculated in the same way.

\subsection{Epidemic transition model}
The epidemic transition model is independent of the network representation. We can model various infectious diseases based on the proposed PEN using different epidemic transition frameworks. For the case study we considered (COVID-19), we employed the Susceptible-Exposed-Infectious-Removed (SEIR) diagram in this study. The SEIR model is generally used to model influenza-like illness and other respiratory infections. For example, \cite{small2005small} used this model to numerically study the evolution of the severe acute respiratory syndrome (SARS), which shares significant similarities with the COVID-19.

\subsubsection{SEIR diagram}
We first divided the population into four different classes/compartments depending on the stage of the disease \citep{anderson1992infectious, diekmann2000mathematical, keeling2007stochastic}: susceptibles (denoted by $S$, those who can contract the infection), exposed ($E$, those who have been infected by the disease but cannot yet transmit it or can only transmit with a low probability), infectious ($I$, those who contracted the infection and are contagious), and removed ($R$, those who are removed from the propagation process, either because they have recovered from the disease with immunization or because they have died). By definition, we have $\mathcal{N} = S \cup E \cup I \cup R$, where $\mathcal{N}$ is the set of the whole population.

\begin{figure}[H]
\centering
\includegraphics[width=4 in]{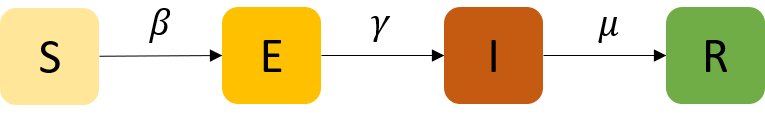}
\caption{Diagram of SEIR}
\label{fig_SEIR}
\end{figure}

The diagram of the SEIR model is shown in Figure \ref{fig_SEIR}. The diagram shows how individuals move through each compartment in the model. The infectious rate, $\beta$, controls the rate of spread and is associated with the probability of transmitting disease between a susceptible ($S$) and an exposed individual ($E$). The incubation rate, $\gamma$, is the rate of exposed individuals ($E$) becoming infectious ($I$). Removed rate, $\mu$, is the combination of recovery and death rates. The SEIR model typically assumes the recovered individuals will not be infected again, given the immunization obtained. It is worth noting that this study focuses on the early stage of an epidemic process, where the impact of outside factors on $\mathcal{N}$ (e.g., birth and natural death) are not considered.

\begin{figure}[H]
\centering
\subfloat[Number of individuals in different classes ($R_0>1$)]{\includegraphics[width=0.4\textwidth]{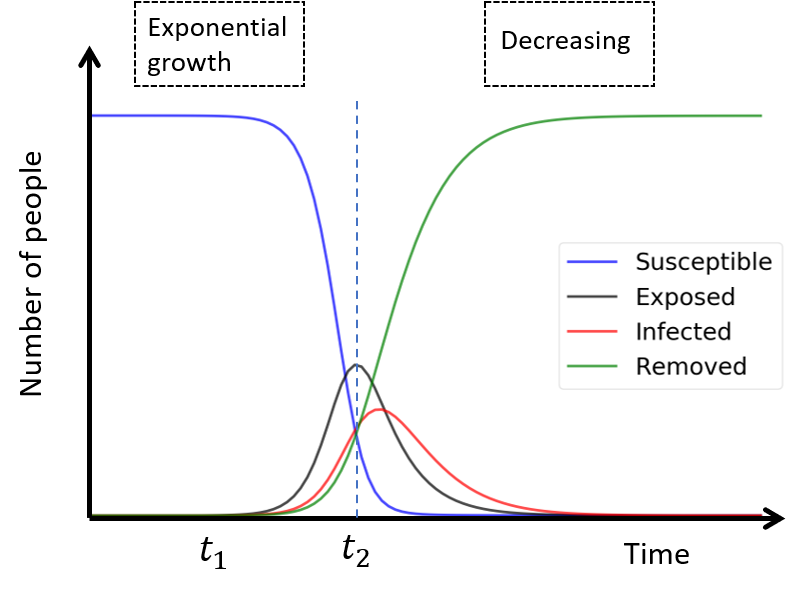}\label{fig_SEIR_infect}}
\hfil
\subfloat[Illustration of epidemic threshold]{\includegraphics[width=0.4\textwidth]{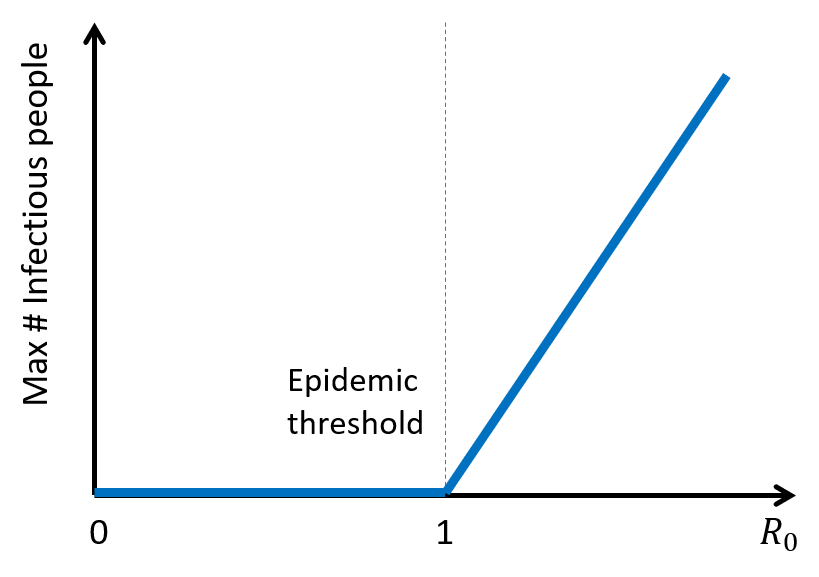}\label{fig_illus_SEIR_infect}}
\hfil
\label{fig_beta_compare}
\caption{Illustration of SEIR model (adapted from \cite{karong2018comparing, pastor2015epidemic})}
\end{figure}

For the epidemic process models, people are concerned about the steady-state, epidemic threshold and reproduction number. According to \cite{pastor2015epidemic}, the number of infected individuals in the SEIR model always tends to zero after a long term (see Figure \ref{fig_SEIR_infect}). This is obvious from the diagram of SEIR (Figure \ref{fig_SEIR}), where there is only one recurrent state $R$. The basic reproduction number, denoted by $R_0$, is defined as the average number of secondary infections caused by a primary case introduced in a fully susceptible population \citep{anderson1992infectious}. In the standard SEIR model, we have $R_0 = \frac{\beta}{\mu}$. Epidemic threshold, in many cases, is defined based on the value of $R_0$. When $R_0 < 1$, the number of infectious individuals tends to decay exponentially; thus, there is no epidemic. However, if $R_0 > 1$, the number of infectious individuals could grow exponentially with an outbreak of epidemic (see Figure \ref{fig_illus_SEIR_infect}). 

\subsubsection{Individual-based approach}
Epidemic modeling can be classified into compartment models and network models. The typical compartment models assume the population are homogeneous and well-mixed, which is not suitable in this study. Thus, the network models are used.

The modeling of epidemic on networks falls into two different categories \citep{pastor2015epidemic}: the individual-based approach and the degree-based approach. Generally, the individual-based approach models the epidemic transmission at the individual level \citep{smieszek2011reconstructing}. While the degree-based approach captures the infection process at the group level, where each group includes a set of nodes (individuals) with the same degree and these nodes are assumed to be statistically equivalent \citep{pastor2015epidemic}. Since in the PEN, passengers with the same degree are not necessarily similar, the individual-based framework is used in this study. 

We denote $S_{i,t}$, $I_{i,t}$, $E_{i,t}$, and $R_{i,t}$ as the Bernoulli random variable that describes whether individual $i$ is in class $S$, $I$, $E$, and $R$ at time interval $t$, respectively (Yes = 1). By definition we have $S_{i,t} + I_{i,t} + E_{i,t} + R_{i,t} = 1$ for all $i$ and $t$. Let $\mathbb{P}(X_{i,t} = 1) = p_{i,t}^X$, where $X \in \{S,I,E,R\}$ and $\sum_X  p_{i,t}^X = 1$. 

Since the contact network is defined in discrete time, we can describe the epidemic process of the SEIR model as a discrete Markov process with specific transition probabilities. To match with the epidemiological characteristics of COVID-19, we assume that the exposed individual can also infect others based on the recent finding \citep{rothe2020transmission}, which may not be the common case in the SEIR model.

Let $\beta_I$ be the probability of a susceptible individual $i\in S$ getting infected by an \emph{infected} individual $j \in I$ at a time interval $t$ if $i$ and $j$ contact each other (either by PT or SA) for the entire time interval. Since the actual transmission probability is related to the interaction duration, we can write the actual probability of $i$ getting infected by $j$ ($\beta^I_{i,j,t}$) as 
\begin{align}
\beta^I_{i,j,t} = a_{ij}^t \cdot h(w_{ij}^t, \beta_I) + \tilde{a}_{ij}^{\ell,t} \cdot h(\tilde{w}_{ij}^{\ell,t}, \beta_I) + \tilde{a}_{ij}^{g,t} \cdot h(\tilde{{w}}^{g,t}_{ij}, \beta_I)  \;\;\;\; \forall i \in S, \; j \in I.
\end{align}
where $h(\cdot, \cdot)$ is a function to describe the transmission probability with respect to CD. It can be a form of survival function (e.g., exponential, Weibull) or a linear function (i.e., $h(w, \beta) = w\beta $, which is used in the case study). $a_{ij}^t$ ($\tilde{a}_{ij}^{\ell,t}$, $\tilde{a}_{ij}^{g,t}$) is an indicator variable showing whether $e_{ij}^t$ ($\tilde{e}_{ij}^{\ell,t}$, $\tilde{e}_{ij}^{g,t}$) exists. It is worth noting that $a_{ij}^t$ is a known constant but $\tilde{a}_{ij}^{\ell,t}$ and $\tilde{a}_{ij}^{g,t}$ are random variables with Bernoulli distribution: $\tilde{a}_{ij}^{\ell,t} \sim \mathcal{B}(L_{ij}^t\theta^{\ell})$ and $\tilde{a}_{ij}^{g,t} \sim \mathcal{B}(\theta^{g})$, where $L_{ij}^t = 1$ if $i$ and $j$ share the same origin or destination at time interval $t$ and $L_{ij}^t = 0$ otherwise (see Section \ref{normal_cont_net} for details). Therefore, we have 
\begin{align}
\beta^I_{i,j,t} = a_{ij}^t \cdot h(w_{ij}^t, \beta_I) + L_{ij}^t\cdot\theta^{\ell} \cdot h(\tilde{w}_{ij}^{\ell,t}, \beta_I) + \theta^{\ell} \cdot h(\tilde{{w}}^{g,t}_{ij}, \beta_I) \;\;\;\; \forall i \in S, \; j \in I.
\label{eq_beta_I_ijt}
\end{align}

Similarly, we define $\beta_E$ as the probability of a susceptible individual $i\in S$ getting infected by an \emph{exposed} individual $j \in E$ at time interval $t$ if $i$ and $j$ contact each other for the entire time interval ($\beta_E \ll \beta_I$). The actual transmission probability considering interaction duration is 
\begin{align}
\beta^E_{i,j,t} = a_{ij}^t \cdot h(w_{ij}^t, \beta_E) + L_{ij}^t\cdot\theta^{\ell} \cdot h(\tilde{w}_{ij}^{\ell,t}, \beta_E) + \theta^{\ell} \cdot h(\tilde{{w}}^{g,t}_{ij}, \beta_E) \;\;\;\; \forall i \in S, \; j \in E.
\label{eq_beta_E_ijt}
\end{align}

Note that if $i$ and $j$ have been in contact, we assume the transmission probability only depends on the CD. The variation of transmission probability due to spatial distribution is neglected in the case study because it requires a more dedicated physical infection model and assumptions about passenger spatial distribution in a vehicle, which is beyond the scope of this study. However, the proposed method can be extended to capture the spatial effect by replacing $h(w_{ij}^t, \beta_I)$ with a dedicated transmission models (e.g., Wells–Riley model \citep{wells1955airborne}) and taking the distance between $i$ and $j$ into consideration. For example, one can replace $h(w_{ij}^t, \beta_I)$ with $h(w_{ij}^t, s_{ij}^t, \beta_I)$, where $s_{ij}^t$ is the distance between $i$ and $j$ in time interval $t$, and it is a function of vehicle loads, vehicle types (e.g., double-decker vs. smallish normal buses), seating distributions, etc. 

\begin{figure}[H]
\centering
\includegraphics[width=3in]{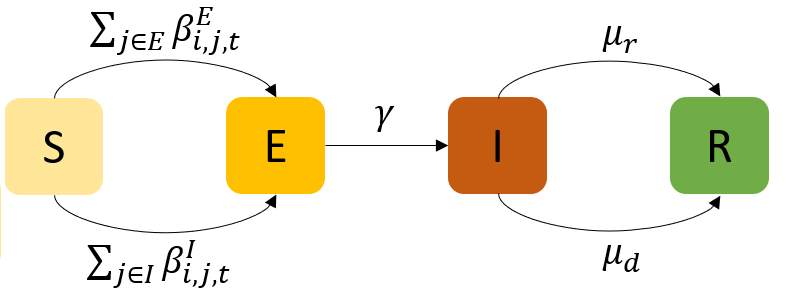}
\caption{Transition graph of individual $i$ at time interval $t$}
\label{fig_transition}
\end{figure}

Let $\gamma$ be the probability of $E \rightarrow I$, which is unrelated to the network. $\mu$ is the probability of $I \rightarrow R$ and $\mu = \mu_r + \mu_d$, where $\mu_r$ and $\mu_d$ are the probability of infectious people getting cured and dying, respectively. Both of them are unrelated to the network. We summarize the transition graph for individual $i$ at time interval $t$ in Figure \ref{fig_transition}.

The notations and epidemic transmission mechanism allow us to write the following system equations:
\begin{align}
p_{i,t+1}^S &= p_{i,t}^S -\sum_{j \in \mathcal{N}} \beta^E_{i,j,t} \mathbb{P}(S_{i,t}=1,E_{j,t}=1) - \sum_{j \in \mathcal{N}} \beta^I_{i,j,t}\mathbb{P}(S_{i,t}=1,I_{j,t}=1) & \label{eq_S}\\ 
p_{i,t+1}^E &= p_{i,t}^E + \sum_{j \in \mathcal{N}} \beta^E_{i,j,t} \mathbb{P}(S_{i,t}=1,E_{j,t}=1) + \sum_{j \in \mathcal{N}} \beta^I_{i,j,t}\mathbb{P}(S_{i,t}=1,I_{j,t}=1) - p_{i,t}^E\gamma & \label{eq_E}\\
p_{i,t+1}^I &= p_{i,t}^I - p_{i,t}^I(\mu_r + \mu_d) + p_{i,t}^E\gamma &\label{eq_I}\\
p_{i,t+1}^R &= p_{i,t}^R + p_{i,t}^I(\mu_r + \mu_d)&\label{eq_R}
\end{align}
Eq. \ref{eq_S} means that the infection probability of an individual in $S$ equals to the probability of infected by surrounding people in $I$ plus the probability of infected by surrounding people in $E$. This comes from the typical assumption of discrete Markov chain that in a time interval the probability of more than one events happening is zero, which holds for this study because the probability of infected by both people in $I$ and people in $E$ in a time interval is small. 

Calculating $\mathbb{P}(S_{i,t}=1,X_{j,t}=1)$ requires the joint distribution of $S_{i,t}$ and $X_{j,t}$, which is usually unavailable. According to the individual-based mean-field approximation, we can assume that the state of neighbors is independent \citep{hethcote2014gonorrhea, chakrabarti2008epidemic, sharkey2008deterministic, sharkey2011deterministic}. Hence, this leads to 
\begin{align}
\mathbb{P}(S_{i,t}=1,E_{j,t}=1) &= p_{i,t}^S p_{j,t}^E \label{eq_ibfm_1}\\ 
\mathbb{P}(S_{i,t}=1,I_{j,t}=1) &= p_{i,t}^S p_{j,t}^I \label{eq_ibfm_2}
\end{align}
By plugging Eqs. \ref{eq_ibfm_1} and \ref{eq_ibfm_2} into Eqs. \ref{eq_S} and \ref{eq_E}, we can get a new group of solvable system equations.

\subsection{Simulation-based solving framework}\label{sim_solve}
Different from the typical SEIR model, the proposed epidemic model with the individual-based PEN has two challenges. First, the infection rate in a typical SEIR model is defined at the population level (i.e., homogeneous network assumption). However, in the proposed framework, we consider one-to-one contagious behaviors at the individual level with heterogeneous contact networks. The heterogeneity is difficult to characterize by probabilistic models (e.g., degree distributions) because contact structures are known from the smart card data. Second, the proposed framework lies on a time-varying network, for which the contagious behaviors and interacted individuals vary over time. 

One of the solution methods for Eqs. \ref{eq_S} - \ref{eq_R} is simulation. Similar to many other complex stochastic process, simulation can output approximate values for $p_{i,t}^X$ for all $X \in \{S,I,E,R\}$ and $t$. The simulation process is described in Algorithm \ref{alg_sim}, where $\bar{p}_{t}^X$ ($X \in \{S,I,E,R\}$) is the proportion of people in class X for time interval $t$. Initialization can assign some seed infectious people in the system. At each time step $t$, we calculate the one-to-one transmission probability $\beta_{i,j,t}^I$ and $\beta_{i,j,t}^E$ for each person in class $S$, where the time complexity is $\mathcal{O}(N^2)$. Therefore, the total time complexity of simulation is $\mathcal{O}(N^2T)$, where $T$ is the total number of time intervals considered. The model also requires to store the network structures and individual states at each time step, where the space complexity is also $\mathcal{O}(N^2T)$. 

\begin{algorithm} 
\caption{Simulation-based solving algorithm} \label{alg_sim}
\begin{algorithmic}[1]
\State Initialize $S_{i,0}$, $E_{i,0}$, $I_{i,0}$, $E_{i,0}$ for all $i=1,2,...,N$.
\For {$t=1,2,...,T$} 
\For {$i=1,2,...,N$} 
    \If {$S_{i,t-1} = 1$}
        \State Calculate $\beta^I_{i,j,t}$ and $\beta^E_{i,j,t}$
        \State Assign $E_{i,t} = 1$ with probability $\sum_{j\in I} \beta^I_{i,j,t} + \sum_{j\in E} \beta^E_{i,j,t}$. 
        \State Let $S_{i,t} = 1 - E_{i,t}$, and $I_{i,t} = 0$, $R_{i,t} = 0$.
    \ElsIf{$E_{i,t-1} = 1$}
        \State Assign $I_{i,t} = 1$ with probability $\gamma$.
        \State Let $E_{i,t} = 1 - I_{i,t}$, and $S_{i,t} = 0$, $R_{i,t} = 0$.
    \ElsIf{$I_{i,t-1} = 1$}
        \State Assign $R_{i,t} = 1$ with probability $\mu = \mu_r + \mu_d$
        \State Let $I_{i,t} = 1 - R_{i,t}$, and $S_{i,t} = 0$, $E_{i,t} = 0$.
    \Else
        \State Assign $R_{i,t} = 1$, and $I_{i,t} = 0$, $R_{i,t} = 0$, $S_{i,t} = 0$.
    \EndIf
    \State $\bar{p}_{t}^X = (1/N) \sum_{i  \in \mathcal{N}} X_{i,t}$ for all $X \in \{S,I,E,R\}$.
\EndFor
\EndFor
\Return $\bar{p}_{t}^X$ for all $t \in \{1,2,..., T\}$, and $X \in \{S,I,E,R\}$.
\end{algorithmic}
\end{algorithm}

Given the $\mathcal{O}(N^2T)$ of time and space complexity, the simulation-based solving framework is hard to scale up to the whole population level (4.7 million in our case study). As we code the calculation process by matrix operation (to improve computing efficiency), large $N$ will easily occupy large memory and makes the matrix operation unavailable. According to the numerical results, $N \geq $ 300k would cause memory errors to a 32 GB RAM personal computer.  

It is worth noting that there exists a sparse representation of the contact network (i.e. only storing the edges $(i,j)$ when $i$ and $j$ are interacted, rather than the whole network). In this way, we only have the space and time complexity of $\mathcal{O}(\langle k \rangle NT)$, where $\langle k \rangle$ is the average degree of the contact network, which may simplify the simulation. However, generating this sparse representation needs a merge or join operation of two lists of all individuals. And this operation still requires a space complexity in $\mathcal{O}(N^2T)$. Hence, future research can be done on optimizing the algorithms of constructing contact networks for large sample sizes.  


\subsection{Theoretical  framework}
Considering computational costs, a scalable and lightweight theoretical model is proposed to handle the epidemic calculation. The theoretical model, though simplified from agent-level simulation, can retain the flexibility to capture the behavioral, mechanical, networked, and dynamical features of the simulation-based models. 

The framework is separated into three steps. 1) We first build up a multi-particle dynamics model for the epidemic process to represent the individual-based model. 2) Considering properties of contact network and multi-particle dynamics \citep{gao2016universal}, an effective model is employed to represent the multi-dimensional dynamics (individual-based) into one-dimension (mean-based). 3) Previous effective models are developed for static network structure. To fit with the time-varying contact network, we innovatively combine the effective model with a temporal network model by adding energy flow into the equations, from which we can capture the impact of the time-varying contact networks on the entire dynamical system. 

For the multi-particle dynamics model, we focus on the early stage of the epidemic process, where the percent of susceptible people is almost 100\%, and recovered people are 0\%. Hence, we could use Taylor expansion to simplify the four-dimension ($S$,$E$,$I$, and $R$) individual dynamical epidemic process described in Eq. (\ref{eq_S} - \ref{eq_R}) to two dimensions ($E$ and $I$):
\begin{align}
p_{i,t+1}^E - p_{i,t}^E &= \sum_{j \in \mathcal{N}} \beta^E_{i,j,t} p_{j,t}^E + \sum_{j \in \mathcal{N}} \beta^I_{i,j,t} p_{j,t}^I - p_{i,t}^E \gamma & \label{eq_E1}\\
p_{i,t+1}^I - p_{i,t}^I &= - p_{i,t}^I(\mu_r + \mu_d) + p_{i,t}^E \gamma &\label{eq_I1}
\end{align}

In this formula, we embedded the dynamical network structure into two tensors $[\beta^E_{i,j,t}]_{i,j,t} \in \mathbb{R}^{N \times N \times T}$ and $[\beta^I_{i,j,t}]_{i,j,t} \in \mathbb{R}^{N \times N \times T}$. These two tensors are non-negative, and each temporal slice of these tensors (i.e., $\beta^I_t =[\beta^I_{i,j,t}]_{i,j}  \in \mathbb{R}^{N \times N}$ and $\beta^E_t =[\beta^E_{i,j,t}]_{i,j}  \in \mathbb{R}^{N \times N}$) are symmetric due to the property of infection. According to the previous studies \citep{gao2016universal, tu2017collapse}, the infectious burst in this canonical system could be captured by a one-dimensional simplification of the individual-based model. This simplification is based on the fact that in a network environment, the state of each node is affected by the state of its immediate neighbors. More details on the simplification can be found in \cite{gao2016universal} and \cite{tu2017collapse}. Therefore, we can characterize the effective state of the system using the average nearest-neighbor activity:
\begin{align}
p_{\text{eff}, t}^E = \frac{\mathbf{1}^T \beta^E_t p^E_t}{\mathbf{1}^T \beta^E_t \mathbf{1}} &= \frac{\sum_{i\in \mathcal{N}}\sum_{j\in \mathcal{N}}\beta_{i,j,t}^{E} p^E_{j,t}}{\sum_{i,j\in \mathcal{N}}\beta_{i,j,t}^{E}} & \label{eq_peeff} \\
p_{\text{eff}, t}^I = \frac{\mathbf{1}^T \beta^I_t p^I_t}{\mathbf{1}^T \beta^I_t \mathbf{1}} &= \frac{\sum_{i\in \mathcal{N}}\sum_{j\in \mathcal{N}}\beta_{i,j,t}^{I} p^I_{j,t}}{\sum_{i,j\in \mathcal{N}}\beta_{i,j,t}^{I}} & \label{eq_pieff} 
\end{align}
where $p^E_t = [ p^E_{i,t}]_i \in \mathbb{R}^N$ ($p^I_t = [ p^I_{i,t}]_i \in \mathbb{R}^N$) is the vector of individuals' exposed (infectious) probability. $\mathbf{1} \in \mathbb{R}^{N}$ is the unit vector. $p_{\text{eff}, t}^E \in \mathbb{R}$ ($p_{\text{eff}, t}^I \in \mathbb{R}$) is the \emph{effective} proportion of exposed (infectious) people in the system at time interval $t$. If we assume that all individuals hold a uniform probability to come into contact with each other, $p_{\text{eff},t}^E$ and $p_{\text{eff},t}^I$ are good proxies for $\bar{p}^E_t$ and $\bar{p}^I_t$, where  $\bar{p}^E_t$ and $\bar{p}^I_t$  are the actual proportion of exposed and infectious population (i.e., $\bar{p}^E_t = \frac{1}{N} \sum_{i \in \mathcal{N}} {p}^E_{i,t}$ and $\bar{p}^I_t = \frac{1}{N} \sum_{i \in \mathcal{N}} {p}^I_{i,t}$). However, this assumption may not hold in reality. The relaxation of the assumption will be described later. $p_{\text{eff},t}^E$ and $p_{\text{eff},t}^I$ allow us to reduce the individual-based equations (Eq. \ref{eq_E1} and \ref{eq_I1}) to an effective mean-based equations:
\begin{align}
p_{\text{eff},t+1}^E - p_{\text{eff},t}^E &= \beta_{\text{eff},t}^E \cdot p_{\text{eff},t}^E + \beta_{\text{eff},t}^I \cdot p_{\text{eff},t}^I - p_{\text{eff},t}^E \gamma & \label{eq_E2}\\
p_{\text{eff},t+1}^I - p_{\text{eff},t}^I &= -p_{\text{eff}, t}^I (\mu_r + \mu_d) + p_{\text{eff}, t}^E \gamma &\label{eq_I2}
\end{align}
where:
\begin{align}
    \beta_{\text{eff},t}^X = \frac{\sum_{i,j\in \mathcal{N}} (\beta_{i,j,t}^{X})^2 }{\sum_{i,j\in \mathcal{N}} \beta_{i,j,t}^{X}}, \qquad \forall X \in \{E, I\} & \label{network_eff}
\end{align}

Considering that people's interaction probabilities are actually heterogeneous, in practice, to relax the uniform contact assumption, we further consider the dynamics of the mobility network on the multi-particle systems based on \cite{li2017fundamental}, which recommends adding the energy flow $f^X_t = (1/N^2) \sum_{i,j\in \mathcal{N}} (\beta_{i,j,t}^{X})^2$ ($ X \in \{E, I\}$) into the general dynamical process:
\begin{align}
p_{\text{eff},t+1}^E - p_{\text{eff},t}^E &= \beta_{\text{eff},t}^E \cdot p_{\text{eff},t}^E + \beta_{\text{eff},t}^I \cdot p_{\text{eff},t}^I - p_{\text{eff},t}^E \gamma + k^E_E \cdot f^E_t + k^E_I \cdot f^I_t & \label{eq_E3} \\
p_{\text{eff},t+1}^I - p_{\text{eff},t}^I &= -p_{\text{eff}, t}^I (\mu_r + \mu_d) + p_{\text{eff}, t}^E \gamma + k^I_E \cdot f^E_t + k^I_I \cdot f^I_t  & \label{eq_I3}
\end{align}
where $\mathcal{K} = [k^E_E$, $k^E_I$, $k^I_E$, $k^I_I]$ are parameters to be estimated. The energy flow and corresponding parameters are expected to capture the heterogeneous contacts in the network.

The theoretical model is calibrated from a two-layer regression method. In the first layer, given a trajectory of epidemic process: $[(\bar{p}^E_t, \bar{p}^I_t)]_{t = 1,...,T}$, we can replace $p_{\text{eff},t}^E, p_{\text{eff},t}^I$ with $\bar{p}^E_t, \bar{p}^I_t$ in Eq. \ref{eq_E3} and \ref{eq_I3}. This leads to a linear regression problem with a total of $T$ samples:
\begin{align}
\bar{p}^E_{t+1} - \bar{p}^E_t &= \beta_{\text{eff},t}^E \cdot \bar{p}^E_t + \beta_{\text{eff},t}^I \cdot \bar{p}^I_t - \bar{p}^E_t \gamma + k^E_E \cdot f^E_t + k^E_I \cdot f^I_t & t=1,2,...,T &\label{eq_linearE} \\
\bar{p}^I_{t+1} - \bar{p}^I_{t} &= -\bar{p}^I_{t} ( \mu_r + \mu_d) + \bar{p}^E_{t} \gamma + k^I_E \cdot f^E_t + k^I_I \cdot f^I_t  & t=1,2,...,T  & \label{eq_linearI}
\end{align}
where the only unknown parameters are $\mathcal{K}$; $\beta_{\text{eff},t}^X$ and $f^X_t$ are calculated from the constructed contact network, and $\bar{p}^X_t$ is given ($ X \in \{E, I\}$). Therefore, $\mathcal{K}$ can be obtained for every given epidemic trajectory. The epidemic trajectory is generated using the simulation-based solving framework for a small sample size (e.g., 100k).

In the second layer, we aim to obtain the relationship between $\mathcal{K}$ and epidemic/mobility parameters (i.e., $\Theta = [\beta_E$, $\beta_I$, $\gamma$, $\mu_r$, $\mu_d$, $\theta^{\ell}$, $\theta^{g}]$). For every combination of $\Theta$, we can use the simulation model to generate a trajectory and thus to estimate $\mathcal{K}$ (as we described above, the first layer regression). Therefore, based on different values of $\Theta$, we can estimate a series of $\mathcal{K}$. Then, we assume a linear relationship between $\Theta$ and $\mathcal{K}$, and use a linear regression model to fit the relationship based on the generated $\Theta$, $\mathcal{K}$ pairs. After the two-layer regression, the theoretical model can be used to flexibly predict the epidemic process under different policy conditions (i.e., different $\Theta$, $\beta^E$, or $\beta^I$). 

The theoretical model can smoothly consider different sizes of the studied population by scaling $\beta_{\text{eff},t}^E$, $\beta_{\text{eff},t}^I$, and energy flows $f^I_t, f^E_t$. Those parameters can be obtained once the contact network was constructed (see Eq. \ref{network_eff}). The scale factor can be computed empirically by comparing the dynamics (i.e. epidemic trajectories) established on sample passengers and the whole population. A time-based pair-wisely scaling technique is applied for each $y_t \in \Phi_t = \{\beta_{\text{eff},t}^E, \beta_{\text{eff},t}^I, f^I_t, f^E_t\}$. Let the corresponding parameters calculated in sample passengers be $y_t(\text{Sample})$ and in whole population be $y_t(\text{Pop})$, where $y_t \in \Phi_t$. Then the empirical scaling factor for $y_t$ can be calculated as: 
\begin{align}
s_t(y_t) = \frac{y_t(\text{Pop})}{y_t(\text{Sample})} \;\;\;\;\text{for all } y_t \in \Phi_t \text{ and } t=1,2,...,T &
\label{eq_scale} 
\end{align}

The empirical scaling factor is not directly the sampling rate, but also depend on the dynamical processes and co-variances between individual travel behaviors. It is worth noting that though the generation of $y_t(\text{Pop})$ may be computational inefficient, it only needs to be calculated once. After obtaining $s_t(y_t)$ and enforcing this scaling to different policy experiments, we can obtain different dynamical behaviors of infectious from a small sampled data set (i.e. using sample data to generate $y_t(\text{Sample})$ and scale it). 

Thus, this model can efficiently test different policy combinations with a low computational cost. According to the numerical test, it can evaluate one-million policy combinations for the full population (4.7 million) within seconds. The memory and computational complexity are both $\mathcal{O}(T)$. This allows us to find the optimal policy to control the contagion.


\subsection{Reproduction number ($R_0$)}
In epidemiology, the basic reproduction number (expressed as $R_0$) of the infection can be viewed as the expected number of cases directly generated by one case in a population where all individuals are susceptible to infection \citep{fraser2009pandemic}. The most important use of $R_0$ is to determine whether an emerging infectious disease would spread throughout the population. In a common infection model, if $ R_0 > 1 $, the infection starts to spread throughout the population, but not if $ R_0 <1 $ (see Figure \ref{fig_illus_SEIR_infect}). In general, the larger the value of $ R_0 $, the more difficult it is to control the epidemic \citep{fine2011herd}.

In the ideal SEIR model where diseases spread uniformly over time and people have a uniform contact probability, $R_0$ is easy to define. To match with the discrete-time expression in this study, let $\hat{\beta}$ be the average number of people infected by one infectious person within one time interval in the ideal SIR system, and $\hat{\mu}$ be the probability that the infectious people (I) are removed (R) within one time interval (Note that in continuous time context, $\hat{\beta}$ and $\hat{\mu}$ represents infectious rate and removal rate, respectively). The basic reproduction number for ideal SEIR model is calculated as 
\begin{align}
     R_{0}=\frac{\beta}{\mu} = \sum_{t = 0}^{\infty} \beta (1-\mu)^t = \sum_{t = 0}^{\infty} \frac{I_{t+1} - I_t}{I_t}(1-\mu)^t \label{eq_R0} 
\end{align}
where $I_t$ is the number of infectious people at time interval $t$. Note that $\frac{I_{t+1} - I_t}{I_t} = \beta, \; \forall t$ only holds under the ideal SEIR system.

In heterogeneous populations, the definition of $R_0$ is more subtle. The definition must take into account the fact that the contact between people is not uniform. One person may only contact a small group of friends and be isolated from the rest of the population. On the temporal evolving side, people's mobility patterns may vary every day, resulting in time-varying contact networks. This defeats ideal SEIR system assumptions. To consider the network heterogeneity \cite{damgaard1995social}, we define $ R_0 $ as "the expected number of secondary cases of a typical infected person in the early stages of an epidemic”, which focuses on the expected \emph{direct} infected population for each time step at the early stage

Let $E_t$ and $I_t$ be the number of exposed and infectious people at time interval $t$. Given a trajectory of epidemic process $[(E_t, I_t)]_{t = 1,...,T}$ (either from the simulation model or the theoretical model), we define the \emph{equivalent} reproduction number for time period $T$ ($R_0(T)$) as:
\begin{align}
    R_0(T) = \sum_{t=0}^{T} \frac{E_{t+1} - E_t }{I_t} \cdot (1-\mu)^t \label{eq_equval_R0}
\end{align}
We assume that the incubation period $\frac{1}{\gamma}$ is much longer than a time interval, which holds true for most of the diseases (e.g., in our case study, 4 days $\gg$ 1 h). Therefore, $E_{t+1} - E_t$ in this study is a good proxy for $I_{t+1} - I_t$ in the ideal SEIR model. Comparing with the original definition in Eq. \ref{eq_R0}, if we take $T$ to $\infty$ and assume that $\frac{E_{t+1} - E_t }{I_t}$ is a constant for all $t$, Eqs. \ref{eq_equval_R0} and \ref{eq_R0} are equivalent. This property illustrates the reasonableness of the defined equivalent $R_0$. 

The equivalent $R_0$ enables flexible and fair comparison of different epidemic processes with heterogeneous contact and time-varying networks. In the following sections, we will use the defined equivalent $R_0$ as a major epidemic measure for different policy discussions.





\section{Case study} \label{case_study}
We use the Singapore bus system as a proximity to demonstrate the dynamics of epidemic spreading through a PT network. The time-varying PEN is constructed based on daily mobility patterns in the bus system and the epidemiological characteristics of COVID-19 is used as the case disease. A series of disease control policies are evaluated to exhibit the sensitivity of the developed approach.\footnote{To facilitate future research, we uploaded the codes of this work to a GitHub repository: \url{https://github.com/mbc96325/Epidemic-spreading-model-on-public-transit}.}

\subsection{Mobility patterns in Singapore bus system}

Singapore is a city-state country where inter-city land transportation is relatively small. This provides an ideal testbed to focus on epidemic spreading through intra-city transportation, especially for bus systems, which count for a high proportion of modes shared in Singapore \citep{mo2018impact,shen2019built}. According to Singapore \cite{LTA2018Public}, the average daily ridership of buses is around 3.93 million, accounting for almost half of all travel modes. There are more than 368 scheduled bus routes operated by four different operators. A total of approximately 5,800 buses are currently in operation. In the case study, the mass rapid transit (MRT) system is neglected because a) passengers' contacts in a bus are more conducive for epidemic transmission compared to the MRT system, given the limited space in a bus; b) smart card data can provide exact bus ID to identify the direct contact of passengers. The direct contact in trains is, however, difficult to obtain from smart card data because the transactions are recorded at the station level. To identify the car that passengers boarded on, a transit assignment or simulation model is required (\cite{zhu2017probabilistic,Mo2020Capacity}, which is beyond the scope of this study.

Though we neglected the MRT system in the case study, our proposed approach focuses on a general modeling framework for virus spreading over PT systems. It is capable for modeling the MRT system once the boarded-trains for each passenger are identified. Moreover, to provide a more meaningful discussion for the omission of the MRT system, we estimate the corresponding impact in \ref{append_train_demand_size}.

\subsubsection{Usage of the system}

\begin{figure}[H]
\centering
\includegraphics[width= 1 \textwidth]{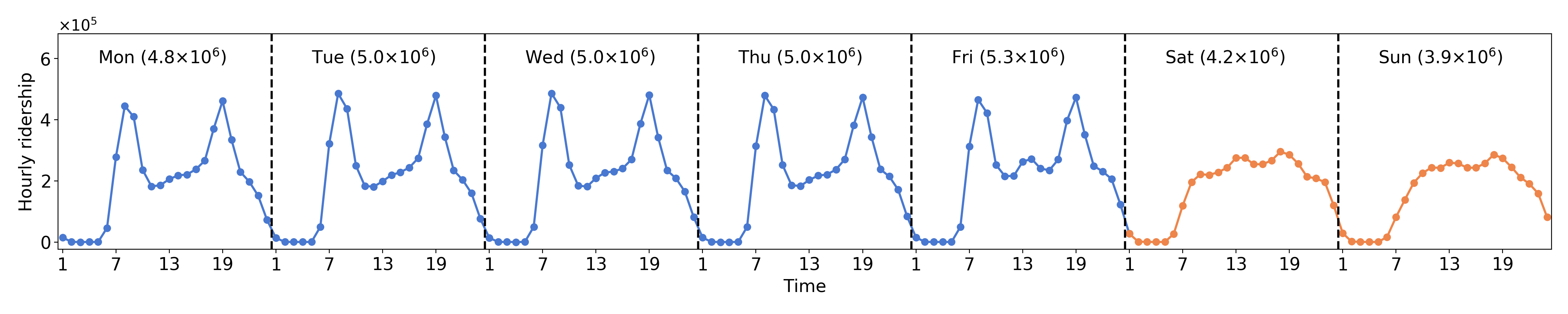}
\caption{Demand distribution. Numbers in x-axis represent the hour ID (e.g., 1 represents 0:00-1:00 AM). The number in brackets at the top of each sub-graph indicates the total daily ridership}
\label{fig_demand}
\end{figure}

The smart card data used in this study are from August 4th (Monday) to August 31st (Sunday), 2014, with a length of four weeks. The dataset contains 109.2 million bus trip transaction records from 4.7 million individual smart cardholders. Given that the population of Singapore in 2014 is around 5.5 million, the smart card data is representative of the population (accounting for 84\% of the population) and can model the epidemic spreading for the whole city.  Figure \ref{fig_demand} shows the hourly ridership distribution for one week (average of four weeks). The ridership of weekdays shows highly regular and recurrent patterns with morning (8:00-9:00 AM) and evening peaks (18:00-19:00). While the ridership distributions on weekends are different from those on weekdays, there are no prominent peaks observed.  

\begin{figure}[H]
\centering
\subfloat[Trip duration distribution]{\includegraphics[width=0.4\textwidth]{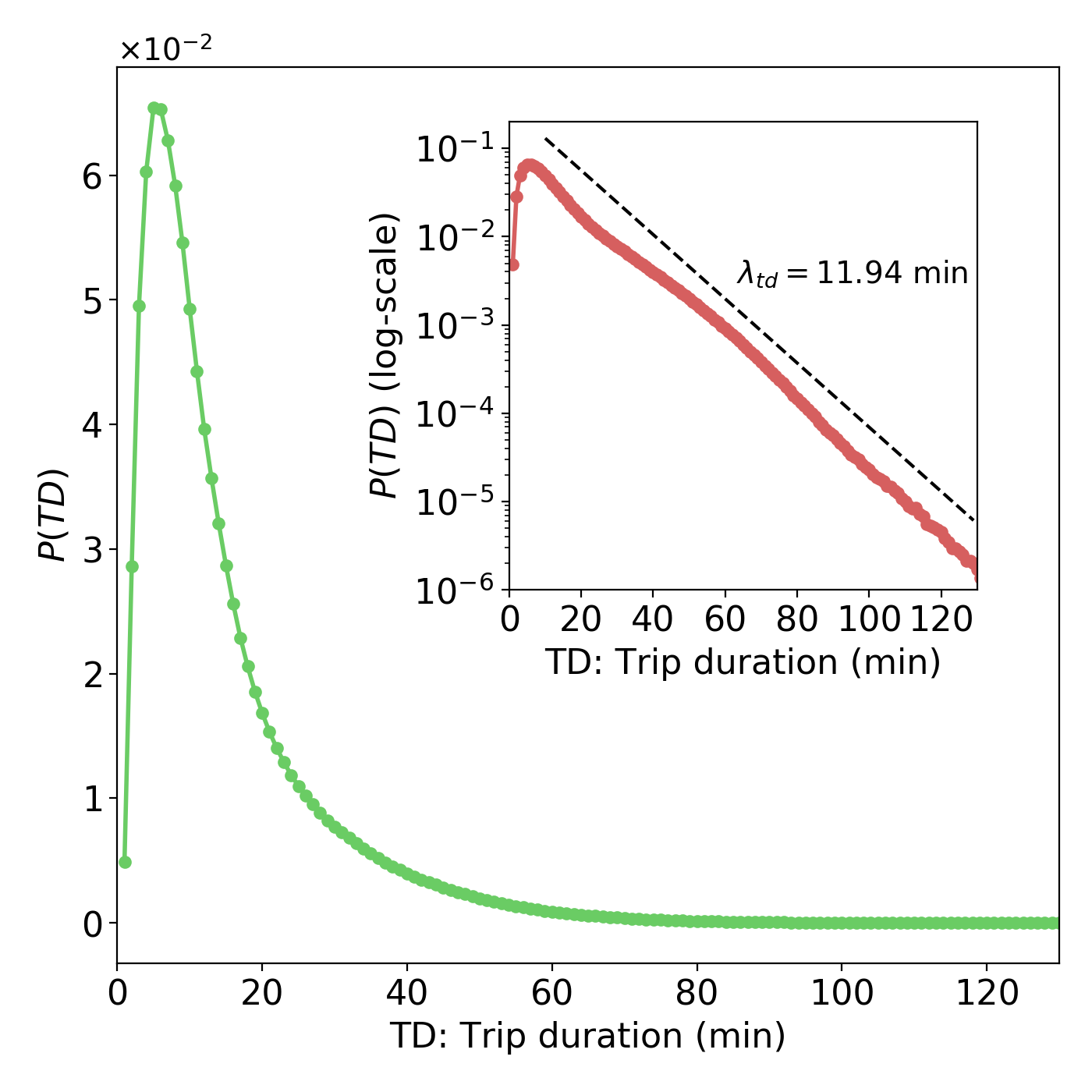}\label{fig_p_td}}
\hfil
\subfloat[Trip frequency distribution]{\includegraphics[width=0.4\textwidth]{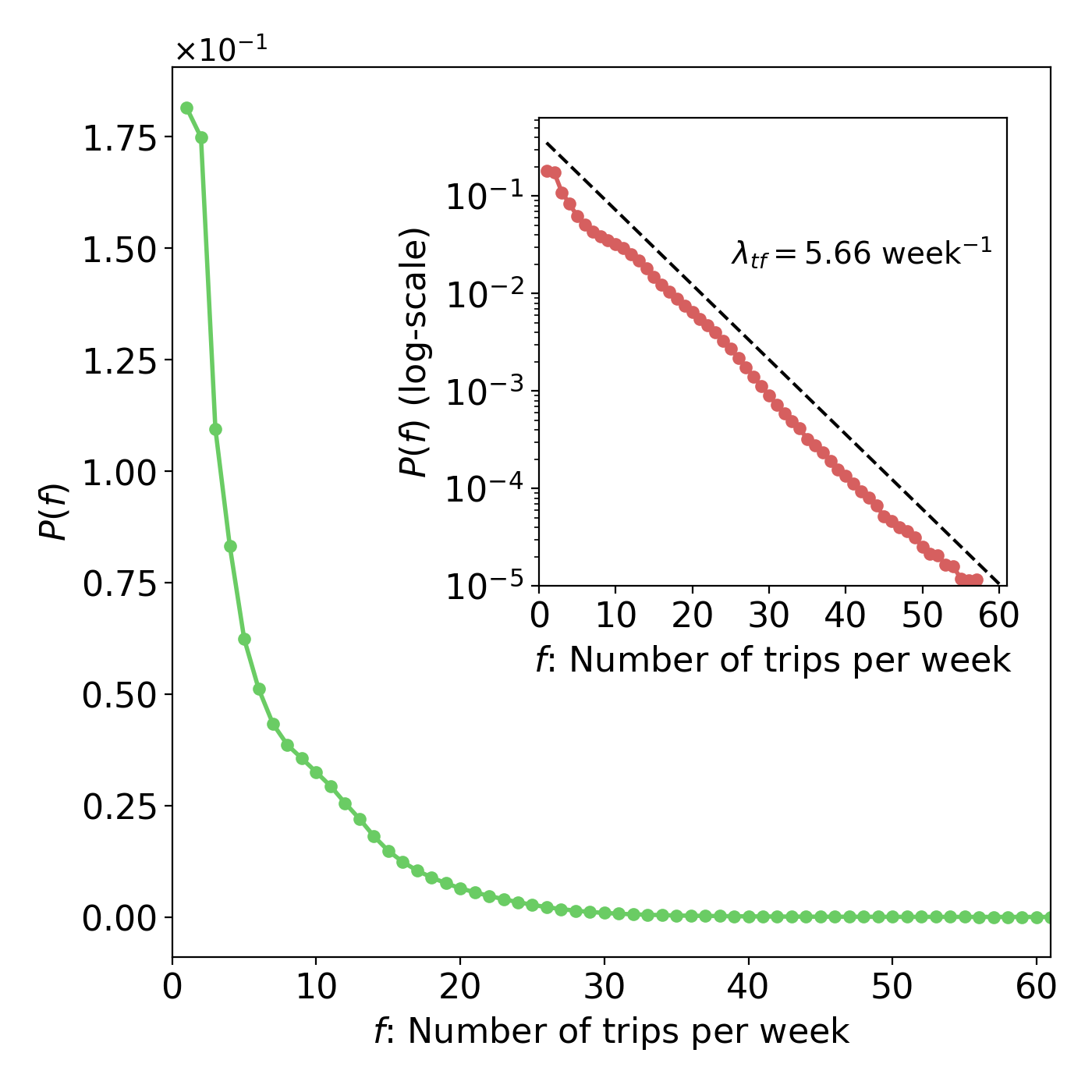}\label{fig_p_tf}}
\label{fig_trip_anal}
\caption{Distribution of trip duration and trip frequency (data of 4 weeks, the inset plots are in semi-log scale). Values in x-axis are rounded up in minute. For example, TD = 20 min includes all values from 19 to 20 min.}
\end{figure}

The usage of bus systems is related to daily activities, which represent mobility patterns of metro travelers in Singapore and can influence the epidemic spreading. Figure \ref{fig_p_td} shows the distribution of trip duration ($P(TD)$) in four weeks, where $TD$ means trip duration. Most trips have a duration of fewer than 40 min. From the inset of Figure \ref{fig_p_td}, we found the tail of $P(TD)$ can be well characterized by an exponential function: when $TD \geq 10$ min, we have $P(TD) \sim e^{-\frac{TD}{\lambda_{td}}}$, where $\lambda_{td} = 11.94$ min calculated by regression. As people tend to use MRT for long-distance travel, the duration of bus trips is relatively short. On average, the duration of bus trips is 14.55$\pm$12.51 min (mean$\pm$standard deviation).

Figure \ref{fig_p_tf} presents the trip frequency (number of trips per week) distribution for all 4.7 million smart card holders. The similar long tail attribute is observed, which can also be quantified by an exponential function: $P(f) \sim e^{-\frac{f}{\lambda_{tf}}}$, where $f$ is the trip frequency and $\lambda_{tf} = 5.66$ per week. On average, trip frequency in Singapore is around 6.34$\pm$5.94 per week

In summary, Singapore has an intense usage of bus systems with high ridership and user frequency, though the trip duration is relatively small. This implies that for highly infectious diseases that can be infected by short-term exposure, the bus system may play a crucial platform for the epidemic spreading.

\subsubsection{Contact networks}
\begin{figure}[H]
\centering
\subfloat[7:00 - 8:00 AM (PT)]{\includegraphics[width=0.33\textwidth]{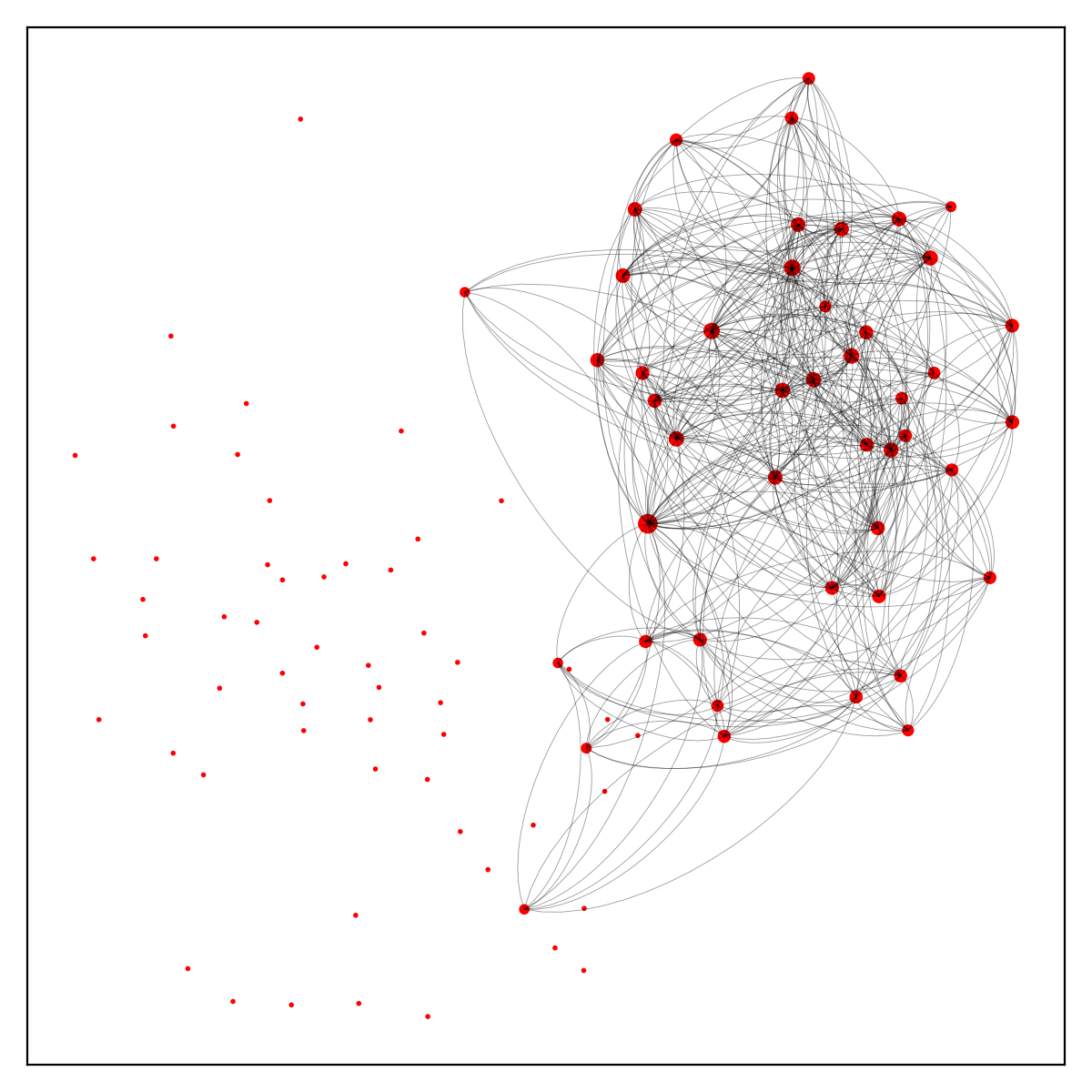}\label{fig_pt_net_1}}
\hfil
\subfloat[8:00 - 9:00 AM (PT)]{\includegraphics[width=0.33\textwidth]{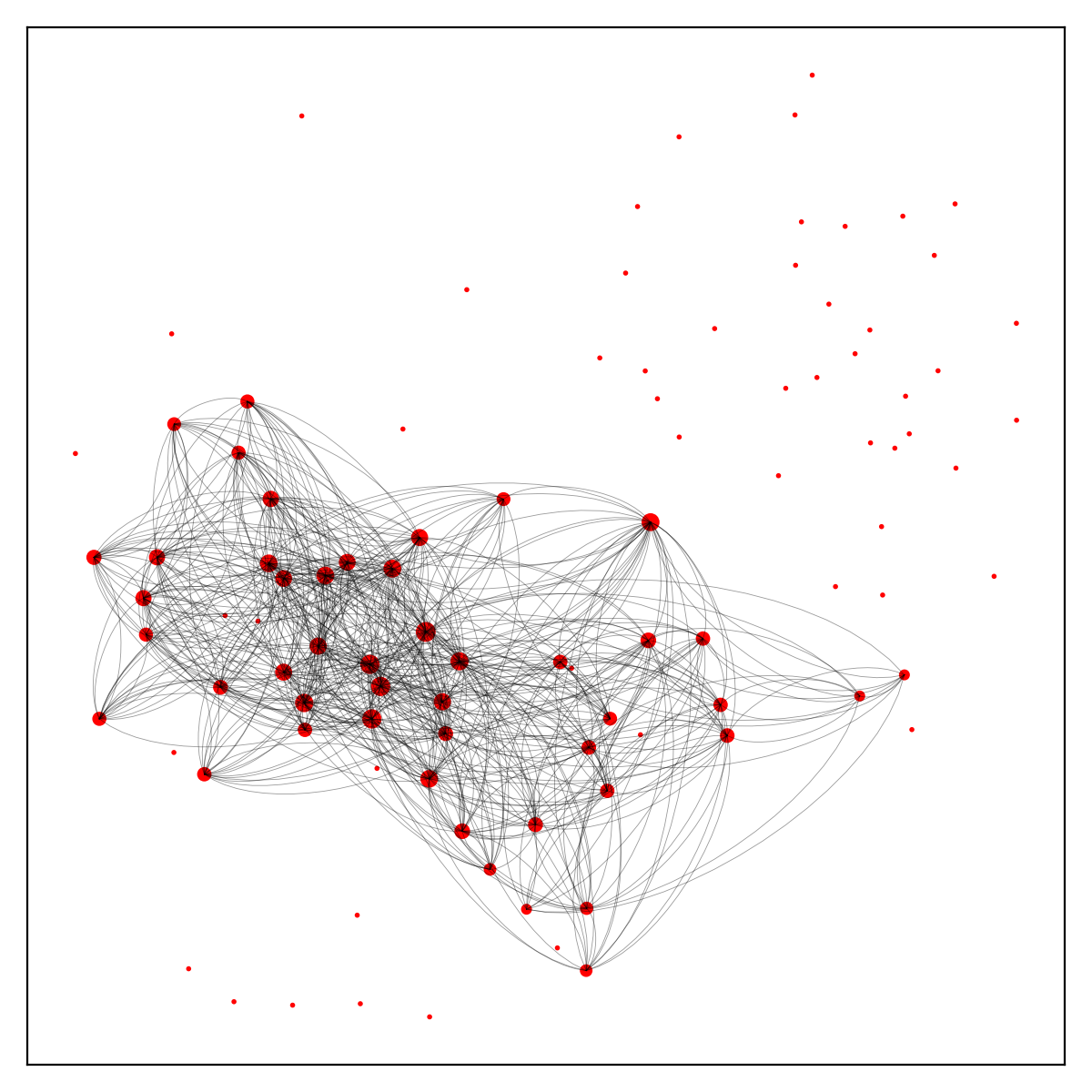}\label{fig_pt_net_2}}
\hfil
\subfloat[9:00 - 10:00 AM (PT)]{\includegraphics[width=0.33\textwidth]{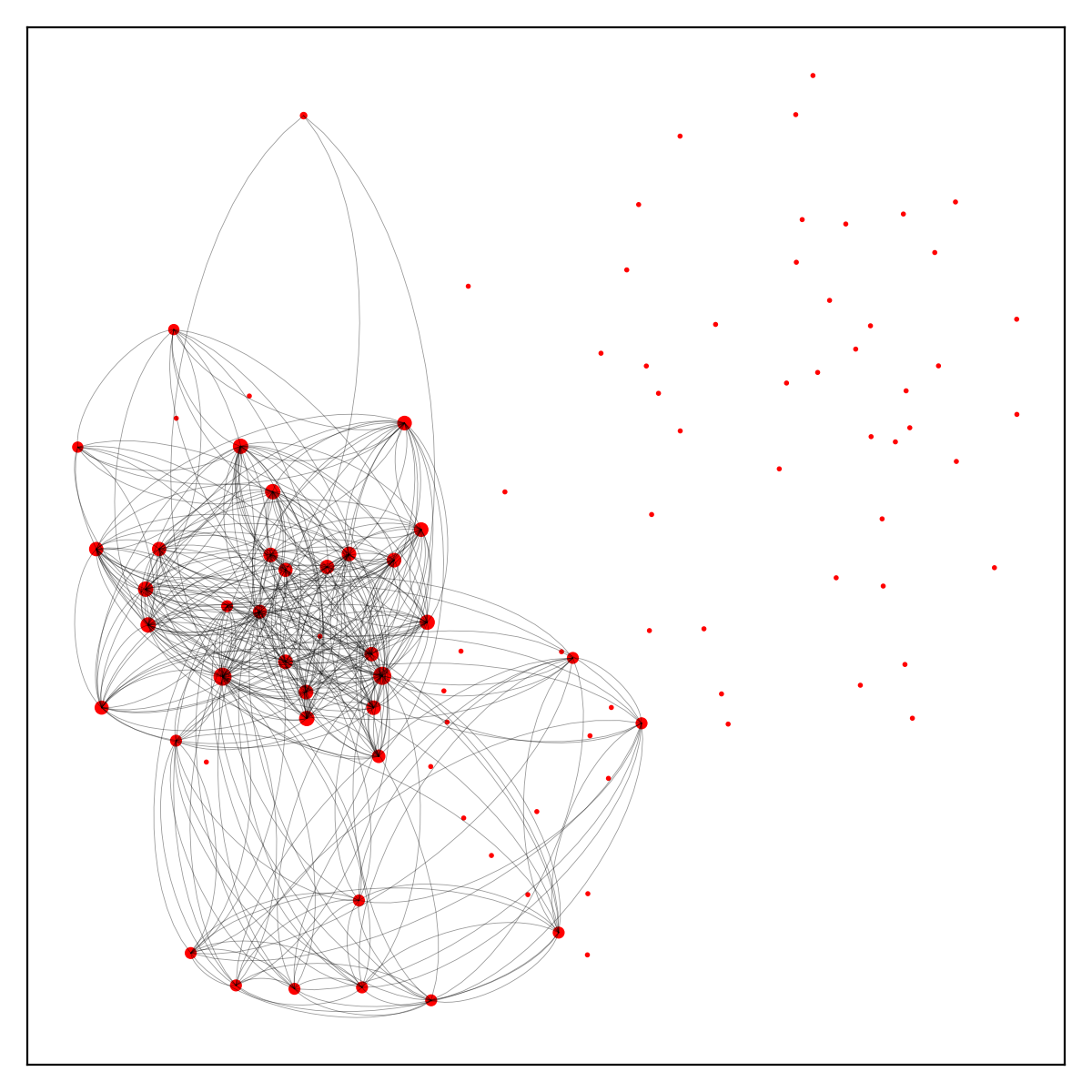}\label{fig_pt_net_3}}
\hfil
\subfloat[7:00 - 8:00 AM (Local)]{\includegraphics[width=0.33\textwidth]{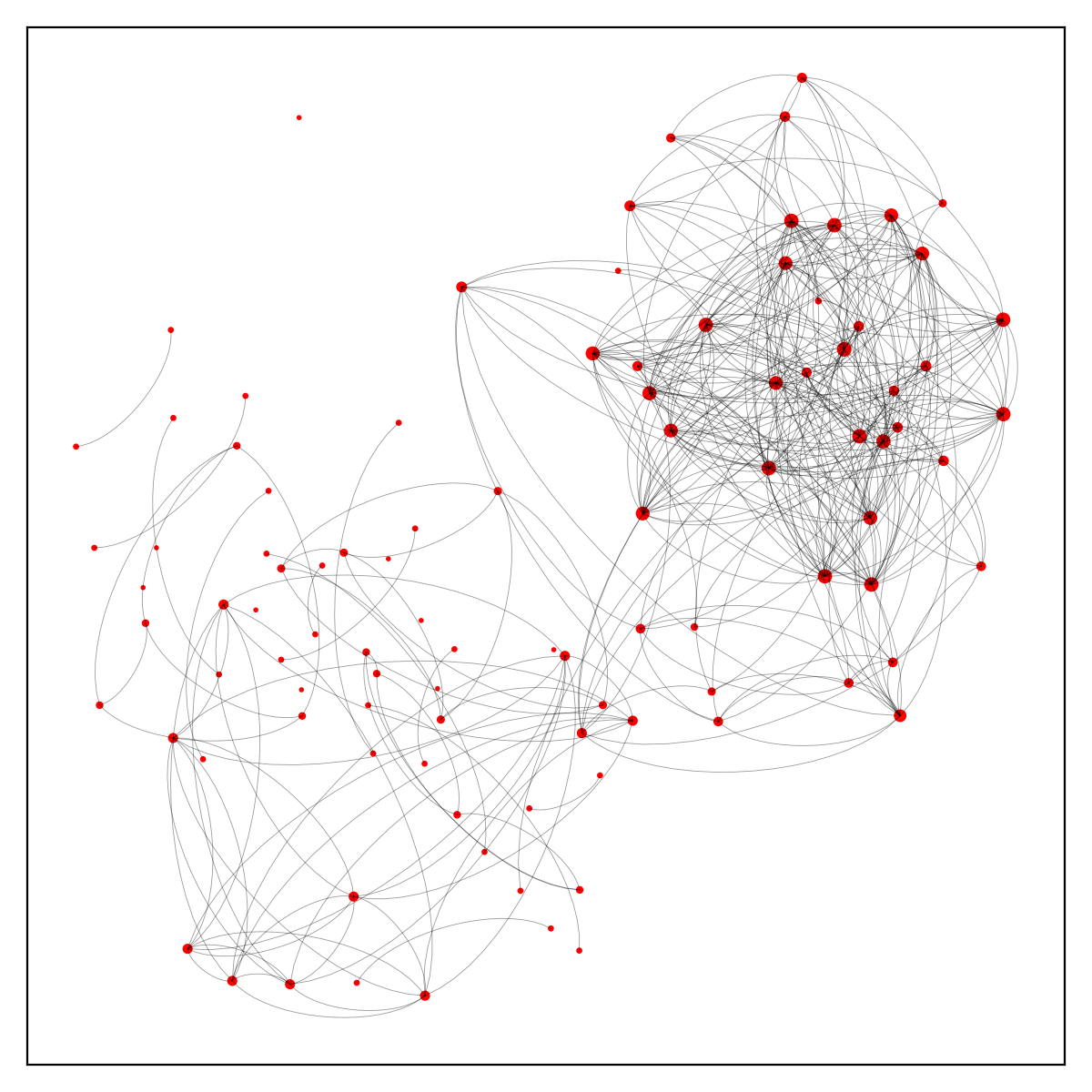}\label{fig_local_net_1}}
\hfil
\subfloat[8:00 - 9:00 AM (Local)]{\includegraphics[width=0.33\textwidth]{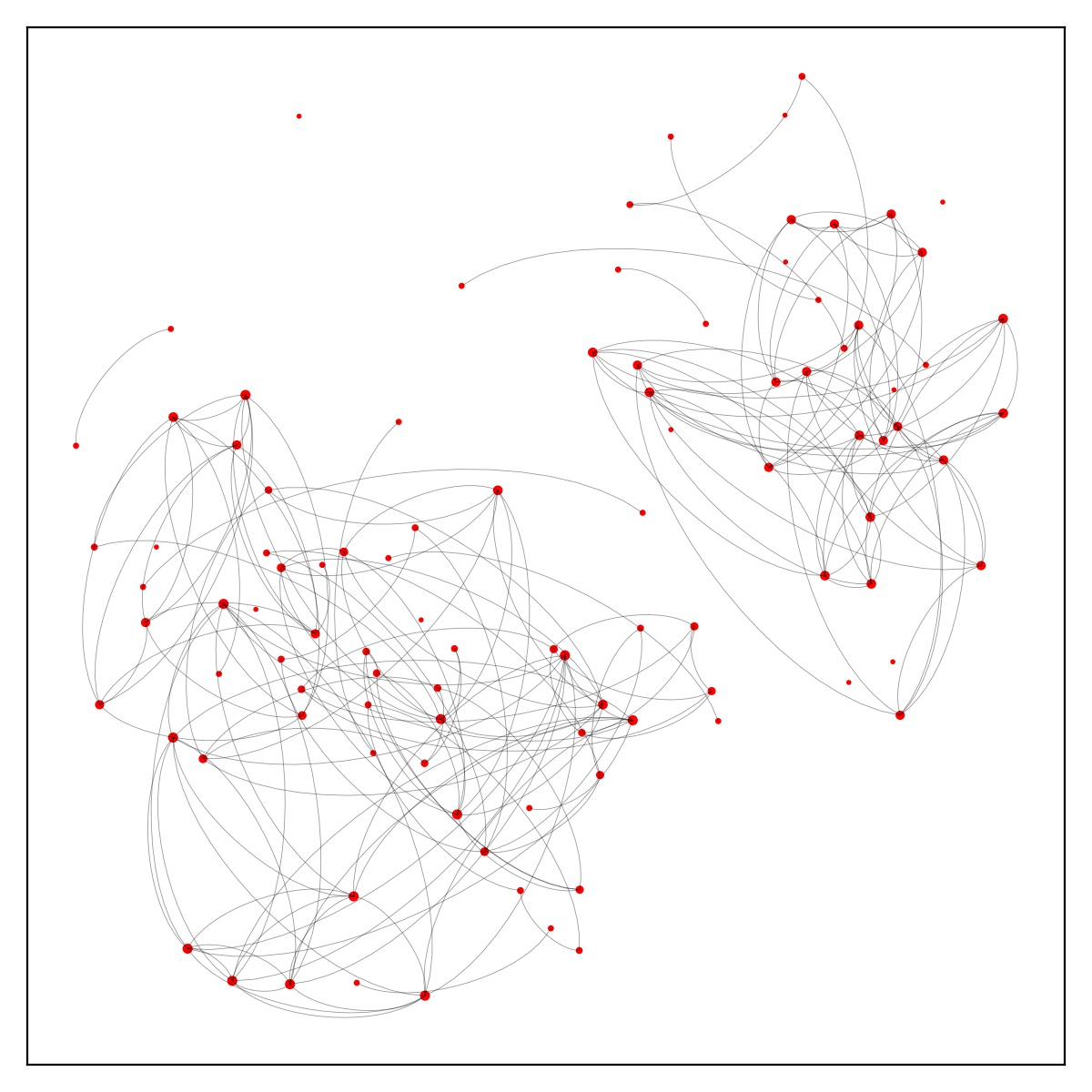}\label{fig_local_net_2}}
\hfil
\subfloat[9:00 - 10:00 AM (Local)]{\includegraphics[width=0.33\textwidth]{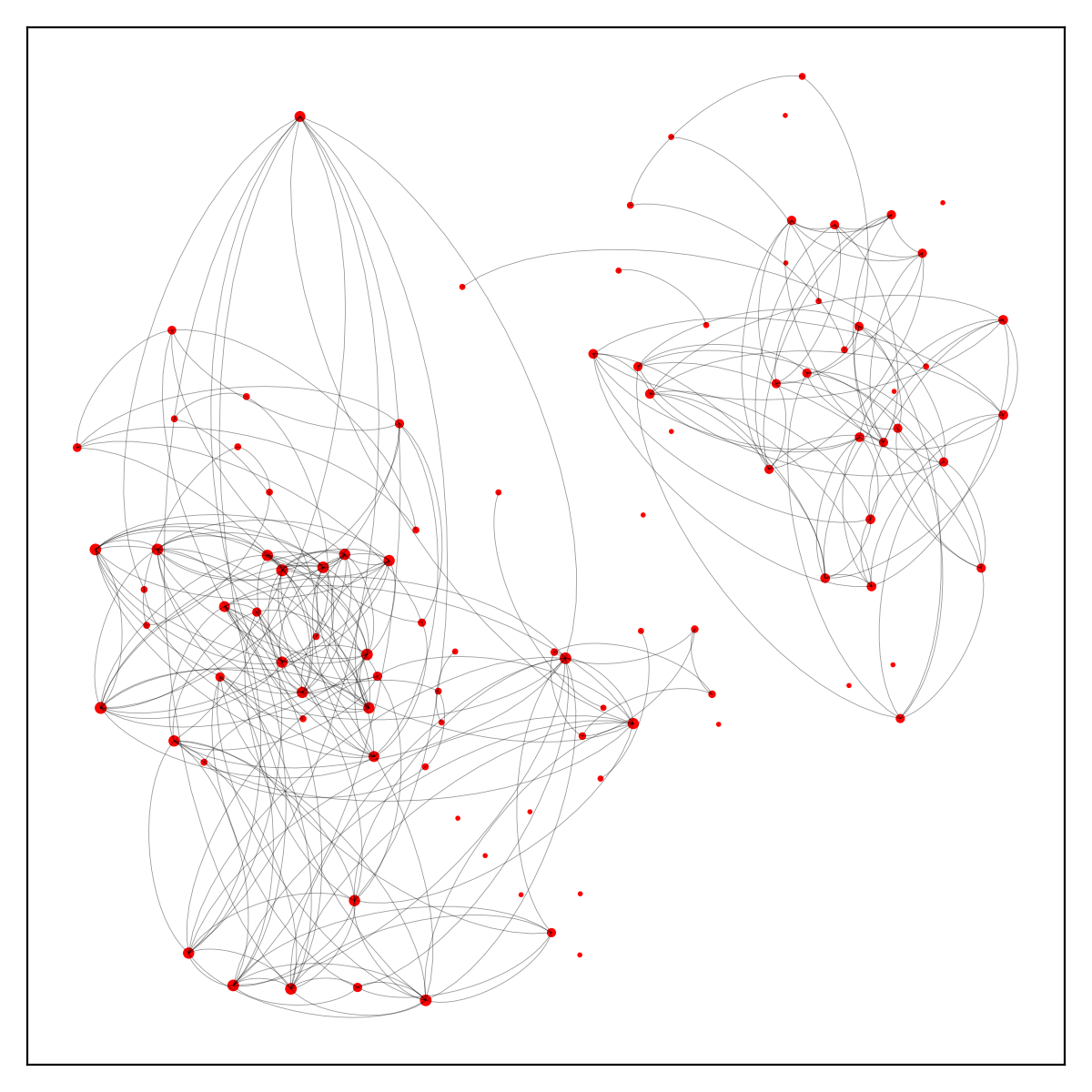}\label{fig_local_net_3}}
\caption{Example of time-varying PEN and LIN (100 sample passengers who have used the same bus, $\theta^{\ell} = 1$)}
\label{fig_network_100_sample}
\end{figure}

As discussed in Section \ref{network_repre}, the PEN and local interaction network (LIN) highly depend on passengers' mobility patterns and present time-varying properties. Figure \ref{fig_network_100_sample} shows example networks of 100 passengers extracted from real-world data (7:00-10:00 AM). The length of time interval $\tau = 1$ h is used in this study. For better visualization, these passengers are chosen from the same bus, and $\theta^{\ell} = 1$ is used for LIN. Both these networks show high dependency on time, with various network structures across time. In PEN, only about half of the passengers are connected during a time interval, and the connectivity changes dramatically over time. This highlights the importance of considering time-varying networks in modeling epidemics. 

\begin{figure}[H]
\centering
\subfloat[Degree (PT, weekday)]{\includegraphics[width=0.25\textwidth]{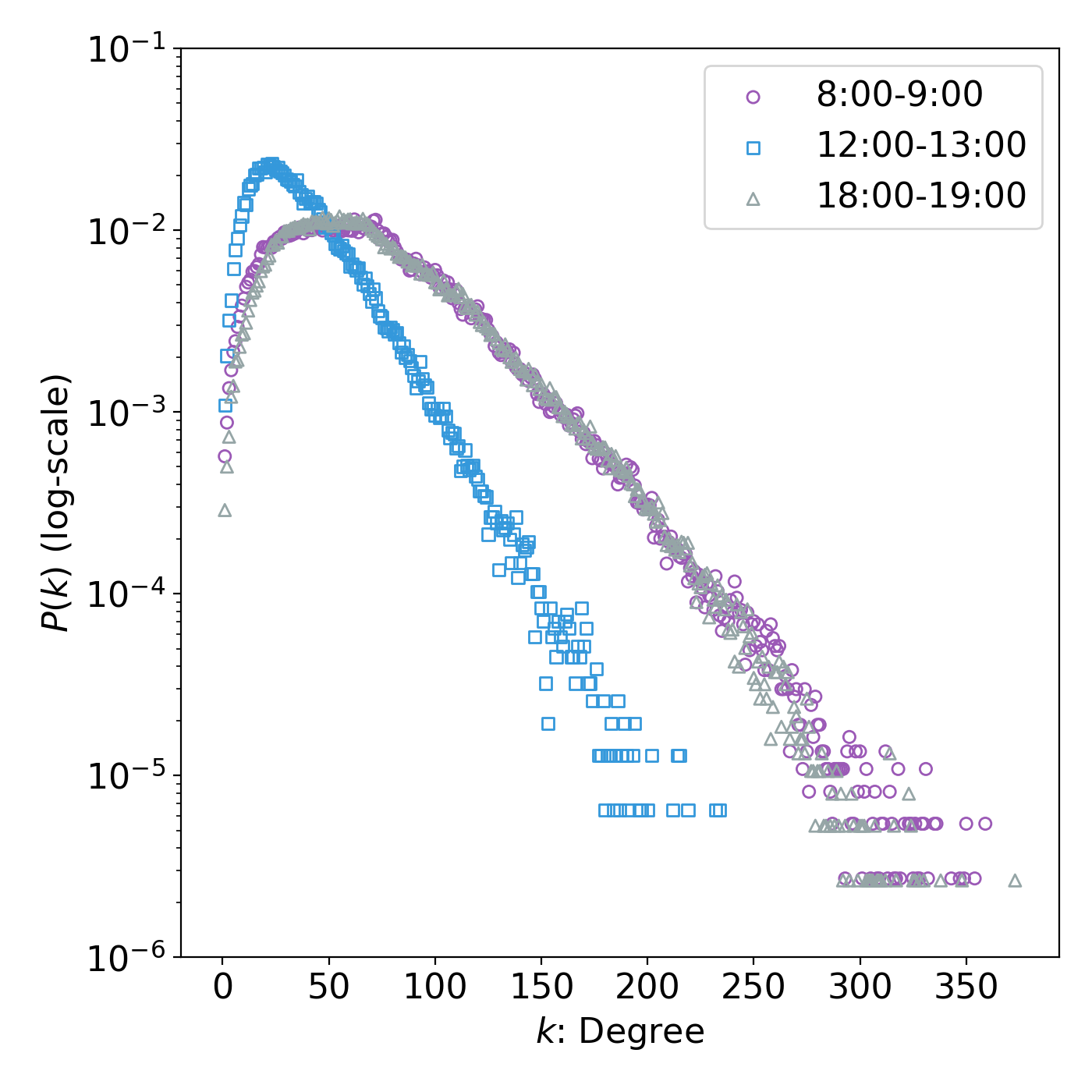}\label{degree_distribution_weekday_pt}}
\hfil
\subfloat[Degree (PT, weekend)]{\includegraphics[width=0.25\textwidth]{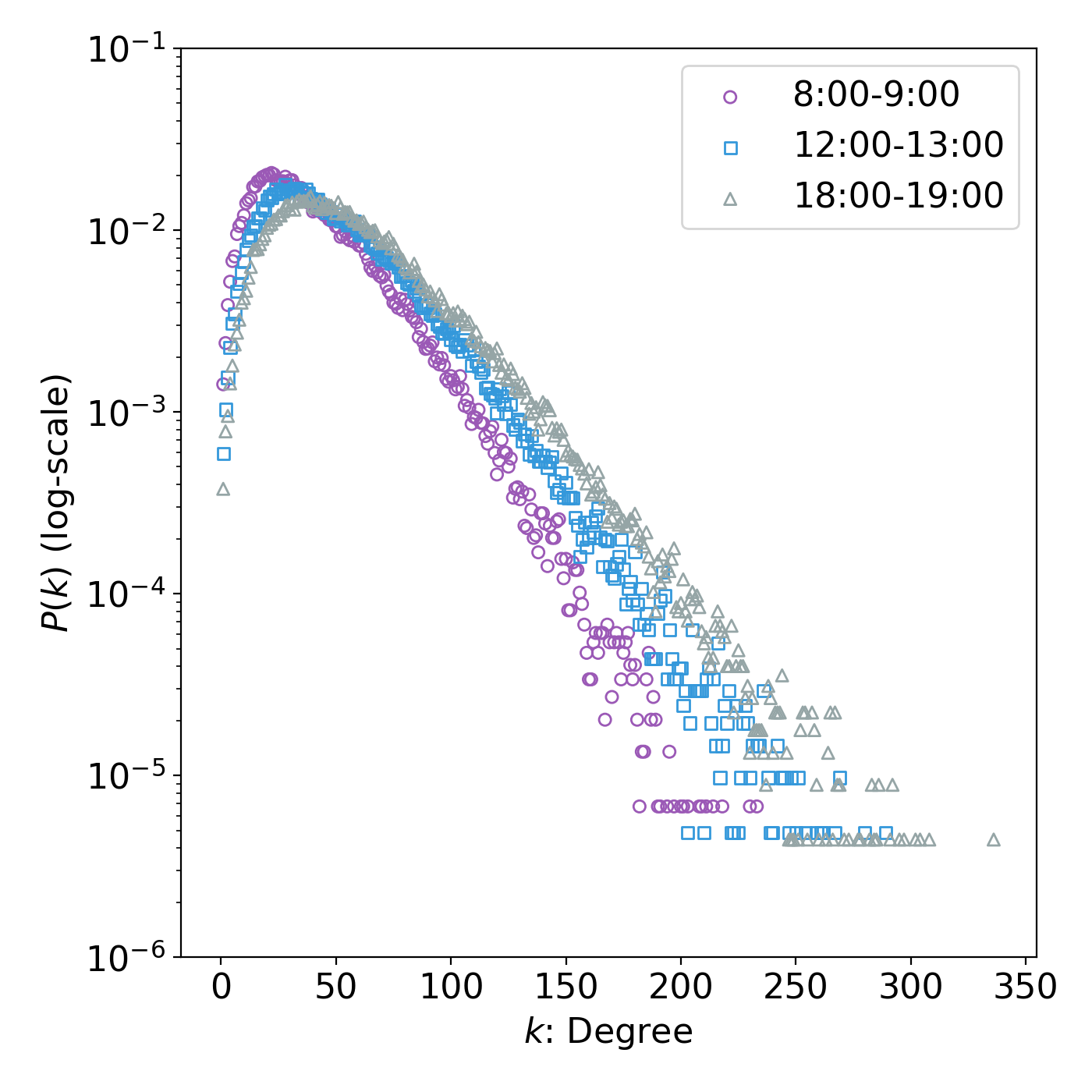}\label{degree_distribution_weekend_pt}}
\hfil
\subfloat[Degree (Local, weekday)]{\includegraphics[width=0.25\textwidth]{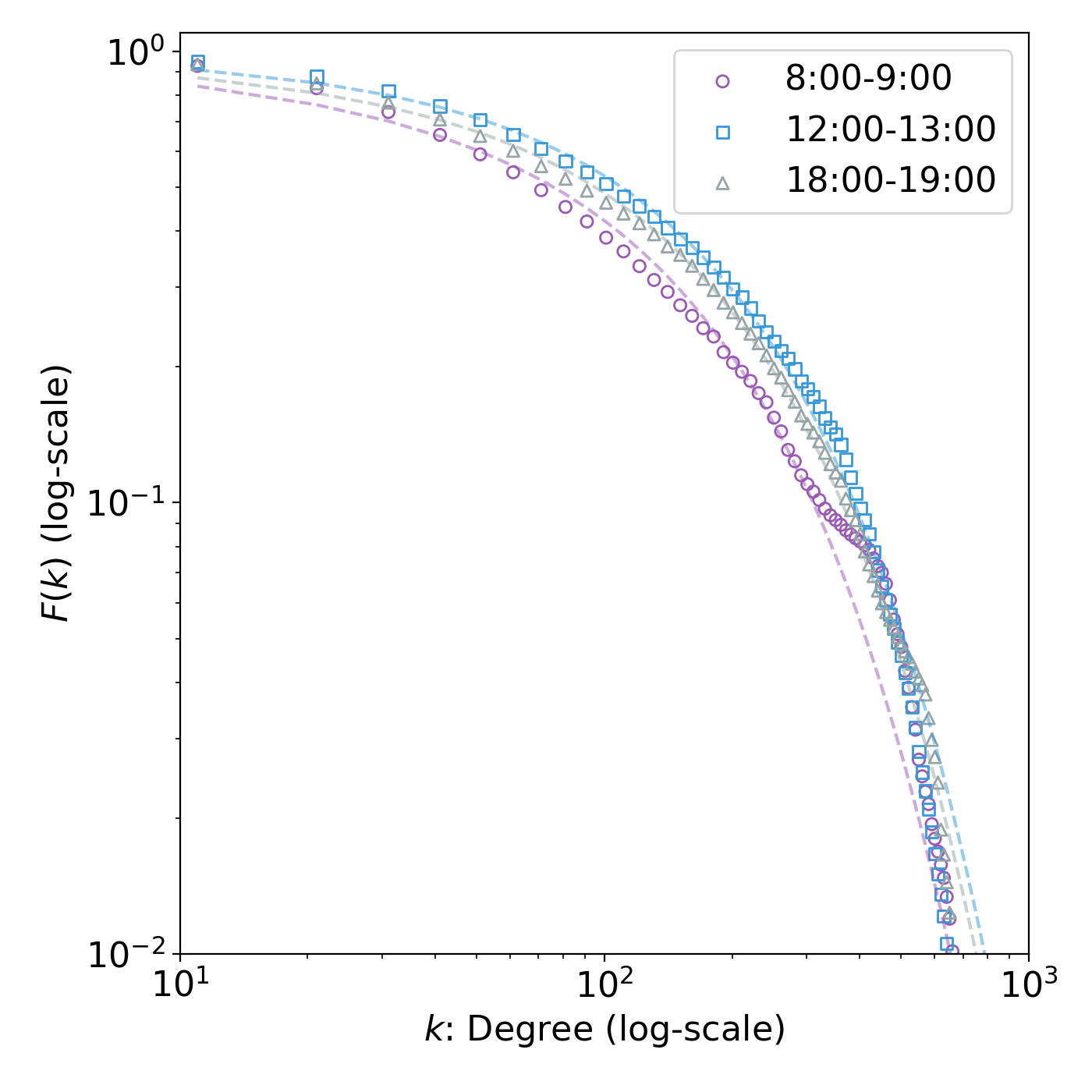}\label{degree_distribution_weekday_local}}
\hfil
\subfloat[Degree (Local, weekend)]{\includegraphics[width=0.25\textwidth]{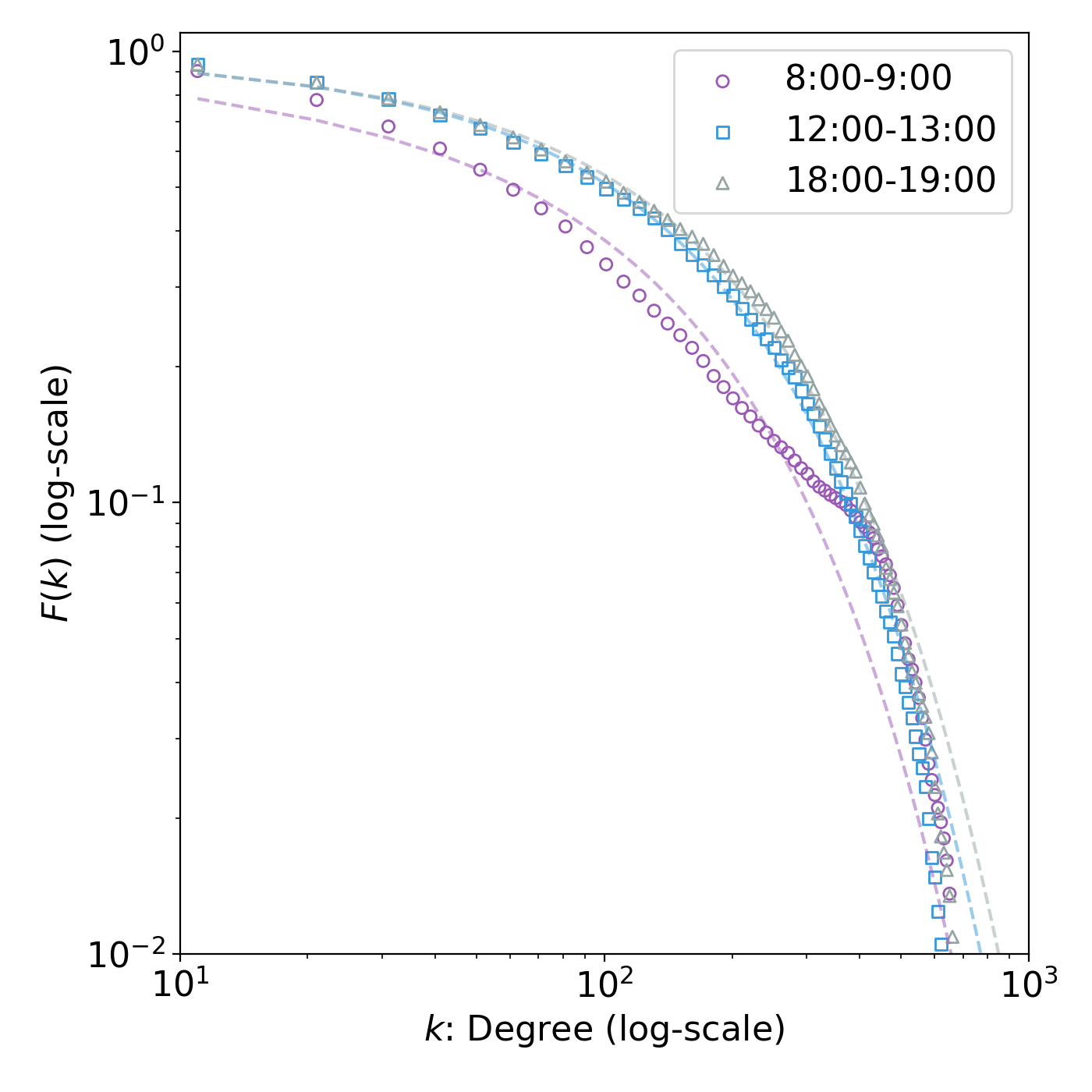}\label{degree_distribution_weekend_local}}
\hfil
\subfloat[CD (PT, weekday)]{\includegraphics[width=0.25\textwidth]{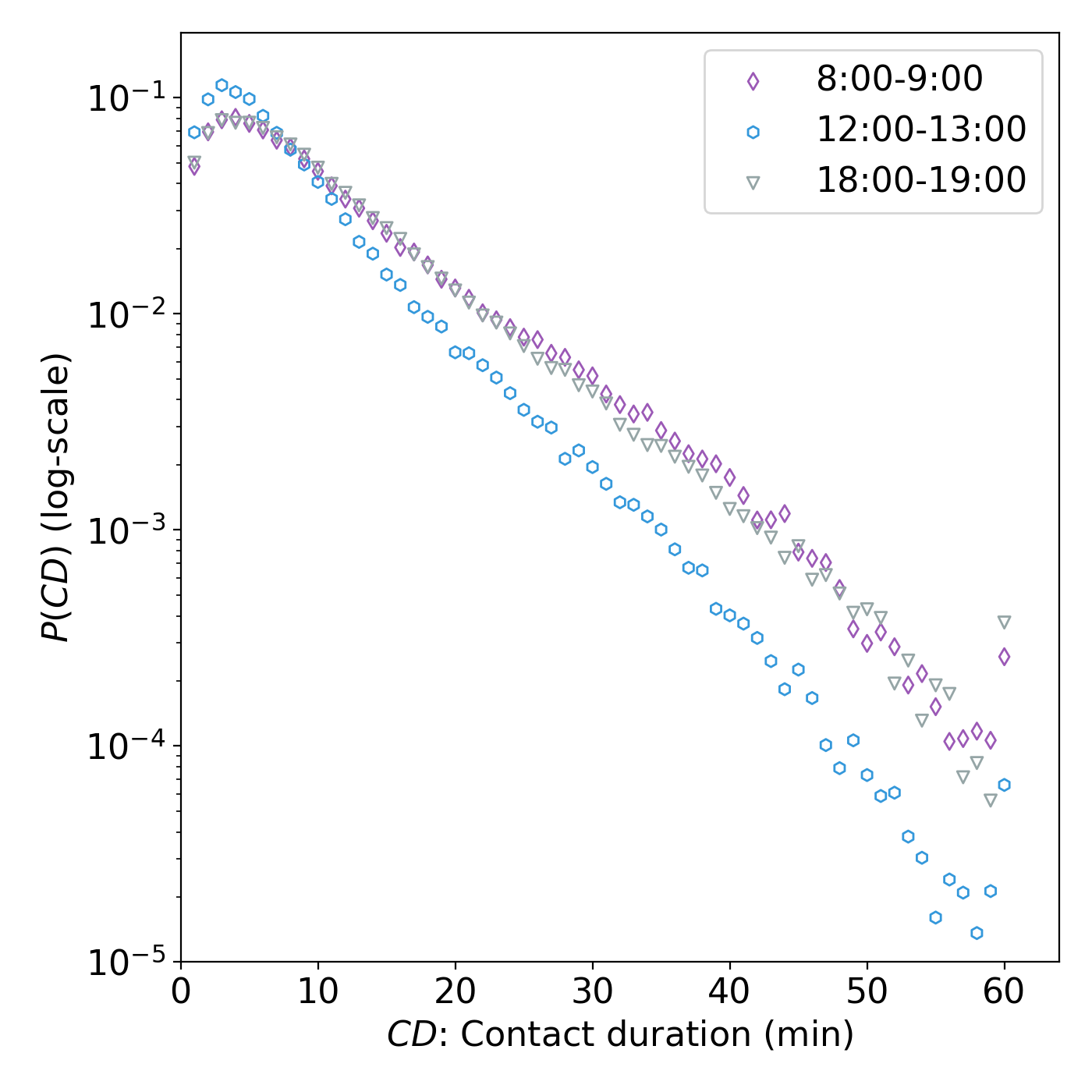}\label{contact_time_distribution_weekday_pt}}
\hfil
\subfloat[CD (PT, weekend)]{\includegraphics[width=0.25\textwidth]{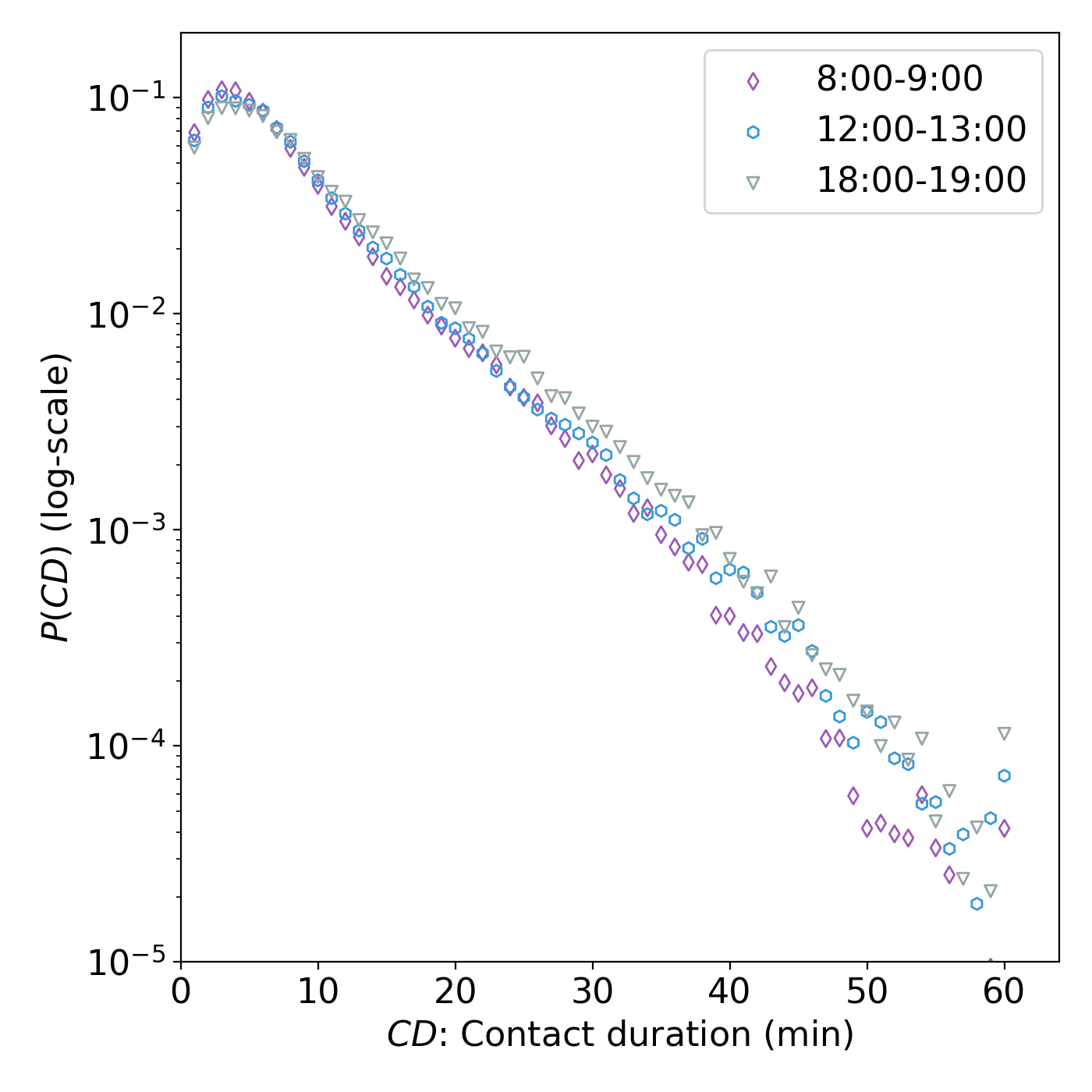}\label{contact_time_distribution_weekend_pt}}
\hfil
\subfloat[CD (Local, weekday)]{\includegraphics[width=0.25\textwidth]{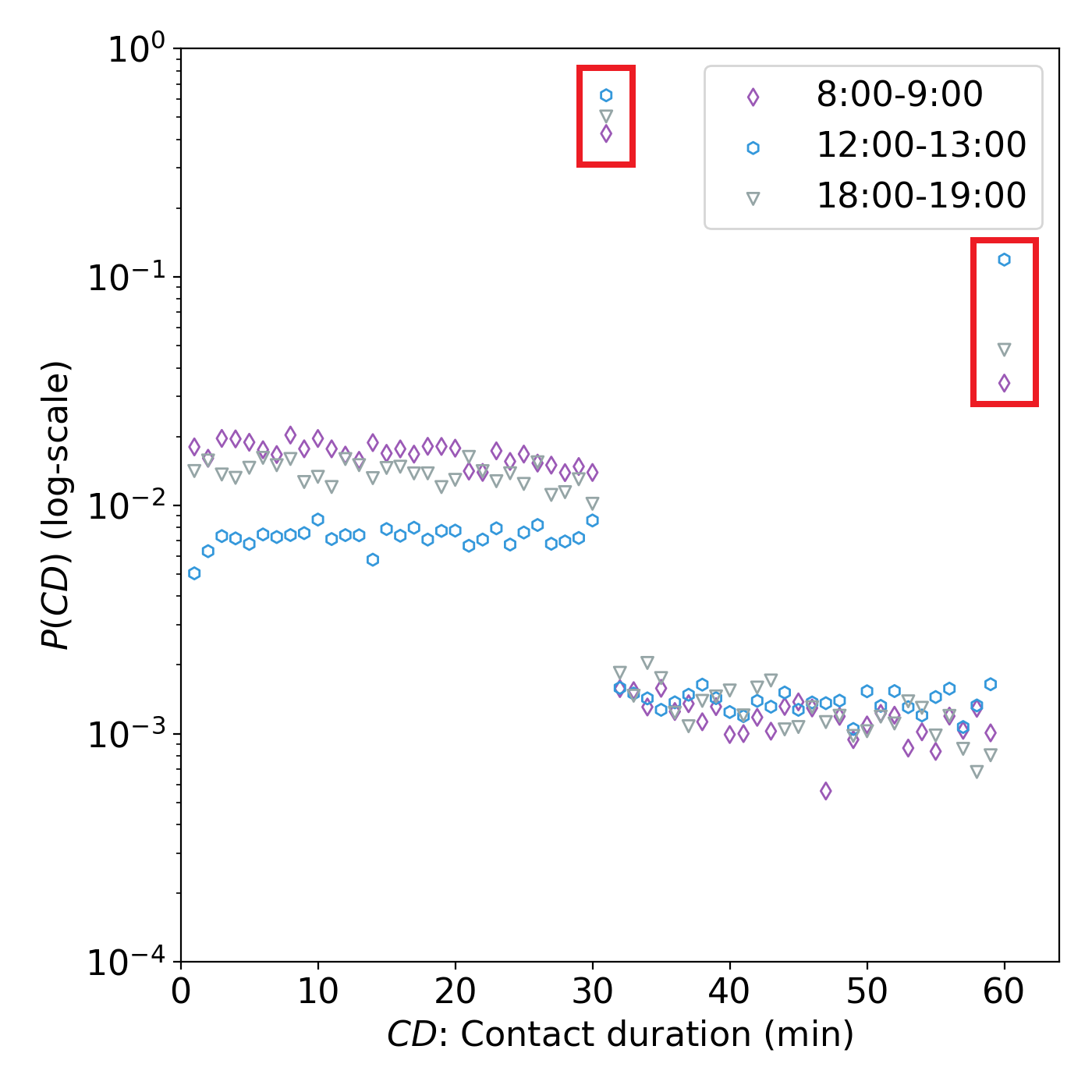}\label{contact_time_distribution_weekday_local}}
\hfil
\subfloat[CD (Local, weekend)]{\includegraphics[width=0.25\textwidth]{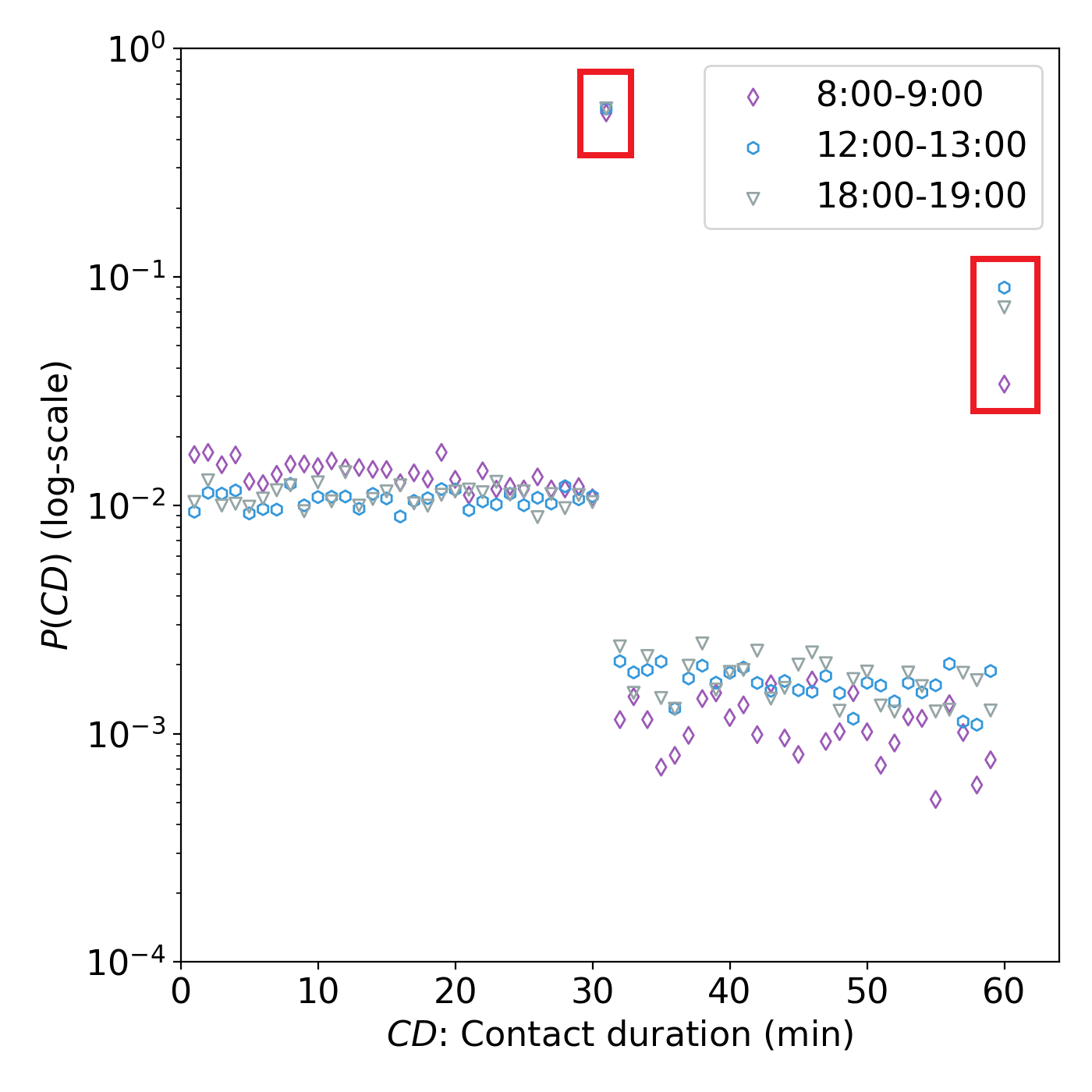}\label{contact_time_distribution_weekend_local}}
\caption{Degree and contact duration (CD) distributions of PEN and LIN (data of all Singapore, $\theta^{\ell} = 1\times10^{-3}$). Note that all figures except for (c) and (d) are empirical probability density functions. While (c) and (d) are CCDFs. The dot and dash lines in (c) and (d) indicate empirical and fitted CCDFs, respectively. The red squares in (g) and (d) represent the value of $P(CD=30 \text{ min})$ and $P(CD=60 \text{ min})$.}

\label{fig_network_property}
\end{figure}

The property of the contact network is essential for analyzing the epidemic spreading. Figure \ref{fig_network_property} summarizes the degree and CD distribution of PEN and LIN. Note that the global interaction network is, by definition, a simple random graph with homogeneous structures. Hence, we did not plot it. Given the time-varying properties of networks, we consider three different time intervals: morning peak (8:00-9:00 AM), noon off-peak (12:00-13:00), and evening peak (18:00-19:00). Figures \ref{degree_distribution_weekday_pt} and \ref{degree_distribution_weekend_pt} show that the degree distribution of PEN displays a exponential tail ($P(k) \sim e^{-\frac{k}{\lambda_k}}$, where $k$ is the degree), implying a significant degree heterogeneity. Most of the nodes are of low or medium degree. The number of super-nodes with a high degree is limited, and the maximum degree is bounded, which is reasonable given the limited capacity of buses. These properties are consistent with the findings in \cite{qian2020scaling}. Although the shapes of $P(k)$ for different times are similar, the exact values are still time-dependent. On weekdays, $P(k)$ for morning and evening peaks are similar but different from the off-peak curve.  PENs in peak hours also have a larger degree of nodes. On weekends, however, the degree distributions in the three time intervals are similar. 

Figures \ref{degree_distribution_weekday_local} and \ref{degree_distribution_weekend_local} show the \emph{complementary cumulative distribution function} (CCDF) of the degree for the LIN ($\theta^{\ell} = 1\times10^{-3}$ is used to correspond to the case study in the following sections). CCDF is defined as $F(k) = \sum_{i=k'}^\infty P(k')$. The reason for showing $F(k)$ instead of $P(k)$ is to avoiding the information loss due to binning and reducing the noise impacts at high degree regions \citep{barabasi2016network}. We observe that the degree distribution can be approximately characterized with a truncated power low: $P(k) \sim (k+k_{\text{sat}})^{-\gamma} \exp({-\frac{k}{k_{\text{cut}}}})$ (dashed lines are the fitted CCDFs), where $k_{\text{sat}}$ accounts for  low-degree saturation and $k_{\text{cut}}$ accounts for high-degree cutoff. Low-degree saturation is a common deviation, which indicates that we have fewer small degree nodes (i.e. isolated individuals in terms of social activities) than expected for a pure power law. High-degree cutoff appears as a rapid drop in $P(k)$ for $k>k_{\text{cut}}$, indicating that we have fewer high-degree nodes (i.e. extremely active individuals in social activities) than expected in a pure power law.

The Figures \ref{contact_time_distribution_weekday_pt} and \ref{contact_time_distribution_weekend_pt}, show that the contact time of PEN follows an exponential tail ($P(CD) \sim e^{-\frac{CD}{\lambda_{cd}}}$), which is similar to the trip duration distribution (Figure \ref{fig_p_td}). There are more contacts with duration less than 10 min in noon off-peak compared to morning and evening peaks on weekdays, while $P(CD)$ for the three time intervals on weekends are similar. By definition, the contact time of time-varying PEN is bounded by 60 min (same for the LIN). 

Figures \ref{contact_time_distribution_weekday_local} and \ref{contact_time_distribution_weekend_local} present the CD distribution of LINs. We observe a huge concentration on $CD=30 \text{ min}$ and $CD=60 \text{ min}$ (see the red squares). This is because people are not traveling most of the day-time. Thus, for a specific time interval, the duration of local interactions, for most individuals, would cover the whole time interval given our definition of LINs. However, since we assume people only spend half of their time on a specific origin or destination of a PT trip (see Section \ref{normal_cont_net}), the local CD for two people without travel within the time interval is either 30 min (if two individuals only share a single origin or destination) or 60 min (if two individuals share both origin and destination). Therefore, there are two concentrations with $CD=30 \text{ min}$ and $CD=60 \text{ min}$. As $P(CD=30) > P(CD=60)$, most people will only locally interact with others in one place (e.g., either at home or workplace). 

We also find that other local CD values are nearly uniformly distributed. Since local interaction duration that does not equal $30$ or $60$ min indicates that the trip occurs or ends in this time interval, the uniform distribution implies a Poisson start and end time of bus trips within the time interval. 


\subsection{Coronavirus disease 2019}
\begin{figure}[H]
\centering
\includegraphics[width= 0.6 \textwidth]{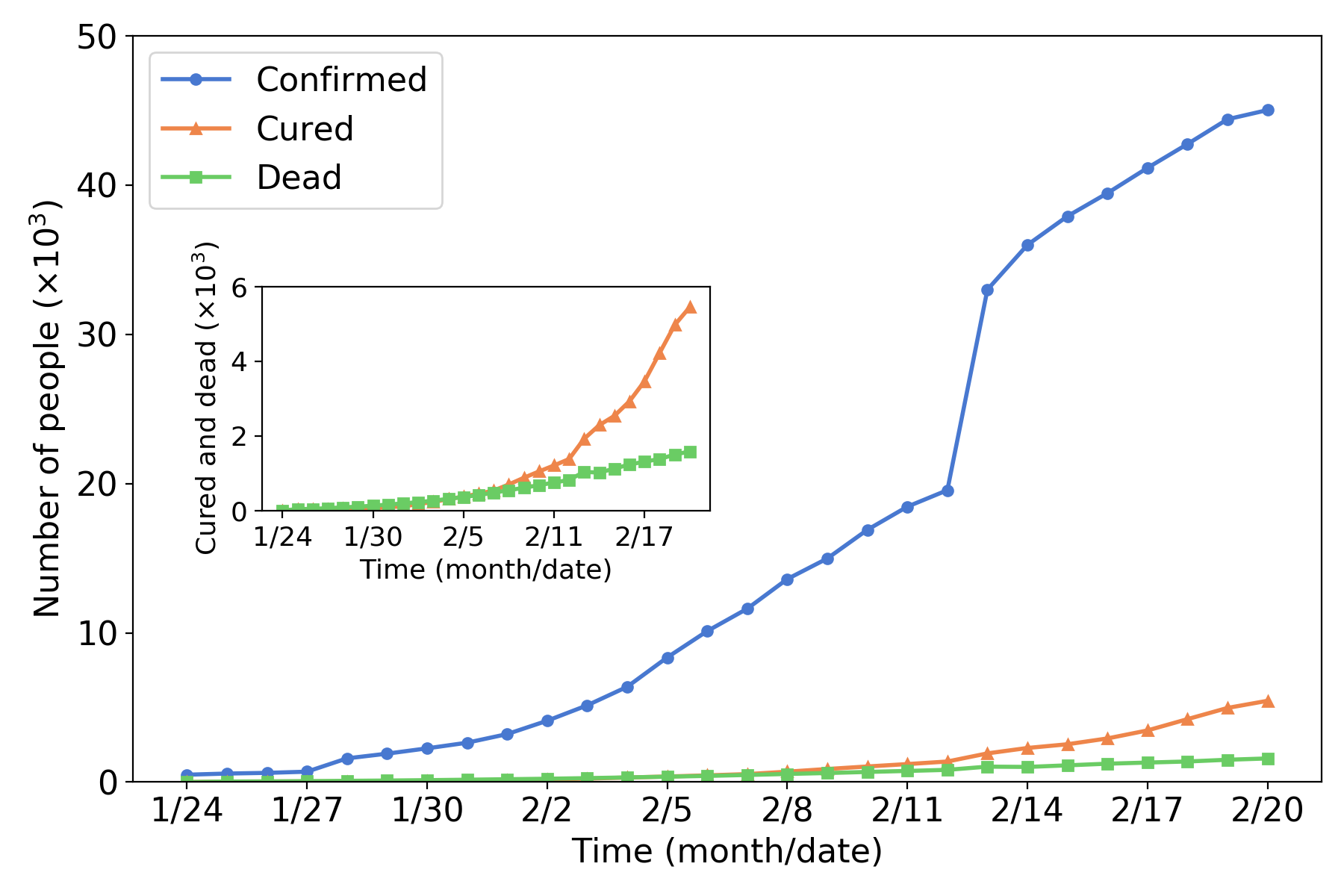}
\caption{Number of confirmed, cured and dead people in Wuhan (Jan 24 to Feb 2020; Data sources: \cite{DXY2020COVIDdata}). The inset plot is the zoom-in of the number of cured and dead people.}
\label{fig_wuhan_COVID}
\end{figure}

COVID-19, also known as 2019-nCoV, is an infectious disease caused by "SARS-CoV-2", a virus closely related to the Severe Acute Respiratory Syndrome (SARS) virus \citep{COVID19}. The disease is the cause of the 2019–2020 coronavirus outbreak. Cases were initially identified in Wuhan, China, and soon spread all over the world. The World Health Organization (WHO) declared the outbreak of COVID-19 as a Public Health Emergency of International Concern (PHEIC) on Jan 30, 2020, and later characterized the coronavirus as a pandemic on Mar 11, 2020. Figure \ref{fig_wuhan_COVID} shows the number of confirmed (infectious), cured, and dead people from Jan 24 to Feb 20, 2020, in Wuhan. Up to Feb 20, there are more than 40 thousand confirmed COVID-19 cases. The total number of healed and dead patients is around 6,000 and 2,000, respectively. The sudden increase in confirmed cases on Feb 12 is due to the revision of diagnosis criteria (adding the cases of clinical diagnosis). 

We select COVID-19 as the case study for the following reasons: a) COVID-19 is extremely contagious. It is primarily spread between people via respiratory droplets from infected individuals when they cough or sneeze. According to \cite{COVID19contact_time}, anyone who has been within approximately 2 m of a person with COVID-19 infection for a prolonged period (more than 1 or 2 min) is considered risky to get infected. Therefore, PT can play a significant intermediary for such a highly contagious disease. b) Even when authors are writing this article, COVID-19 is a big threat to global public health. Singapore is also experiencing the impact of COVID-19 \citep{MOHCOVID19}. The case study of COVID-19 can provide disease control suggestions from the transportation side, which adds real-time value to this research.

\begin{figure}[H]
\centering
\includegraphics[width= 0.6 \textwidth]{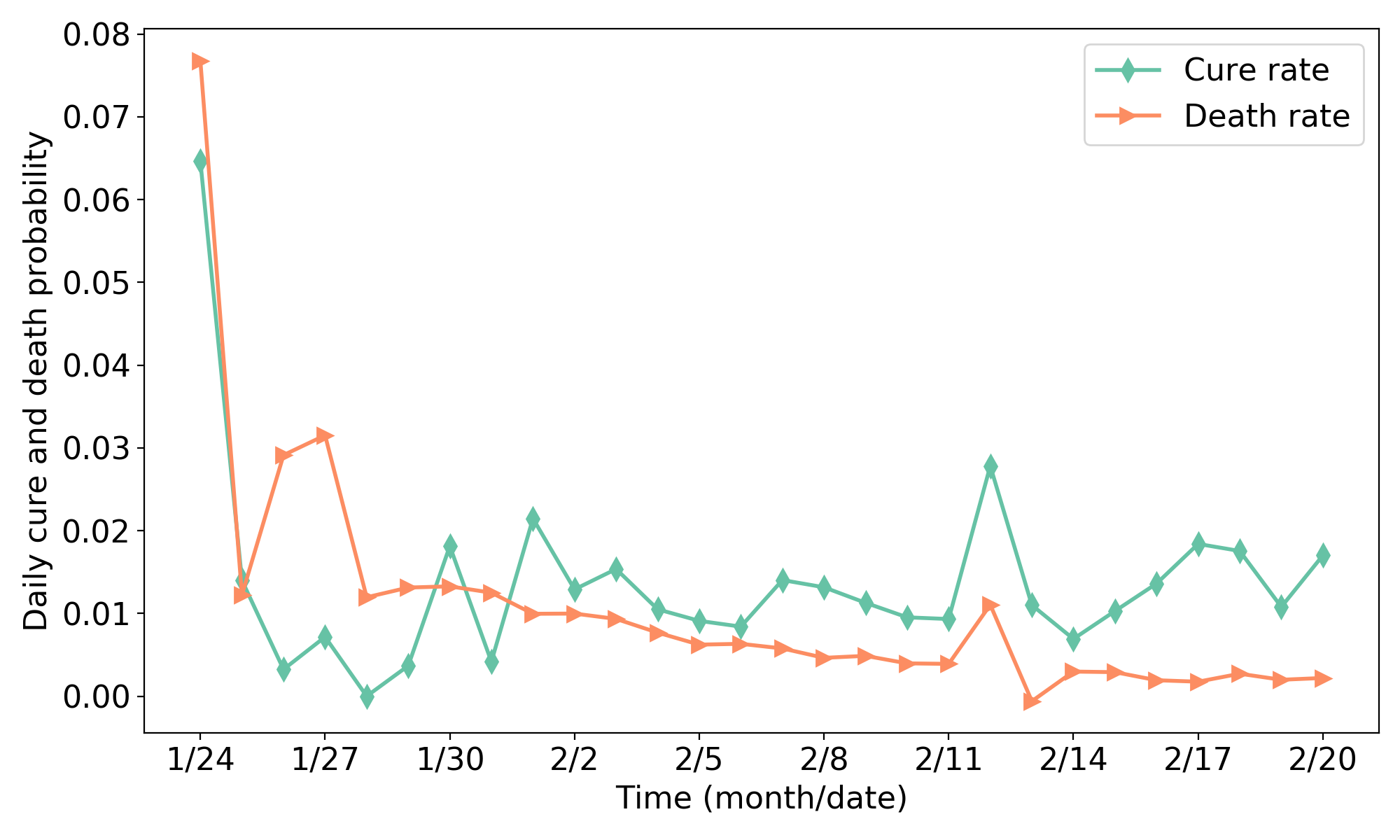}
\caption{Daily cure and death rate in Wuhan (data sources: \cite{DXY2020COVIDdata})}
\label{fig_wuhan_cure_death_rate}
\end{figure}

The SEIR model parameters are chosen based on the epidemiological characteristics of COVID-19. Time from exposure to onset of symptoms (latent or incubation period) is generally between 2 to 14 days for COVID-19. \cite{read2020novel} suggested setting the latent period as 4 days. We, therefore, have $\gamma = \frac{1}{24\times4} = 0.0104$ (probability from $E$ to $I$ per h). According to \cite{read2020novel}, the transmission rate of COVID-19 in the \emph{static} SEIR model is 1.96 day$^{-1}$, which can be seen as the number of people that one infectious person can infect per day in a \emph{well-mixed} network. Therefore, assuming one person, on average, has close contact with 100 others per day, we can calculate the hourly one-to-one infectious probability as $\beta_I = \frac{1.96}{24\times100} = 8.17 \times 10^{-4}$. Although recent studies show $\beta_E >0$ for COVID-19 \citep{rothe2020transmission}, calibrating the exact value of $\beta_E$ is difficult due to lack of data. Since people in the latent period (group $E$) usually have extremely lower probability of transmission, we arbitrarily set $\beta_E = 0.01 \beta_I$. We calculate $\mu_r$ and $\mu_d$ using data from Wuhan. Figure \ref{fig_wuhan_cure_death_rate} shows the daily cure and death rate (number of cured/dead people per day divided by the total number of confirmed people on that day) in Wuhan. The reason for the high value on the first day may be the inaccurate data. From Figure \ref{fig_wuhan_cure_death_rate}, we observe the average daily cure and death rate are approximately $1\%$ at the early stage. Therefore, we can calculate the hourly cured and death probability as $\mu_r = \mu_d = \frac{0.01}{24} = 4.17 \times 10^{-4}$. 

$\theta^{\ell} = 1 \times 10^{-3}$ is used for the status quo analysis. This value is calculated as follows. Consider a community with 10,000 people, assume each person may have close contact with another 10 people on average per hour locally. We therefore have $\theta^{\ell} = \frac{10}{10,000} = 1 \times 10^{-3}$. The global interaction captures individual's probability of close contact with people outside his/her community. Given that the population of Singapore is around 5.6 million, we assume one person, on average, can closely contact 10 people per \emph{day} globally. Then $\theta^{g} = \frac{10}{5.6 \times 10 ^6 \times 24} = 7.44 \times 10 ^{-8}$. Note that the number $24$ in the denominator is used to get the \emph{hourly} probability.

Table \ref{tab_para} summarizes all parameters of the status quo analysis, which can be seen as the reference scenario. The sensitivity analysis column indicates whether this value will be changed in the following policy analysis sections. Specifically, the sensitivity analysis for $\beta_I$, $\beta_E$, $\mu_r$ and $\mu_d$ is shown in Section \ref{impact_of_beta_mu}, for $\theta^{g}$ and  $\theta^{\ell}$ is shown in Section \ref{impact_of_trip_occ_rate}.

\begin{table}[H]
\centering
\caption{Parameters value for status quo analysis (reference scenario)}
\begin{tabular}{@{}ccccc@{}}
\toprule
Category                        & Parameters & Value                 & Sources                  & Sensitivity analysis \\ \midrule
\multirow{5}{*}{Epidemic}       & $\beta_I$  & $8.17 \times 10^{-4}$ & \cite{read2020novel} & True                 \\
                                & $\beta_E$  & $8.17 \times 10^{-6}$ & Authors assumption       & True                 \\
                                & $\gamma$   & $0.0104$              & \cite{read2020novel} & False                \\
                                & $\mu_r$    & $4.17 \times 10^{-4}$ & \cite{DXY2020COVIDdata} & True                 \\
                                & $\mu_d$    & $4.17 \times 10^{-4}$ & \cite{DXY2020COVIDdata} & True                 \\ \midrule
\multirow{2}{*}{Human mobility} & $\theta^{\ell}$ & $1 \times 10^{-3}$    & Authors assumption       & True                 \\
                                & $\theta^{g}$ & $7.44 \times 10^{-8}$ & Authors assumption       & True                 \\ \bottomrule
\end{tabular}
\label{tab_para}
\end{table}

\subsection{Theoretical model calibration}\label{cc1}

\begin{figure}[H]
\centering
\subfloat[Comparison on infectious people]{\includegraphics[width=0.4\textwidth]{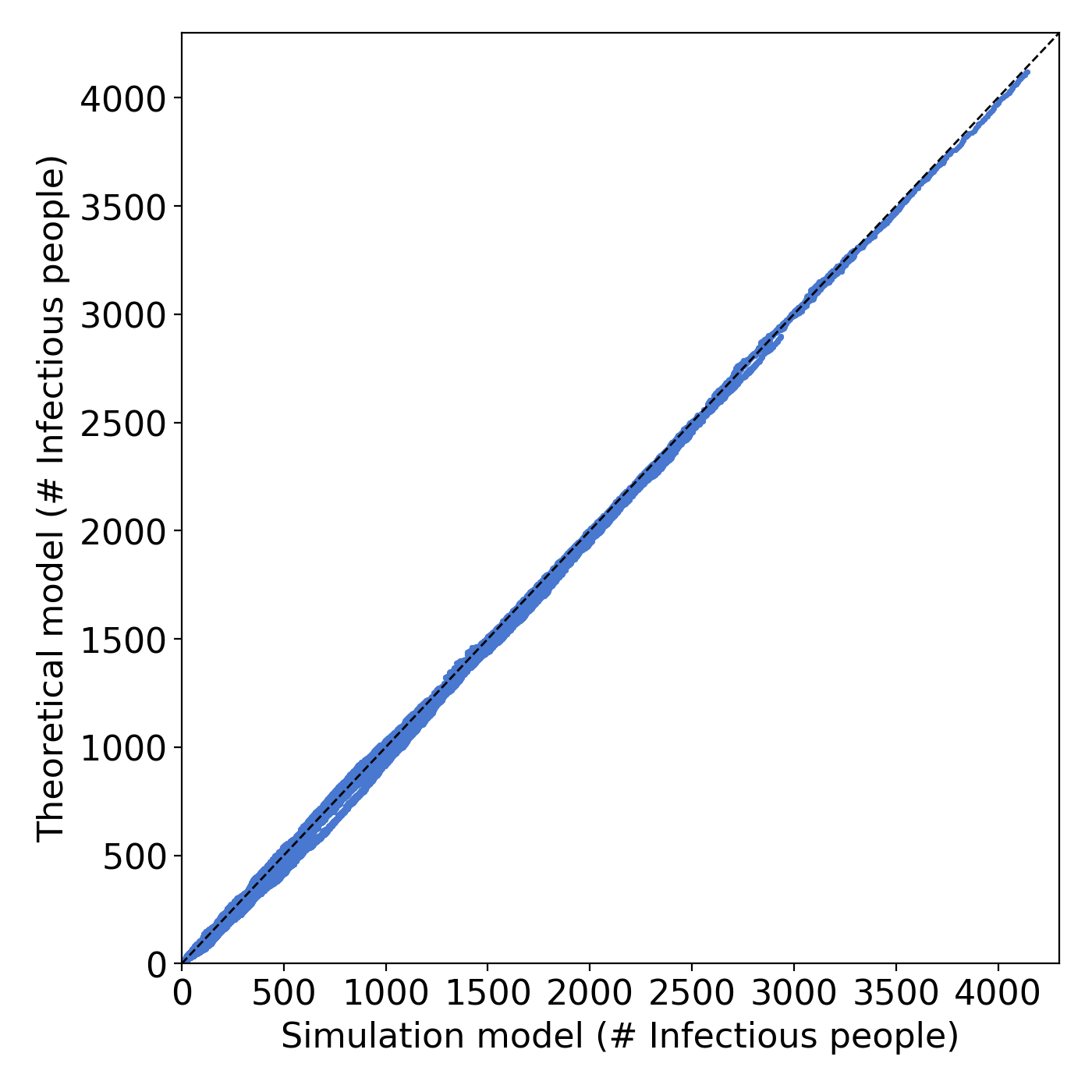}\label{fig_compare_I}}
\hfil
\subfloat[Comparison on exposed people]{\includegraphics[width=0.4\textwidth]{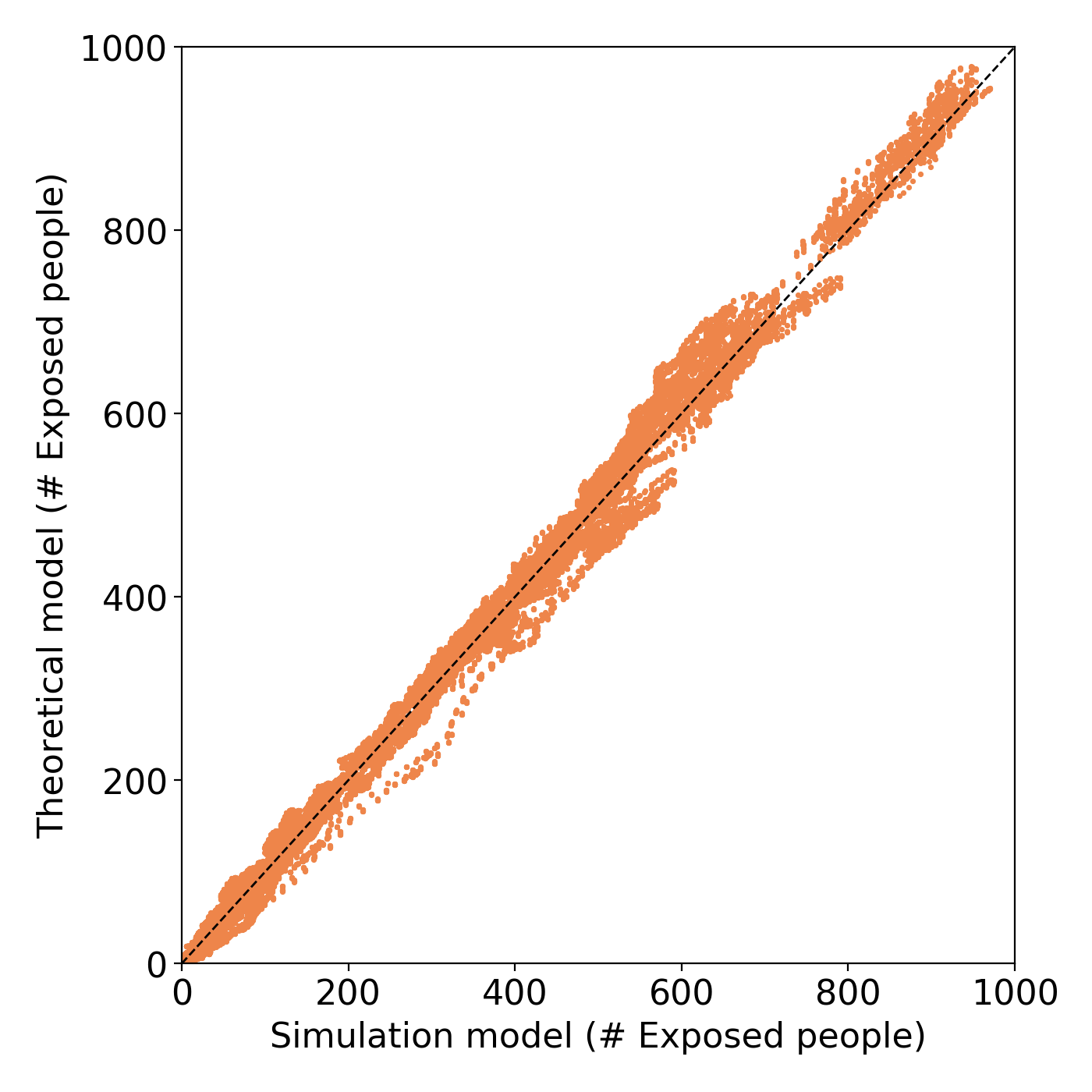}\label{fig_compare_E}}
\caption{Comparison between simulation model and calibrated theoretical model (100k sample passengers). One dot represents \# of infectious/exposed people in a specific time interval.}
\label{fig_compare_scatter}
\end{figure}

We first calibrate the theoretical model using the generated epidemic dynamics from the simulation model. 320 cases with different combination of parameters (Table \ref{tab_para_calibration}) for 100k sample passengers are simulated. These cases are fed into a regression model to obtain the parameters for the theoretical model.
Figure \ref{fig_compare_scatter} shows the comparison of the number of infectious and exposed people between the simulation and the calibrated theoretical model. We observe a high goodness-of-fit for the theoretical models, which implies the proposed theoretical framework can capture the epidemic spreading through PT and SA contacts. 

\begin{figure}[H]
\centering
\subfloat[Comparison on infectious people]{\includegraphics[width=0.8\textwidth]{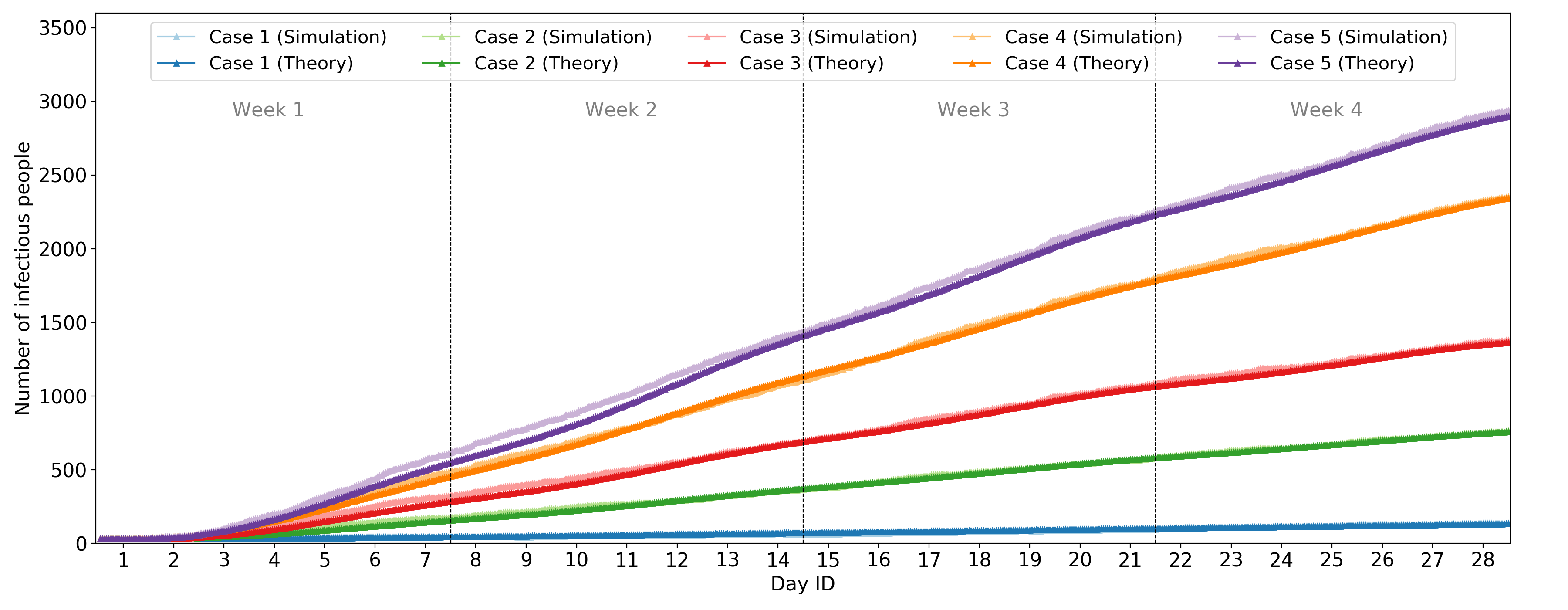}\label{fig_compare_I_time}}
\hfil
\subfloat[Comparison on exposed people]{\includegraphics[width=0.8\textwidth]{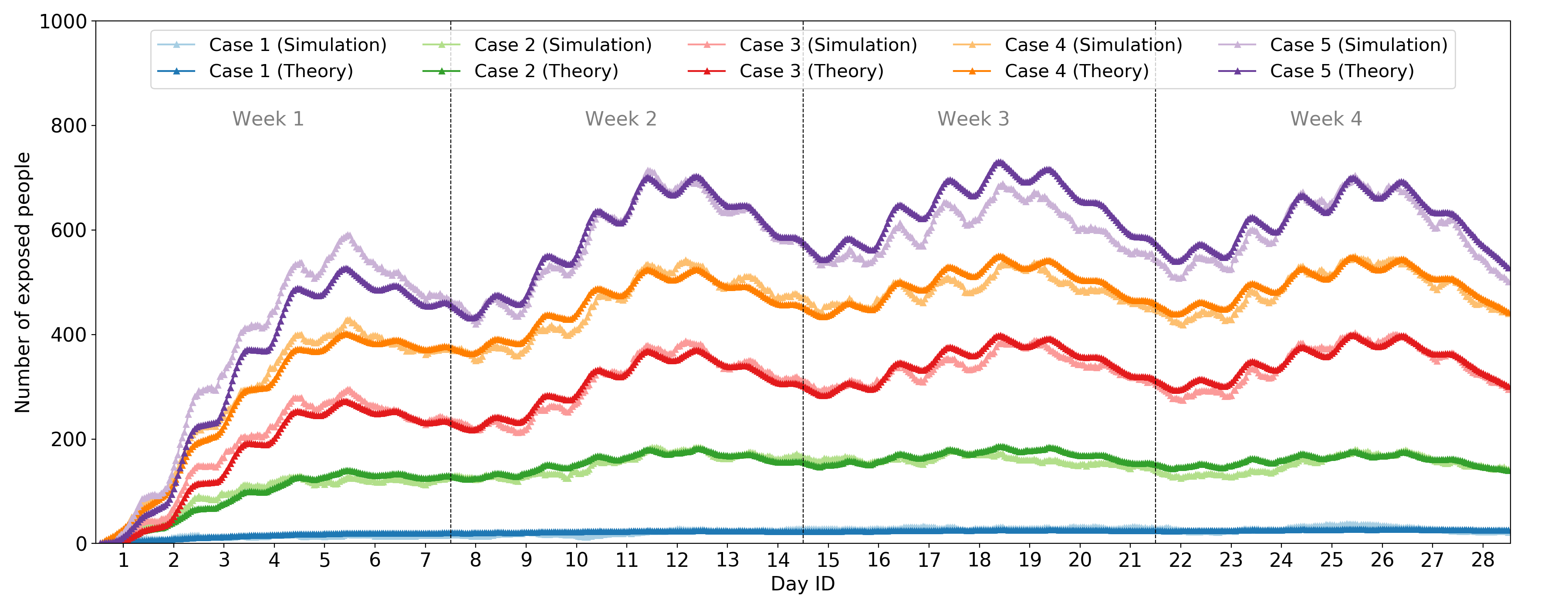}\label{fig_compare_E_time}}
\caption{Comparison of infectious/exposed people by time for 5 cases (100k sample passengers)}
\label{fig_compare_plot}
\end{figure}

Figure \ref{fig_compare_plot} shows the comparison of the number of infectious and exposed people by time for the five selected cases. Generally, the simulation and theoretical models show a similar number of infectious people over time. For the number of exposed people, the two models show similar dynamic fluctuations with only a slight difference for some periods.

After model calibration, we also calculate the scaling factors based on Eq. \ref{eq_scale}. The individual-level policy experiments (Section \ref{sec_impact_depature},\ref{close_bus},\ref{bus_cap}, and \ref{sec_impact_kcore}) are evaluated based on scaled 100k sample passengers.

\begin{table}[H]
\centering
\caption{Parameters value space for theoretical model calibration}
\begin{tabular}{lll}
\hline
Category                        & Parameters & Value space                                                                                                                                                \\ \hline
\multirow{5}{*}{Epedimic}       & $\beta_I$  & {[}$8.17 \times 10^{-4}$, $1 \times 10^{-4}, 1 \times 10^{-5}$, $1.5\times 10^{-3}$, $1 \times 10^{-6}${]}                                                 \\
                                & $\beta_E$  & 0.01$\beta_I$                                                                                                                                              \\
                                & $\gamma$   & $0.0104$                                                                                                                                                   \\
                                & $\mu_r$    & [$4.17 \times 10^{-4}$, $8 \times 10^{-4}$, $1 \times 10^{-3}$, $5\times 10^{-3}$] \\
                                & $\mu_d$    & $4.17 \times 10^{-4}$                                                                                                                                      \\ \hline
\multirow{2}{*}{Human mobility} & $\theta^{\ell}$ & {[}$1 \times 10^{-3}$, $1 \times 10^{-2}$, $1 \times 10^{-4}$, 0]                                                     \\
                                & $\theta^{g}$ & {[}$7.44 \times 10^{-8}$, $1 \times 10^{-7}$, $1 \times 10^{-9}$, 0{]}                                                                                        \\ \hline
\end{tabular}
\label{tab_para_calibration}
\end{table}

\subsection{Status quo analysis}

\begin{figure}[H]
\centering
\includegraphics[width= 0.8 \textwidth]{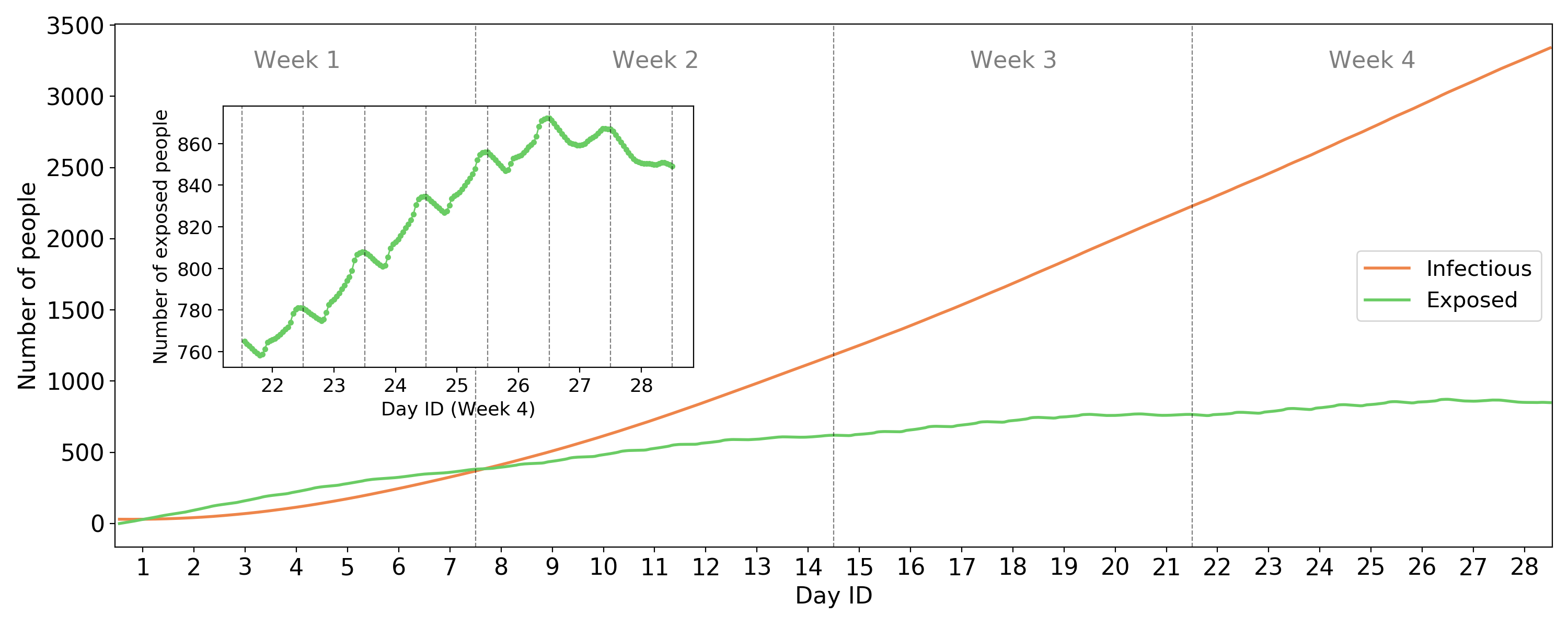}
\caption{Epidemic process in status quo scenario (whole population). The inset plot shows the zoom-in of the number of exposed people from Day 22 (Monday) to Day 28 (Sunday).}
\label{fig_status_quo}
\end{figure}

Based on the parameters in Table \ref{tab_para}, we evaluate the epidemic process in Singapore for all smart cardholders (4.7 million) using the calibrated theoretical model. Figure \ref{fig_status_quo} shows the dynamics of the number of infectious and exposed people. We randomly assign 30 initial infectious passengers in the system. Results show that if there are no control policies for the disease, the number of infectious people will increase to more than 3,000 (100 times the initial value) after four weeks. This is consistent with \cite{liu2020time}'s results on early human-to-human transmission of COVID-19 in Wuhan. The equivalent $R_0$ is $2.667$, which is consistent with many previous estimates:  2.0–3.3 \citep{majumder2020early}; 2.6 \citep{imai2020report}; 2.92 \citep{liu2020time}.

The inset plot shows the intra-day dynamics of the number of exposed people in week 4, from which we observe the high intensity of infections from 7:00 to 22:00. The sudden increases in morning and evening peaks for weekdays (Day 22-26) highlights the transmissibility through PT. During the weekends (Day 27 and 28), the number of exposed people shows a decreasing trend, which implies lower transmission rates on weekends. 

It is worth noting that, the increasing trends of infection is not as dramatic as many standard SEIR models with exponential rates. The reasons may be as below. First, this case study focuses on the early stage of the epidemic process (first month). According to the Taylor seris, $\exp(x) \approx 1+x$ when $x$ is close to zero. This explains the approximately linear trend at the early stage. Second, we consider the specific individual-level passengers contacts. According to Sun et al., (2013), passengers usually meet with repeated PT encounters. Therefore, the infected passengers may continue to meet with same passengers every day. These ``repeated encounters'' delayed the actual transmission of virus because the probability of meeting new people is smaller than the typical SEIR model where the population is assumed to be well-mixed.

\subsection{Evaluation of control policies}\label{sec_policy}

Motivated by current epidemic control strategies worldwide in PT systems, especially the control policies of COVID-19 exemplified in \ref{ww_strategies}, we can hardly find the general criteria to determine whether, when, and where to suspend the urban PT services. The individual-based PEN developed in this work contributes to better understanding of the spatiotemporal impacts of various PT operation strategies for epidemic control, to facilitate the decision-making process for PT operation adjustments.

\subsubsection{Impact of $\beta_I$ and $\mu$}\label{impact_of_beta_mu}
We first evaluate the impact of $\beta_I$ and $\mu$. $\beta_I$ is related to people's preventive behavior, such as wearing masks and sanitizing hands, which results in a decrease of $\beta_I$. $\mu$ is related to the hospital's medical behavior, such as increasing cure rate and developing vaccines, which leads to an increase in $\mu$. 

\begin{figure}[H]
\centering
\sbox{\measurebox}{%
\begin{minipage}[b]{.45\textwidth}
  \subfloat
    [Joint impact of $\beta_I$ and $\mu$]
    {\label{fig_joint_impact_beta_mu}\includegraphics[width=1.1\textwidth]{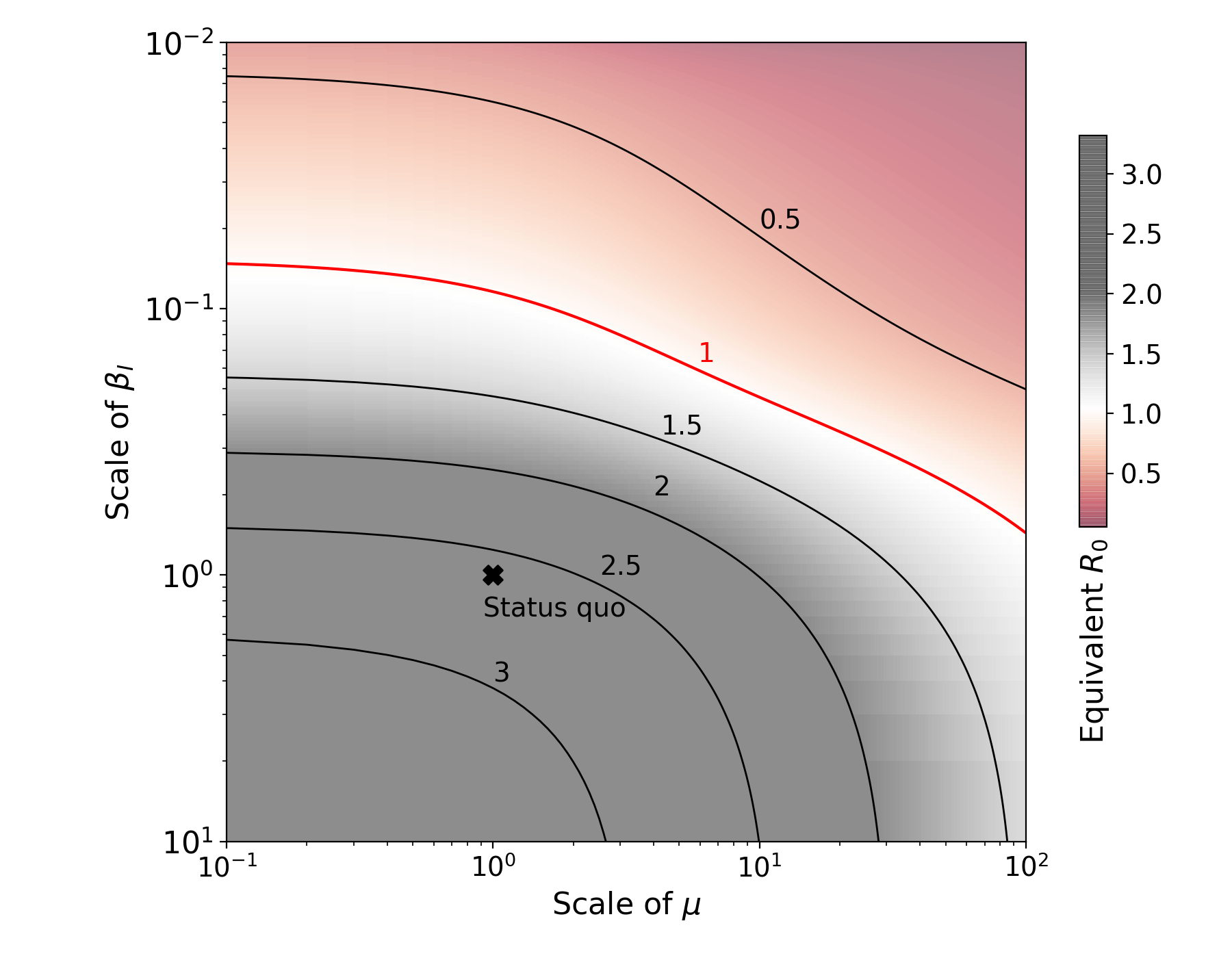}}
  \end{minipage}}
\usebox{\measurebox}\qquad
\begin{minipage}[b][\ht\measurebox][s]{.45\textwidth}
\centering
\subfloat
  [Impact of $\beta_I$]
  {\label{fig_impact_beta}\includegraphics[width=0.7\textwidth]{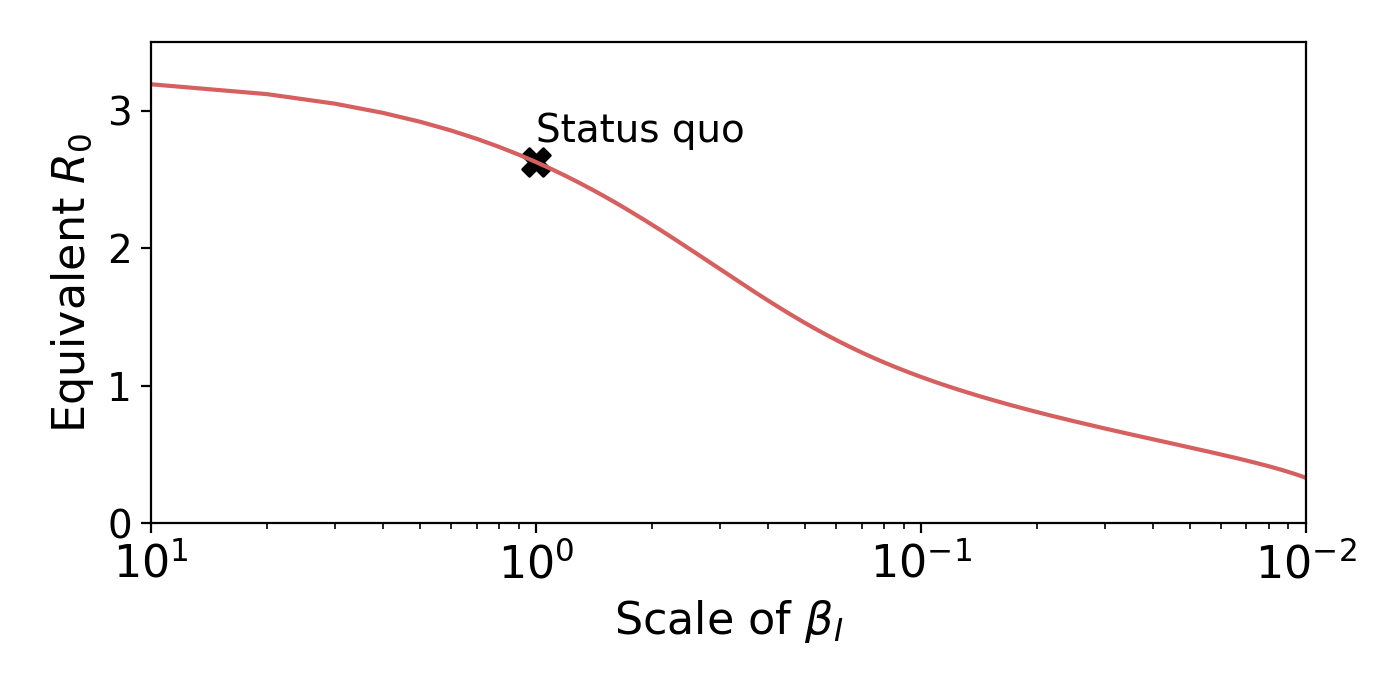}}
 \vfill
\subfloat
  [Impact of $\mu$]
  {\label{fig_impact_mu}\includegraphics[width=0.7\textwidth]{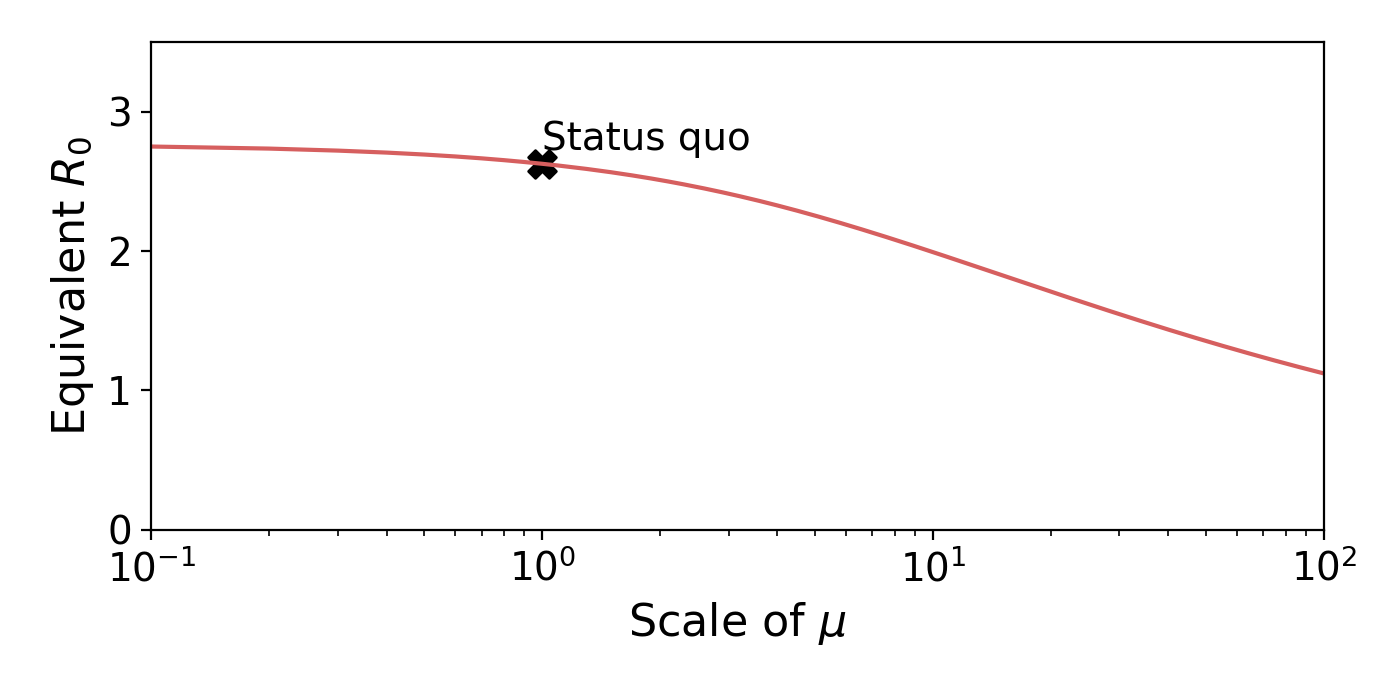}}
\end{minipage}
\caption{Impact of $\beta_I$ and $\mu$ on equivalent $R_0$. The scale of a parameter means multiples of the parameter with respect to the status quo scenario. For example, "scale of $\beta_I" = 10$ means 10 times $\beta_I$ of the status quo scenario (i.e., the new $\beta_I = 10 \times 8.17 \time 10^{-4} =  8.17 \time 10^{-3}$)}
\label{fig_impact_mu_beta}
\end{figure}

Figure \ref{fig_impact_mu_beta} shows the impact of $\beta_I$ and $\mu$ on equivalent $R_0$. $\beta_I$ is scaled from $10^{-2}$ to $10^1$ and $\mu$ is scaled from $10^{-1}$ to $10^2$. We fixed $\beta_E = 0.01\beta_I$ for all testing. Figure \ref{fig_impact_beta} shows that the epidemics would fade out (equivalent $R_0 < 1$) if transmissibility was reduced to less than $10^{-1}$ of current value. However, Figure \ref{fig_impact_mu} suggests that even if the cure rate is increased by 100 times, the epidemics would still happen, though the process would be lagged (with smaller $R_0$). This implies reducing transmissibility is more effective than increasing the cure rate for COVID-19. In Figure \ref{fig_joint_impact_beta_mu}, we show the joint impact of $\beta_I$ and $\mu$ and the critical bound of equivalent $R_0$. If $\beta_I$ was decreased to $32\%$ and $\mu$ was enlarged tenfold, the equivalent $R_0$ would be decreased to less than 1.  If the cost of controlling each parameter is given, this graph can help to optimize the controlling strategies with limited costs.  

\subsubsection{Impact of trip occurrence rate}\label{impact_of_trip_occ_rate}
\begin{figure}[H]
\centering
\includegraphics[width= 0.6 \textwidth]{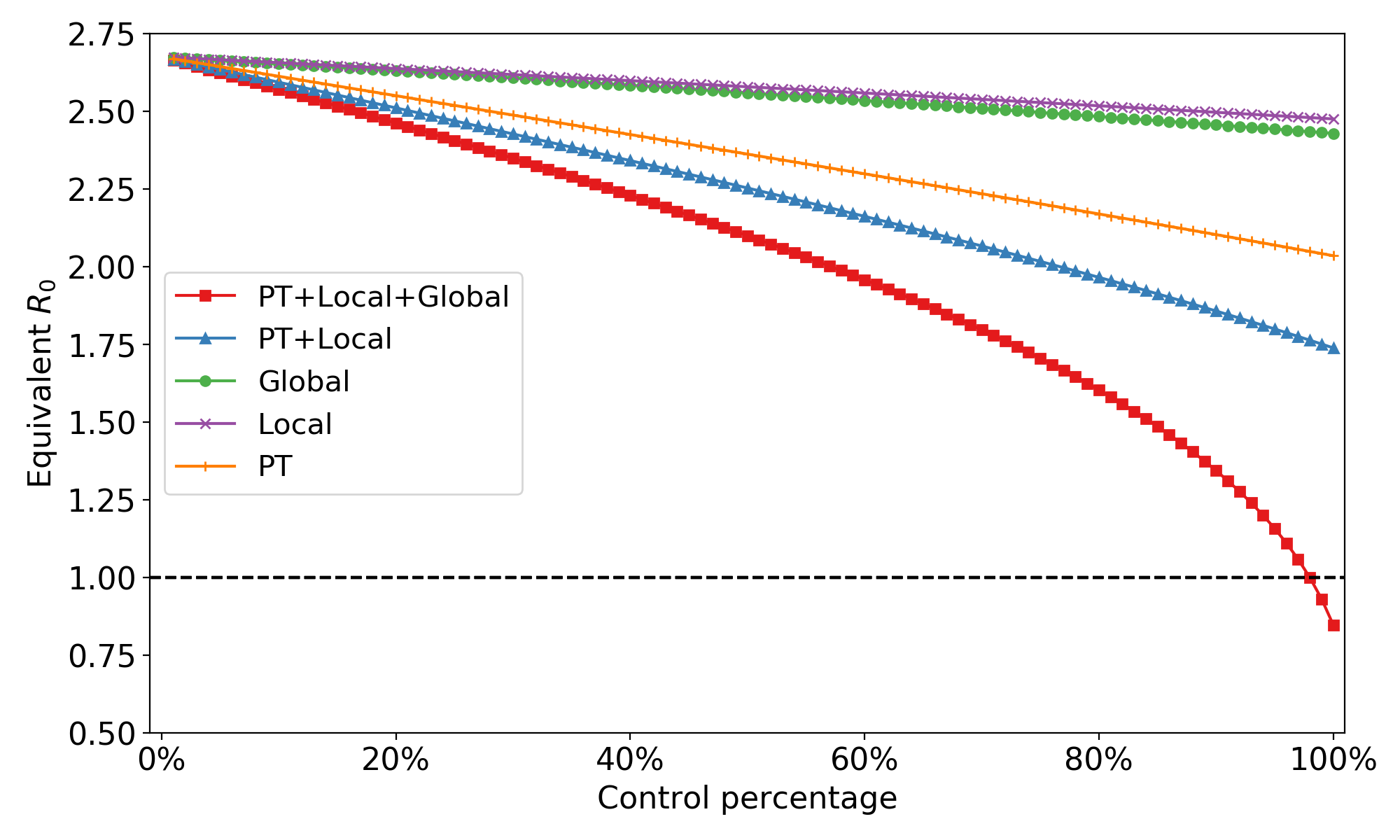}
\caption{Impact of controlling trip occurrence rate (whole population)}
\label{fig_control_trip}
\end{figure}

One of the typical control strategies for the epidemic is decreasing the trip occurrence rate in a city. At the average level, this is equivalent to reducing the total contact time, and total squared contact time\footnote{Let $x_t$, and $y_t$ be the average number of trips and the average contact time per trip at time interval $t$, respectively. Then the total contact time at $t$ is $x_t y_t$; total squared contact time is $c \times x_ty_t^2$, where $c$ is used to capture the difference of squared sum and sum of squares. If we reduce the number of trips to $\alpha x_t$ ($\alpha<1$), the total contact time will be reduced to $\alpha x_t y_t$. The total square contact time will be reduced to $\alpha c \times x_t y_t^2$. Hence, the trip occurrence rate is roughly linear with the total contact and squared contact time.}. Figure \ref{fig_control_trip} shows the impact of different control percentage of trips (i.e., reducing the percentage of total contact and squared contact time). We observe that controlling one trip (PT or local or global) cannot eliminate the epidemic. Based on current parameter settings, reducing PT trips contribute more to the control of the epidemic process than the other two. The impact of reducing trips to $R_0$ is generally linear unless the reduction percentage is sufficiently large. When all trips are reduced by more than 80\%, the reduction rate for $R_0$ starts to accelerate.  When all trips were decreased by 98\%, the spreading process would fade out, which implies travel control can only be effective at an extreme level. This is corresponding to \cite{wu2020nowcasting}'s statement that a 50\% reduction in inter-city mobility in Wuhan has a negligible effect on the COVID-19 epidemic dynamics.

\subsubsection{Impact of distributing departure time}\label{sec_impact_depature}
From a transportation perspective, the trip concentration in morning and evening peaks can strengthen people's interaction in PT. It is possible to reduce people's contact time on PT by distributing their departure time. The distributing process is as follows. Given the departure time flexibility constraints (e.g., $\pm 30$ min), we generate new departure time for passengers so that passengers with the same boarding stops are assigned to different buses as dispersed as possible. 

\begin{figure}[H] 
\centering
\includegraphics[width= 0.6 \textwidth]{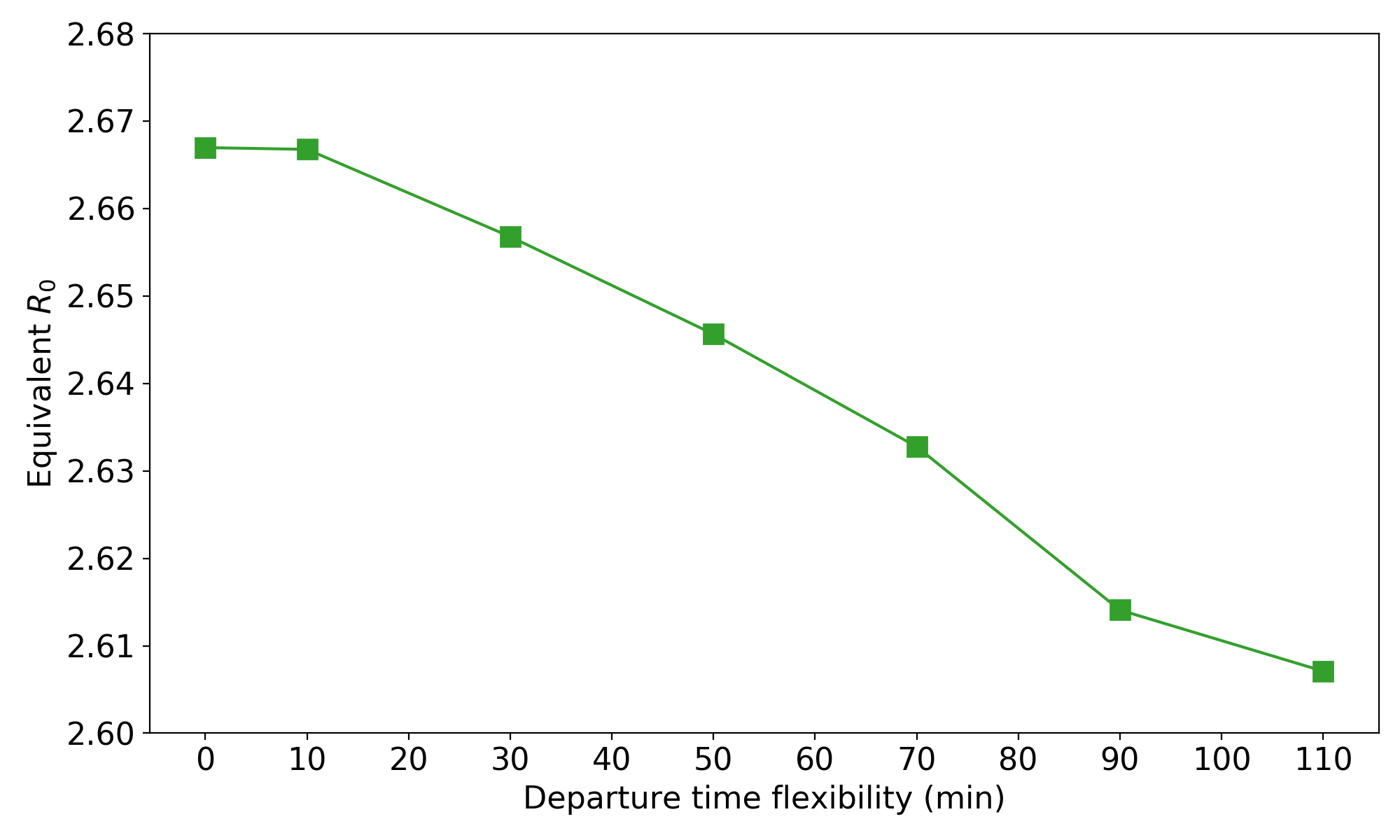}
\caption{Impact of spreading departure time}
\label{fig_impact_of_spreading}
\end{figure}

Figure \ref{fig_impact_of_spreading} shows the influence of distributing departure time with different flexibility (from 0 to $\pm 110$ min). We observe a decline in equivalent $R_0$ as the departure time flexibility increases. This is because higher flexibility allows more dispersed riding on the bus; thus, fewer contacted passengers. However, the effectiveness of distributing passengers is very limited. With $\pm 110$ min flexibility, there is only a 2.2\% decrease in $R_0$ (from 2.667 to 2.607).

\subsubsection{Impact of closing bus routes}\label{close_bus}
As what has been summarized in Section \ref{intro}, the closure of bus routes is an in-practice implemented strategy from the transportation side to reduce people's close-contacts during the outbreak of COVID-19. We assume that the suspension of bus service is a sign of travel restrictions on the corresponding bus routes. While keeping the SA contact network the same, we evaluate four different strategies of closing various percentages of bus routes (from 10\% to 90\%): a) Close from high demand to low demand routes (H-L). b) Close from low demand to high demand routes (L-H). c) Close by randomly picking bus routes (Random). d) Close by different local planning areas. We assume that passengers who originally take these closed bus routes will change to alternative routes if available; otherwise, they will cancel their trips. 

\begin{figure}[H]
\centering
\includegraphics[width= 0.6 \textwidth]{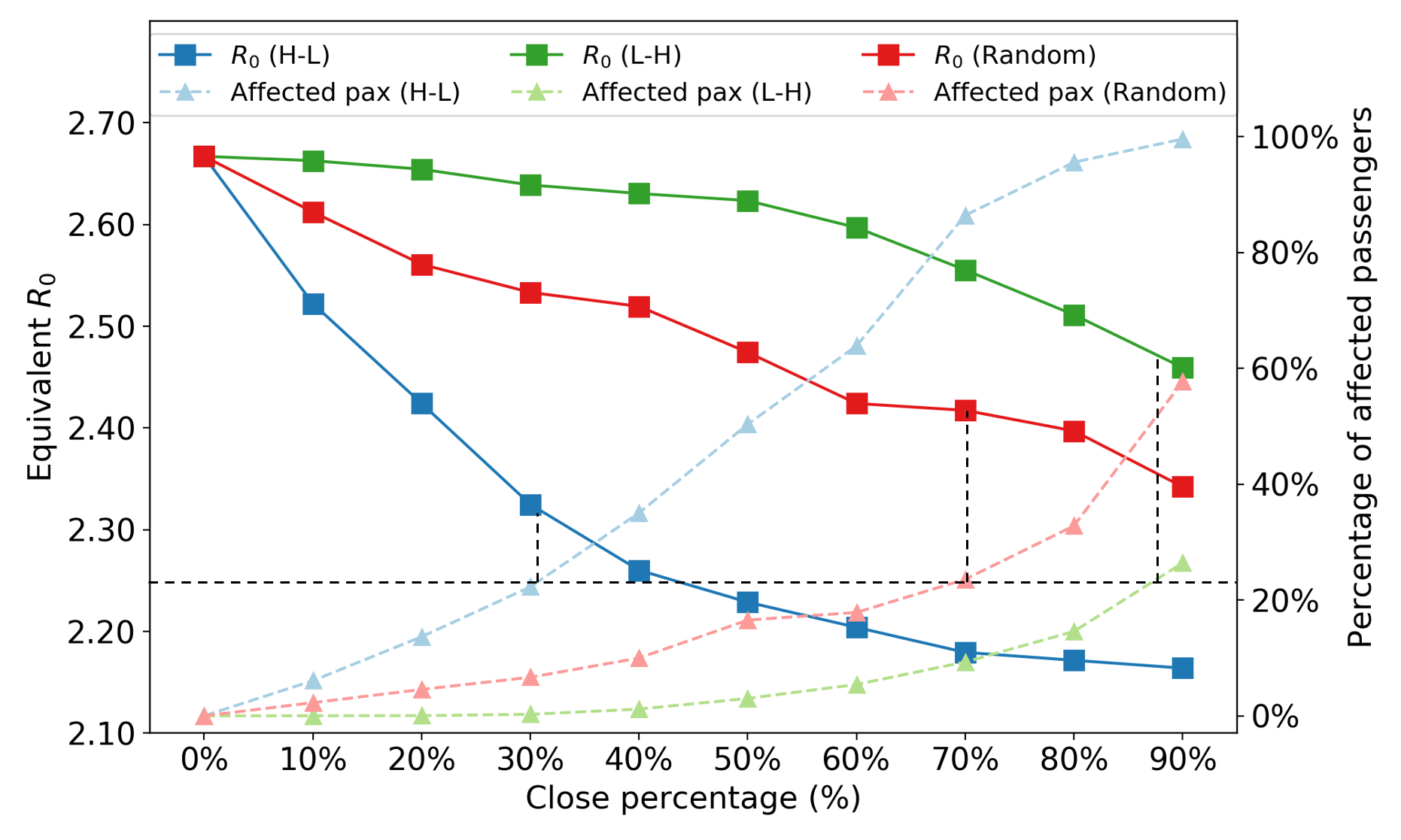}
\caption{Impact of closing bus routes. The black dashed lines are auxiliary lines to analyze the effectiveness of different polices under same number of affected people}
\label{fig_impact_of_close_bus}
\end{figure}

Figure \ref{fig_impact_of_close_bus} shows the effect of closing different percentage of bus routes. We observe that the H-L policy shows a convex curve on $R_0$ while the L-H policy shows a concave curve. Under the same percentage of closed bus routes, closing high demand routes brings a larger controlling effect on $R_0$. Closing the top 40\% high-demand bus routes can reduce the equivalent $R_0$ by 15.3\%. 

It is also important to look at how many passengers are affected by each strategy. The affected passengers are defined as those who cannot find alternative bus routes when the original routes are closed. As expected, for the same closing percentage of bus routes, the H-L strategy affected more passengers than the other two policies. However, we find that the H-L strategy is more effective in terms of $R_0$ reduction per affected passengers. If we fix the percentage of affected passengers at approximately 22\% (black dashed line), the H-L strategy can reduce $R_0$ to 2.25, while random and L-H strategies can only reduce $R_0$ to 2.41 and 2.50, respectively. This may be because passengers in high demand bus routes are more influential (e.g., with a higher degree in the PEN) in the system.  Therefore, to control the epidemic with fewer people getting affected, PT agencies should close buses from high to low demand routes. 

\begin{figure}[H]
\centering
\subfloat[Reduction in percentage of equivalent $R_0$]{\includegraphics[width=0.8\textwidth]{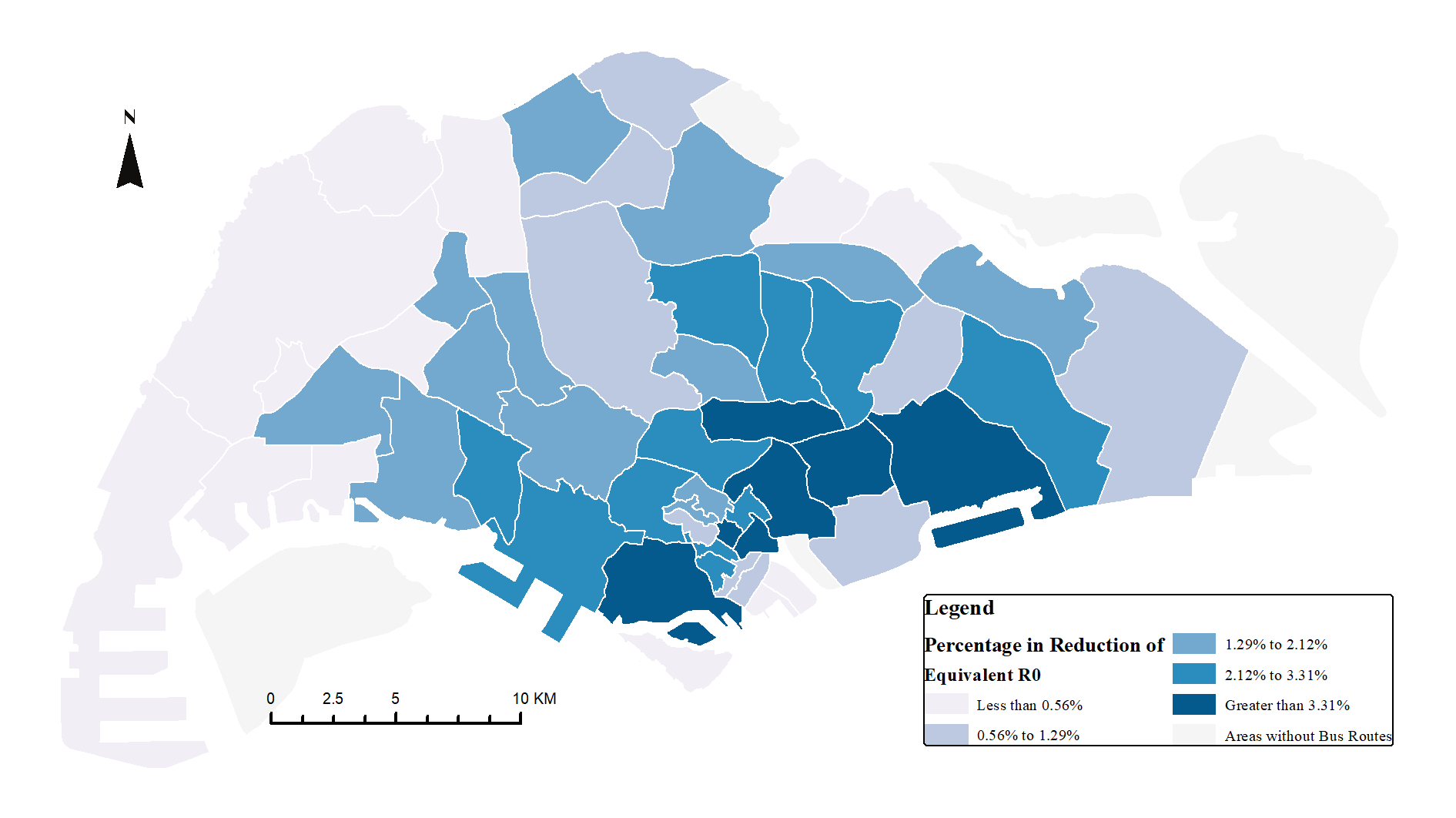}\label{fig_GIS_R0}}
\hfil
\subfloat[Percentage of affected PT passengers]{\includegraphics[width=0.8\textwidth]{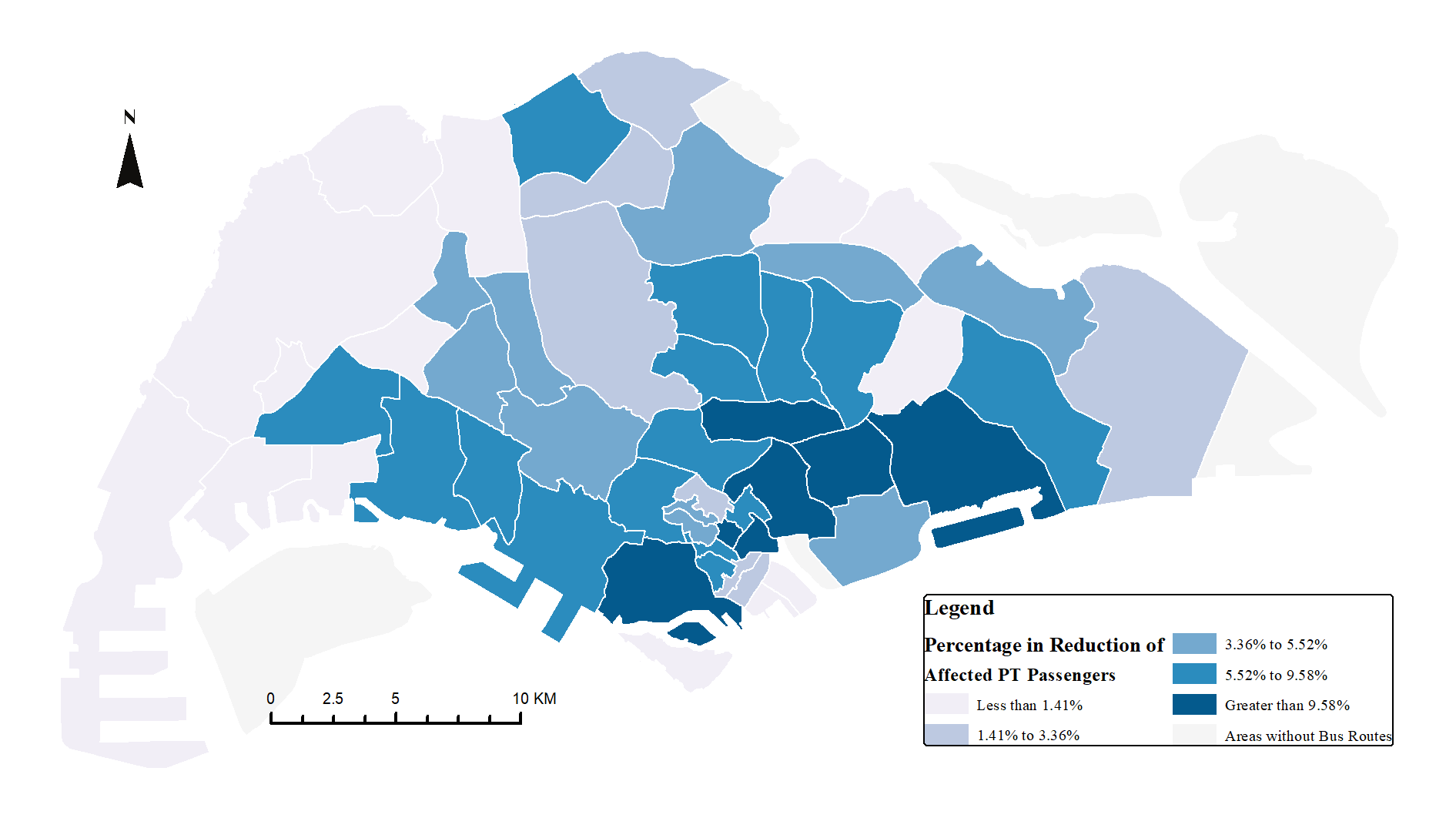}\label{fig_GIS_pass}}
\caption{Impacts of closing bus routes by planning areas}
\label{fig_impact_of_close_bus_by_pa}
\end{figure}

Figure \ref{fig_GIS_R0} shows the percentage of reduction in $R_0$ resulting from the closure of bus routes by planning areas, while Figure \ref{fig_GIS_pass} shows the corresponding percentage of affected PT passengers. Generally, the high reduction in $R_0$ is the result of a high number of affected passengers. Closing bus routes in the main business and residential areas in the Southern part of Singapore Island leads to higher controlling effects of epidemics than closing other areas. However, to minimize the impact on passengers' daily travel, PT agencies should first close bus routes in regions with relatively high $R_0$ reduction and a low number of affected passengers, such as core central business district (CBD) areas. Passengers who take buses crossing core CBD areas can easily find alternative routes. Thus, the concentrated demands at CBD areas can be distributed to other less crowded routes, which leads to a $R_0$ reduction and less number of affected passengers.

\subsubsection{Impact of limiting maximum bus load}\label{bus_cap}
\begin{figure}[H]
\centering
\includegraphics[width= 0.6 \textwidth]{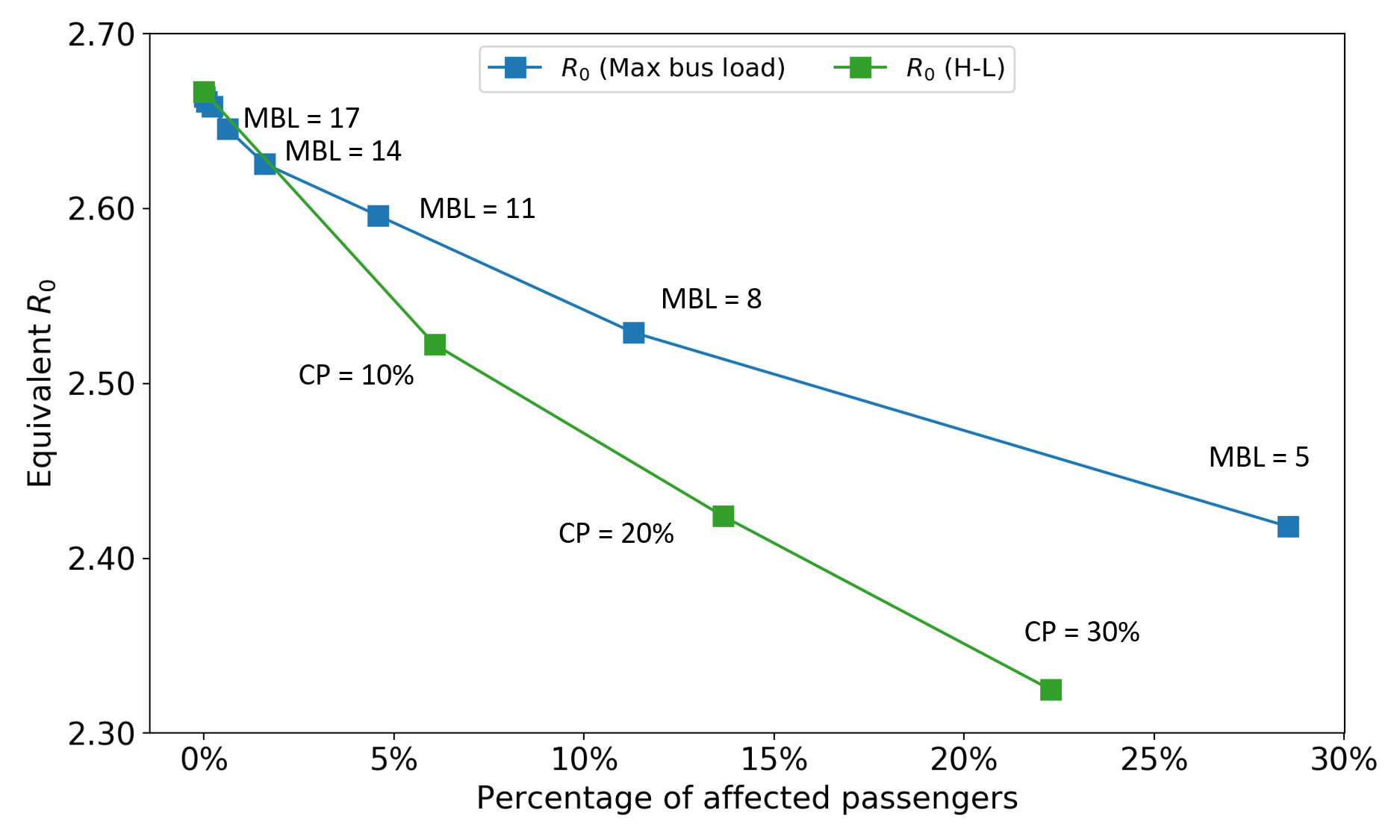}
\caption{Impact of limiting maximum bus load. MBL: maximum bus load. CP: close percentage of the H-L strategy}
\label{fig_impact_of_max_bus_load}
\end{figure}

Although closing bus routes can postpone the epidemic spreading, it also brings huge inconvenience to society, for example, decreasing people's accessibility to hospitals. A moderate alternative way is preserving the PT supplies but limiting the maximum bus load to reduce passengers' interaction. Figure \ref{fig_impact_of_max_bus_load} shows the impact of this policy. We test the maximum bus load from 26 to 5. Passengers who cannot board the bus due to this policy are assumed to cancel their trips (these passengers are called affected passengers). To compare with closing bus routes strategies, the x-axis is set as the percentage of affected passengers. Only the H-L strategy is plotted as it is the most effective one among all closing bus routes strategies. We find the policy of limiting the maximum bus load can only take effects when the available maximum bus load is small. With the same percentage of affected passengers, it is not as effective as closing bus routes from high demand to low demand. However, since this policy preserves the city's mobility capacity, it can be seen as a more moderate way to control epidemics compared with directly closing bus routes. 

\subsubsection{Impact of isolating critical passengers}\label{sec_impact_kcore}
Ideally, a more precise pandemic policy goes into the individual-level. The government could find out those influential passengers who are potential to spread viruses considerably and get them isolated at an early stage. Individual-based policies can outperform region-based or population-level policies in effectiveness and flexibility. However, due to the computational cost, it is hard to optimize the isolation options for each individual directly. Hence, we employed an isolation method based on $k$-core decomposition. Different from traditional degree-based methods, $k$-core method shows higher impacts on the dynamics of multi-particle systems \citep{kitsak2010identification,morone2019k,borge2012absence,yang2017small}.  

A $k$-core of a graph G is a maximal connected subgraph of G in which all vertices have a degree of at least $k$. Each vertex in a $k$-core is connected to at least $k$ other nodes (i.e., has a degree of at least $k$). A high $k$ number represents the highly concentrated structure of the local network, which indicates the most clustered part in the whole network. Nodes in a core with larger $k$ usually have larger degrees on average \citep{dorogovtsev2006k}. In the context of infectious diseases, if one node in a high $k$-core is infected, it has, on expectation, $>k^2$ times chances to spread the disease to other nodes in that core in one-time step compared to an arbitrary node (if the network is under scaling law) \citep{serrano2006percolation}. Therefore, we designed the policy by first limiting the nodes in high $k$-cores, which means limiting the more influential nodes. The population in a higher $k$-core is always lower than that in a lower $k$-core. Thus, by applying this policy, we can isolate a small portion of people to limit the spreading of the disease. 

\begin{figure}[H] 
\centering
\includegraphics[width= 0.6 \textwidth]{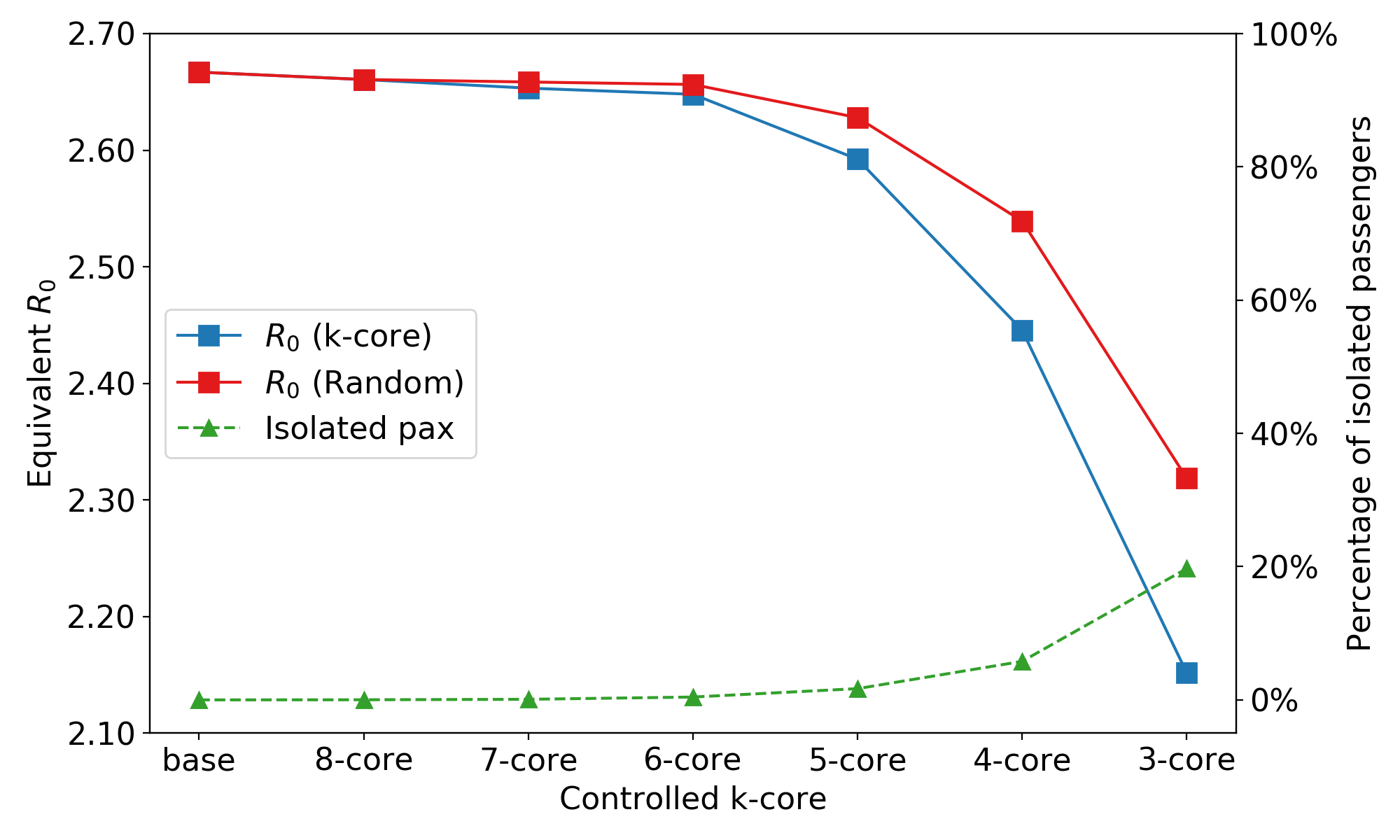}
\caption{Impact of isolating critical passengers. "base" and "$k$-core" scenarios indicate no isolation and isolation of people of core number $\geq k$ in the PEN, respectively.}
\label{fig_impact_of_isolate_pax}
\end{figure}

Figure \ref{fig_impact_of_isolate_pax} shows the impact of isolating passengers in different $k$-cores and the corresponding number of isolated passengers. For comparison purposes, we also evaluate a random policy. For each $k$-core, the random policy isolates the same number of randomly picked passengers in the system, which corresponds to implementing isolation at the population level. Since the number of passengers with a core number greater than 5 is low, the reduction in $R_0$ is not significant. However, isolating all $4$-core passengers, which accounts for 5\% of the whole population, the equivalent $R_0$ is reduced from 2.667 to 2.446 (8.3\%), which shows higher effectiveness than any other region-based or route-based policies in Section \ref{close_bus}. We also observe that the $k$-core isolating method can outperform the benchmark random isolating method.  

\section{Conclusion and discussion} \label{conclusion}

This paper proposed a general time-varying weighted PEN to model the spreading of infectious diseases over the PT system. The social-activity contacts at both local and global level are also considered. The network is constructed using smart card data as an undirected graph, in which a node refers to a PT passenger; an edge refers to a pair of passengers staying in the same vehicle; the weight of an edge captures the CD. We employ the SEIR diagram---a general diagram to model the influenza-like disease---to model the disease dynamics using the recent global outbreak of COVID-19 as a case study. A scalable and lightweight theoretical framework is derived to capture the time-varying and heterogeneous network structures, which enables to solve the problem at the whole population level with low computational costs.

We use the PT smart card data from Singapore as a proximity to understand the general spatiotemporal dynamics of epidemic spreading over the PT networks in one month. The status-quo analysis shows that the COVID-19 infected population is expected to increase by 100 times of their initial value by the end of the month without any disease control enforcement. A series of disease control and prevention scenarios are envisioned from both public health policy and PT operations sides. From the public health side, the model sheds light on people's preventative behavior. Wearing face masks and sanitizing hands, are considered the most effective measures to control the spreading of the epidemic; however, an increased cure rate can only postpone the outbreak of the disease. From the perspective of PT operation adjustments, several policies are evaluated, including reducing trip occurrences, enlarging departure time flexibility, closing bus routes, limiting maximum bus loads, and isolating critical passengers. In general, the control of the epidemic process starts to take effect, with over 80\% of all trips being canceled. The equivalent $R_0$ can be reduced to 1 if over 98\% of trips are banned. In terms of bus operation policies, distributing departure times and limiting maximum bus loads can slightly decelerate the spreading process. Closing high-demand bus routes, especially in the main business areas, is more effective than the closure of low-demand bus routes. The most effective approach is isolating influential passengers at the early stage, in which the epidemic process can be significantly reduced with a small proportion of people being affected. 

\subsection{Policy implications}
Many policy implications can be derived from the case study. On the public health side, the government should encourage people to take preventative behaviors, such as wearing face masks and sanitizing hands, to reduce the transmission probability. The travel restriction policy can take effect---with an equivalent $R_0$ less than 1---only at the extreme travel, such as in Hubei Province, China, where all travels were banned during the outbreaks of COVID-19 in Feb 2020. 

On the PT operation side, according to our models, partial closure of bus routes and limiting the maximum bus load can postpone the spreading of epidemics. The most effective way is closing bus routes with high demands, especially those crossing the CBD areas. In practice, a (partially) shutdown of PT services is a serious decision for authorities, many related issues, such as equity and accessibility, should be considered to determine how to design the suspension of PT services in the pandemics. 

The influential passengers with large core numbers can be identified based on smart data and they should be suggested to isolate themselves or reduce travels for prevention purpose. In addition, all entities should cancel events with a high number of participants to avoid generating large $k$-core contact networks. However, the identification and privacy projection of influential passengers are the two ends of the balance for epidemic control. The government should also carefully achieve the trade-off between privacy and urgency in fighting against the spreading of epidemics.

\subsection{Limitations and future works}
Several limitations of this study are as follows: 1) Some parameters of the model (e.g., $\theta^{\ell}$, $\theta^{g}$) are determined by the authors' assumptions, which hurts the credibility of the results. Although we end up with a reasonable $R_0$, suggesting the values of these parameters are reasonable, more parameter calibration jobs should be done in the future. 2) Policy evaluations are based on some ideal assumptions. In the real world, many unexpected results can happen. For example, closing bus routes may decrease people's accessibility to the hospital, thus, decrease the cure rate. The government should think cautiously from multiple perspectives before applying any control strategies. 3) This study did not model the contacts of passengers at trains due to difficulties in identifying the vehicles that passengers belong to. Future research can incorporate a transit assignment model \citep[\textit{e.g.,}][]{zhu2017probabilistic,Mo2020Capacity} to infer passengers' boarding trains and construct the PEN by trains. Meanwhile, due to the large space in a train, the variation of transmission probability due to passengers' spatial distribution should also be captured. 4) Other modes of transmission than contact transmission are not considered (e.g., infectious passengers contaminating surfaces), which may result in under-estimation of transmissibility of PT systems. 5) Population infected and mobility heterogeneity is neglected in this study. In reality, the infectious probability may depend on age, gender and health conditions. And people with different occupations may have various mobility characteristics (e.g., in Singapore, working pass holders and maids may not use PT frequently). The demographics distribution can be incorporated in the future. 

Future works include the following: 1) Elaborate on the SA contacts based on other data sources (e.g., mobile phone data) and extend $\mathcal{N}$ to the whole population. Due to a lack of data, the contacts of social activities are simplified and $\mathcal{N}$ is assumed to be PT users in this paper. Though PT users in Singapore account for 84\% of the population, future research can combine different data sources to model the SA contacts for the whole population in more details \citep{wang2018inferring}. Given the challenges in collecting real-world trajectory data at the population level, future research can also model the epidemic spreading based on synthetic trajectory data. The synthetic data can be generated by combining household travel surveys and mobile phone data \citep{muller2020using}. And the data can include individual's socio-demographic information to model infection heterogeneity. 2) Incorporate spatial effect and model transmission probability more finely. The current transmission probability between two individuals only depends on the contact duration. The contact distance, passenger density, and distribution on a vehicle can be considered in future research. 
3) Conduct case studies in cities with COVID-19 outbreaks (e.g., Wuhan, New York City) to validate the model. These case studies can calibrate the model based on ground truth data, quantify the contribution of disease transmission by PT systems, predict the epidemic spreading, and evaluate the effects of different policies. 4) Incorporate the time-varying epidemic and mobility parameters to better predict the reality. Although this study does not attempt to predict and reproduce the COVID-19 spreading, the proposed model can potentially better fit the epidemic process, given the more fine-grained framework. However, the complexity in the real-world lies on the time-varying mobility and epidemic patterns. Future research can make the epidemic parameters ($\Theta$) as time-dependent ($\Theta (t)$), instead of constant, to better fit the reality. 5) Propose a generative contact network construction model. One of the computational bottle network in this paper is to construct individual-level contact network using smart card data. Future study can propose a generative network construction model with compressed demand and supply information. This can be used to easily conduct PT-based policy evaluation and scenario testing without heavily constructing actual contact network.

\section{Author Statement }
The authors confirm contribution to the paper as follows: study conception: B. Mo, K. Feng, Y. Shen; experiments design: B. Mo, K. Feng, J. Zhao, D. Li, C. Tam, Y. Yin; data collection: B. Mo, Y. Shen; analysis and interpretation of results: B. Mo, K. Feng; draft manuscript preparation: B. Mo, K. Feng Y. Shen. All authors reviewed the results and approved the final version of the manuscript.

\section{Acknowledgements}
The research is sponsored by the Natural Science Foundation of China (71901164) and the Natural Science Foundation of Shanghai (19ZR1460700). The study is also supported by the National Research Foundation, the Prime Minister’s Office of Singapore under the CREATE programme, and the SMART’s Future Urban Mobility IRG. Yu Shen is sponsored by Shanghai Pujiang Program (2019PJC107). Yafeng Yin would like to acknowledge partial support from National Science Foundation, U.S. (CMMI-1854684). The authors thank Prof. Caroline Buckee from Harvard T.H. Chan School of Public Health for her comments on this work.

\appendix

\section{Worldwide COVID-19 control policies in urban PT systems}\label{ww_strategies}
In practice, since the outbreak of COVID-19 in late January 2020, a variety of epidemic control strategies in PT systems, such as the requirement of PT riders to wear face masks, the sterilization of bus and metro carriages, the adjustment of PT operation schedules, the closure of bus routes, etc., have been implemented during the outbreak of COVID-19. 

In China, the requirement of wearing face masks in PT systems has been successively implemented in many provinces since late January of 2020. In addition, a variety of PT operation control strategies have also been enforced in many cities. In cities of Hubei Province, especially in Wuhan, along with the lockdown policies to control the spreading of COVID-19, almost all PT services have been shut down since Jan. 23rd and 24th, whereas the patients with severe symptoms are transported by ambulance. After the lockdown and travel restrictions of Hubei Province, the PT operation adjustment strategies implemented in other Chinese cities were largely diverse. Different from Wuxi with only 8\% of arterial bus routes kept running, in Nanjiang, another major city of Jiangsu Province, the PT services were still in operation but with shortened operation hours and dispatching frequencies. In Shanghai, both inter-provincial PT services and the bus services between rural districts like Qingpu and Jinshan were closed from Jan. 27th, but most of the urban PT services remained in operation by limiting the maximum passenger loads. 

Outside China, in Italy, where the reported COVID-19 cases dramatically increased in March 2020, the suspension of PT has been officially proposed in the Lombardy area, where the urban PT service (such as in Milan) was still in normal operation. In London, The Transport for London (TfL) is running reduced service across the network in London, closing up to 40 stations. No services of Waterloo and City line was provided since March 30, 2020. Many US cities have seen reduced services, including Boston and Washington D.C., In cities of other countries with fewer COVID-19 reported cases (such as Singapore), the PT services were kept normal with more frequent sterilization. 

\section{Impact of neglecting the MRT system}\label{append_train_demand_size}
In the case study, we neglect the passengers' interaction in the MRT system since the smart card data cannot provide the direct contacts of passengers in the train. To capture the contact at the train, an detailed transit assignment model is needed to identify which train the passengers boarded given the origin, destination, and train schedule information. This is beyond the scope of this study. However, we can analyze the demand patterns of the MRT system and provide a simplified estimation for the impact of MRT system in the epidemic spreading. 

\begin{figure}[H]
\centering
\includegraphics[width= 1 \textwidth]{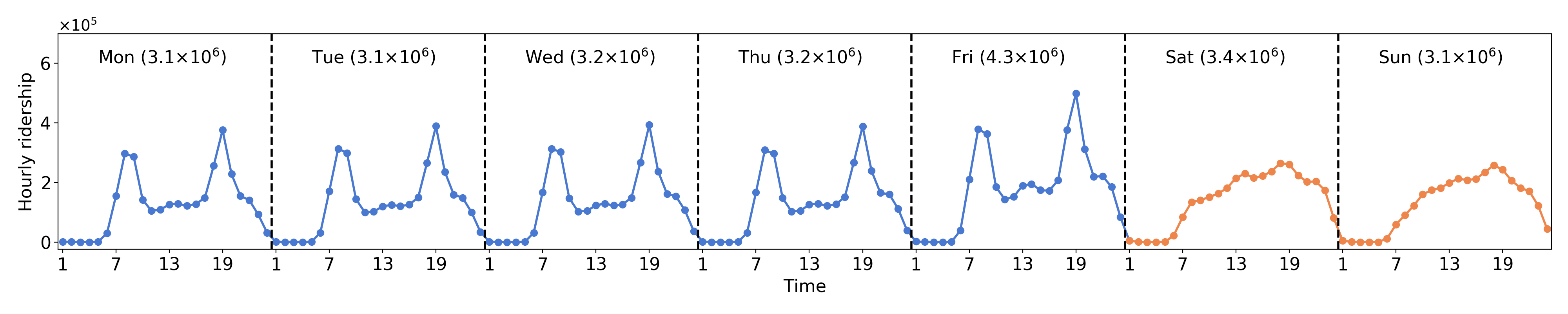}
\caption{MRT demand distribution. Numbers in x-axis represent the hour ID (e.g., 1 represents 0:00-1:00 AM). The number in brackets at the top of each sub-graph indicates the total daily ridership}
\label{fig_demand_MRT}
\end{figure}

\begin{figure}[H]
\centering
\subfloat[MRT Trip duration distribution]{\includegraphics[width=0.4\textwidth]{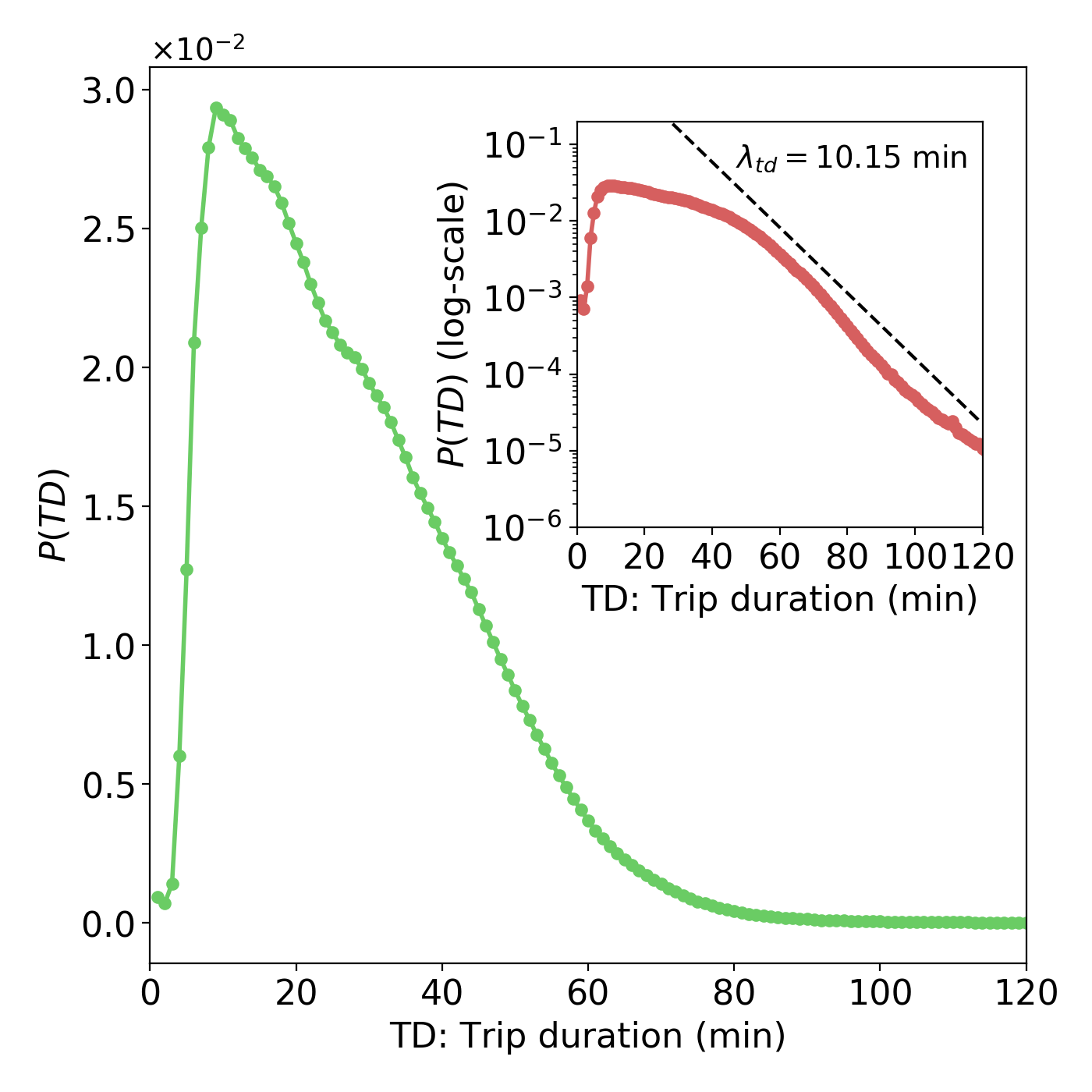}\label{fig_p_td_MRT}}
\hfil
\subfloat[MRT Trip frequency distribution]{\includegraphics[width=0.4\textwidth]{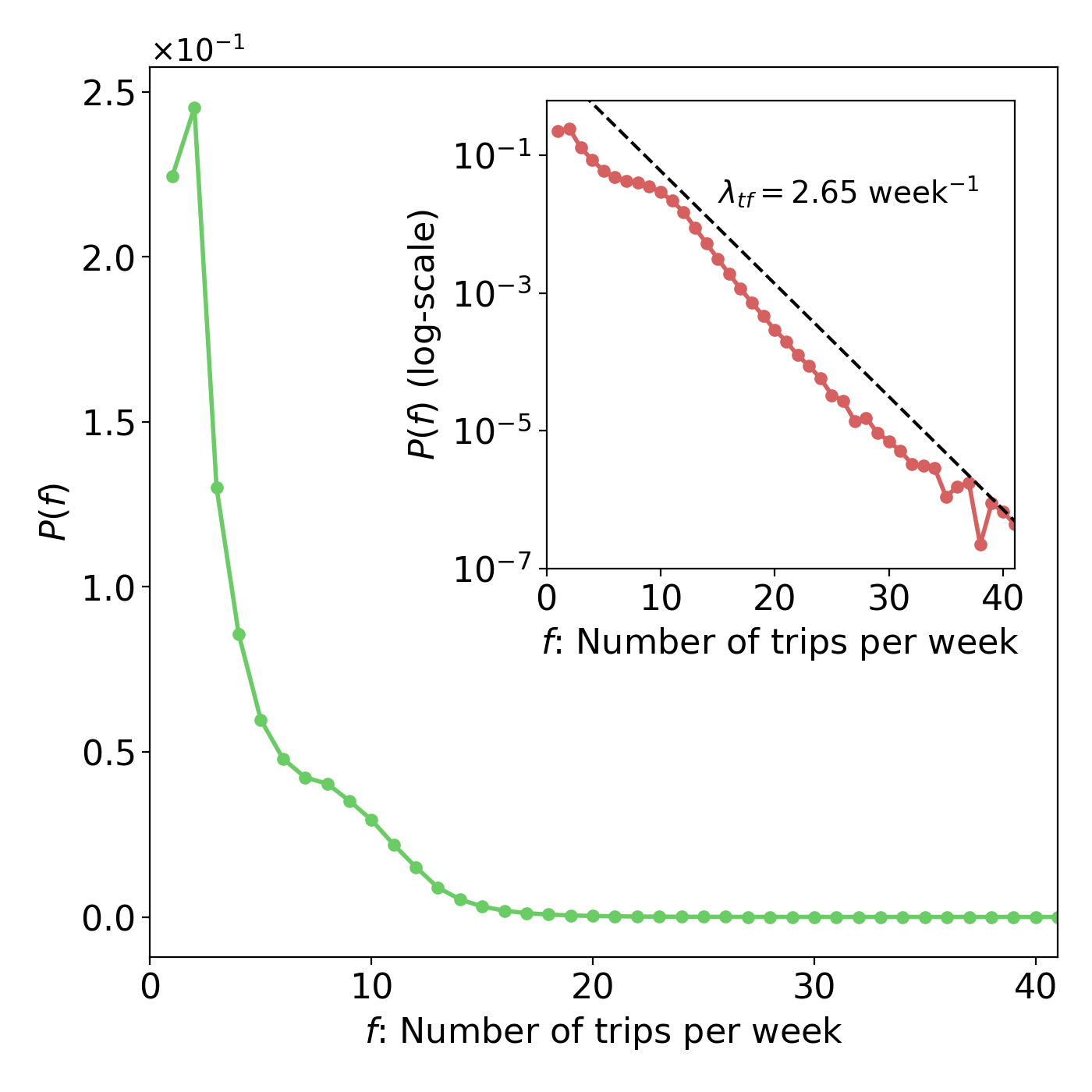}\label{fig_p_tf_MRT}}
\caption{Distribution of MRT trip duration and trip frequency (data of 4 weeks, the inset plots are in semi-log scale)}
\label{fig_trip_anal_MRT}
\end{figure}

Figure \ref{fig_demand_MRT} shows the demand distribution of MRT trips. Compared to the bus trips (Figure \ref{fig_demand}), the temporal distribution patterns are very similar while the average demand of MRT is around 67\% of the bus demand. For the MRT trip duration and frequency distributions (Figure \ref{fig_trip_anal_MRT}), both of them follow the exponential tail. The average trip duration for MRT is 26.88 min (14.55 min for bus), and the average trip frequency is 4.26 per week (6.34 per week for bus).

According to Eq. \ref{eq_beta_E_ijt} and \ref{eq_beta_I_ijt}, the infection probability between two individuals are proportional to the contact duration (i.e., $h( w,\beta)$ is a linear function). Based on the theoretical solution method (Eq. \ref{eq_E3} and \ref{eq_I3}), the epidemic spreading process can be quantified by the total contact duration ($\sum_{i,j\in \mathcal{N}} \beta_{i,j,t}^{X}$) and total squared contact duration ($\sum_{i,j\in \mathcal{N}} (\beta_{i,j,t}^{X})^2$). As the demand patterns of MRT and bus trips are similar, we assume the contact time of passengers in the MRT is proportional to the trip duration (i.e., the average contact duration of MRT is 1.85 times that of bus). Therefore, the total contact duration in MRT is around $0.67\times 1.85 = 1.24$ times that of bus, and the total squared contact duration in MRT is around $0.67\times 1.85^2 = 2.29$ times that of bus. 

To estimate the impact of ignorance MRT, we add the inferred passengers' contact duration in MRT to the total contact and squared contact duration, and calculate the new $R_0$ for the status quo scenario. We test the add-MRT scenario. The estimated $R_0$ is 3.192 (compared to 2.667 for the scenario without considering MRT). Therefore, the omission of MRT reduces the $R_0$ by around 16.3\% based on the simplified estimation. Future research should capture passengers' interaction on MRT to better model the epidemic process.

\bibliography{mybibfile}


\end{document}